\documentclass[fleqn,usenatbib]{mnras}

\usepackage{newtxtext,newtxmath}
\usepackage{ulem}

\usepackage[T1]{fontenc}
\usepackage{ae,aecompl}
\usepackage{color}
\usepackage{subfig}

\usepackage{graphicx}	
\usepackage{amsmath}	
\usepackage{longtable}
\usepackage{lscape}

\defcitealias{Parikh2018}{P18}
\defcitealias{Parikh2019}{P19}
\defcitealias{Johansson2012}{J12}
\defcitealias{ThomasD2011b}{TMJ}

\title[Spatially-resolved stellar population properties]{SDSS-IV MaNGA: Radial Gradients in Stellar Population Properties of Early-Type and Late-Type Galaxies}

\author[T. Parikh et al.]{
Taniya Parikh$^{1, 2}$\thanks{E-mail: tparikh@mpe.mpg.de},
Daniel Thomas$^{2, 3}$,
Claudia Maraston$^{2}$,
Kyle B. Westfall$^{4}$,
\newauthor{Brett H. Andrews$^{5}$},
Nicholas Fraser Boardman$^{6}$,
Niv Drory$^{7}$,
Grecco Oyarzun$^{8}$
\\
$^{1}$Max-Planck-Institut f\"ur extraterrestrische Physik, Giessenbachstrasse 1, 85748 Garching bei M\"unchen, Germany\\
$^{2}$Institute of Cosmology and Gravitation, University of Portsmouth, 1-8 Burnaby Road, Portsmouth PO1 3FX, UK\\
$^{3}$School of Mathematics and Physics, University of Portsmouth, Lion Gate Building, Portsmouth, PO1 3HF, UK\\
$^{4}$University of California Observatories, University of California, Santa Cruz, 1156 High St., Santa Cruz, CA 95064, USA\\
$^{5}$PITT PACC, Department of Physics and Astronomy, University of Pittsburgh, Pittsburgh, PA 15260, USA\\
$^{6}$Department of Physics \& Astronomy, University of Utah, Salt Lake City, UT, 84112, USA\\
$^{7}$McDonald Observatory, The University of Texas at Austin, 1 University Station, Austin, TX 78712, USA\\
$^{8}$Astronomy Department, University of California, Santa Cruz, 1156 High St., Santa Cruz, CA 95064, USA\\
}

\date{Accepted XXX. Received YYY; in original form ZZZ}

\pubyear{2021}

\begin{document}
\label{firstpage}
\pagerange{\pageref{firstpage}--\pageref{lastpage}}
\maketitle

\begin{abstract}
We derive ages, metallicities, and individual element abundances of early- and  late-type galaxies (ETGs and LTGs) out to 1.5~R$_e$. We study a large sample of $1900$ galaxies spanning $8.6 - 11.3\;\log M/M_{\odot}$ in stellar mass, through key absorption features in stacked spectra from the SDSS-IV/MaNGA survey. We use mock galaxy spectra with extended star formation histories to validate our method for LTGs and use corrections to convert the derived ages into luminosity- and mass-weighted quantities. We find flat age and negative metallicity gradients for ETGs and negative age and negative metallicity gradients for LTGs. Age gradients in LTGs steepen with increasing galaxy mass, from $-0.05\pm0.11$~log Gyr/R$_e$ for the lowest mass galaxies to $-0.82\pm0.08$~log Gyr/R$_e$ for the highest mass ones. This strong gradient-mass relation has a slope of $-0.70\pm0.18$. Comparing local age and metallicity gradients with the velocity dispersion $\sigma$ within galaxies against the global relation with $\sigma$ shows that internal processes regulate metallicity in ETGs but not age, and vice versa for LTGs. We further find that metallicity gradients with respect to local $\sigma$ show a much stronger dependence on galaxy mass than radial metallicity gradients. Both galaxy types display flat [C/Fe] and [Mg/Fe], and negative [Na/Fe] gradients, whereas only LTGs display gradients in [Ca/Fe] and [Ti/Fe]. ETGs have increasingly steep [Na/Fe] gradients with local $\sigma$ reaching $6.50\pm0.78$~dex/log km/s for the highest masses. [Na/Fe] ratios are correlated with metallicity for both galaxy types across the entire mass range in our sample, providing support for metallicity dependent supernova yields.
\end{abstract}




\section{Introduction}
\label{sec:intro}
Stellar population analysis is a powerful tool for extracting physical parameters from galaxy spectra, often referred to as extragalactic archeology. Results can be obtained via modelling the full spectrum \citep{Fernandes2005, Ocvirk2006, Koleva2009, Conroy2014, Cappellari2017, Goddard2017, Conroy2017b, Wilkinson2017} or absorption index measurements \citep{Trager2000, Proctor2002, Thomas2005, Shiavon2007, Thomas2010, Johansson2012} which respond to a combination of stellar population parameters. Through this we have learnt that more massive galaxies contain older, more metal rich populations \citep{Kuntschner2000, Thomas2005} and appear to have formed their stars earliest \citep{Trager1998, Thomas2005, Bernardi2006, Clemens2006}.

High resolution data and state-of-the-art stellar population models have given way to detailed studies of abundance patterns in early type galaxies. A wealth of information can be obtained by modelling the abundances of individual elements. Furthermore, spatially resolved spectroscopy has led to studies of these properties as a function of galaxy radius. Abundance gradients within galaxies provide additional information on the processes which regulate the growth and assembly of galaxies. Through this we can learn about quenching, mergers, and inside-out or outside-in formation. Stellar yields outline the amount of each element produced by stars of different masses, and is a basic parameter for chemical evolution models which can be compared to observations in order to place constraints. Gradients also provide information on internal processes such as stellar migration, inflow of pristine gas into galaxies, outflows of enriched material via supernovae and stellar winds into the ISM \citep[see][for a review]{Maiolino2019}

The IMF of galaxies is a hot topic in recent literature. Studies are increasingly finding evidence for the IMF becoming more bottom heavy with increasing galaxy mass for ETGs \citep{Cappellari2012, Conroy2012b, Spiniello2012, Ferreras2013, LaBarbera2013, Lyubenova2016}. Additionally, measuring the IMF within massive ETGs points towards a bottom-heavy IMF in galaxy centres, and a Kroupa IMF at larger radii \citep{MartinNavarro2015a, LaBarbera2016, vanDokkum2016, LaBarbera2017, Parikh2018, Dominquez2019, Bernardi2019}. \citet{Vaughan2017} find that their data cannot conclusively rule out IMF gradients, but that trends can also be explained by abundance variations, while \citet{Alton2017, Alton2018} also find no IMF gradients for a small sample of galaxies. Furthermore, constraints from four massive lensed galaxies, with velocity dispersions of \textgreater 400 km/s, reveal a mass-to-light ratio corresponding to a Kroupa IMF \citep{Smith2015a, Collier2018}.

Focussing on possible systematics in stellar population analysis, for near-infrared gravity-sensitive features that are highly sensitive to changes in the IMF, the difference between a Kroupa and a Salpeter IMF can be on the percent level, and the degeneracies between different parameters remains the largest uncertainty. Hence the certainty in determining the age, metallicity and chemical abundances directly impacts the accuracy of the determined IMF.

Most of literature has focused on ETGs to carry out detailed stellar population analysis. This is because they can be better approximated by simple star formation histories with no ongoing star formation. ETGs are known to be composed of mostly old stars, with little or no current star formation while LTGs can have complicated star formation histories \citep{Kauffmann2003}. Additionally, absorption features in LTGs can be affected by moderate to severe contribution to absorption features from the interstellar medium, depending on the viewing angle. Still, there are studies using full spectrum fitting, which have extracted the luminosity-weighted (LW) ages and metallicities of LTGs \citep{Sanchez-Blazquez2014, GonzalezDelgado2015, Goddard2017, Zheng2017}.

When considering LTGs, which are known to be composed of different stellar populations, it would be prudent to consider the star formation history, and represent the galaxies using composite stellar populations (CSPs). However, it is non-trivial to derive accurate star formation histories and computationally expensive to explore the full posterior distribution. Spectral fitting codes have attempted to find work-around options, e.g. VESPA includes models with increasing levels of complexity that are used as and when data require them \citep{Tojeiro2009}, STARLIGHT fits using full set of models and then restricts the parameter space to a coarser grid \citep{CidFernandas2005}, and FIREFLY uses liberal parameter searching with a convergence test while obtaining linear combinations of best-fit models, to avoid falling into local minima \citep{Wilkinson2017}. However, none of these fitting approaches include a detailed account of element abundance ratios, which is instead the scope of our paper.

Studies have shown that simple stellar population (SSP)-equivalent parameters can still provide useful information when analysing complex stellar populations \citep[e.g.][]{Serra2007, Trager2009, Pforr2012, Citro2016, Leethochawalit2018}. We should however apply the correct interpretation of the age. The age derived through the SSP-based analysis will not correspond to the age of the whole galaxy, rather it will be closer to the one of the latest stellar generation. Therefore the SSP-based analysis quantifies the relative proportion of young populations in the galaxy. Similarly, the derived chemical composition may reflect the one of the latest generations, thereby hiding the oldest, more metal-poor components. Abundance-ratio analysis may actually allow us to pull out these hidden generations. In this work we use composite models to quantify bias in our analysis, as described below.

\citet[][P18]{Parikh2018}; \citet[][P19]{Parikh2019} obtained stellar population properties for a sample of 366 ETGs out to the half-light radius using data from Mapping Nearby Galaxies at Apache Point Observatory \citep[MaNGA][]{Bundy2015}, part of the Sloan Digital Sky Survey IV \citep{Blanton2017}. In the present work, we make use of a subsequent date release to extend results from \citetalias{Parikh2019} in mass, radius, and morphology. For the first time, we attempt to include LTGs in our analysis. We measure absorption features for LTGs and derive SSP-equivalent stellar population parameters. Since the light is dominated by young stars, we expect to be biased towards lower ages. We perform extensive testing using mock CSP models with an exponentially decaying SFH to show that luminosity-weighted (LW) and mass-weighted (MW) ages can be obtained after a correction, while derived metallicities and abundances are good tracers of the underlying parameters. We derive ages and metallicities consistent with literature, and use these to constrain abundances of chemical elements. Although considerable progress is required on the modelling to be able to accurately determine such parameters for LTGs, we present this work as an indication of detailed stellar population signatures in spiral galaxies.

This paper is outlined as follows: the new sample selection, changes to the stacking, and a revised method of deriving abundances, are described in \autoref{sec:data}. Next, the results of our stellar population model fittings to the measured absorption indices on ETG and LTG stacked spectra are presented in \autoref{sec:results}. We discuss our findings in \autoref{sec:discussion}, focussing on differences between the two galaxy types, and spatial variations. This Section also includes a comparison of our results for LTGs to the abundances obtained from studying our own Galaxy. Finally, a summary is given in \autoref{sec:conclusions}.

\begin{figure*}
\centering
  \includegraphics[width=.49\linewidth]{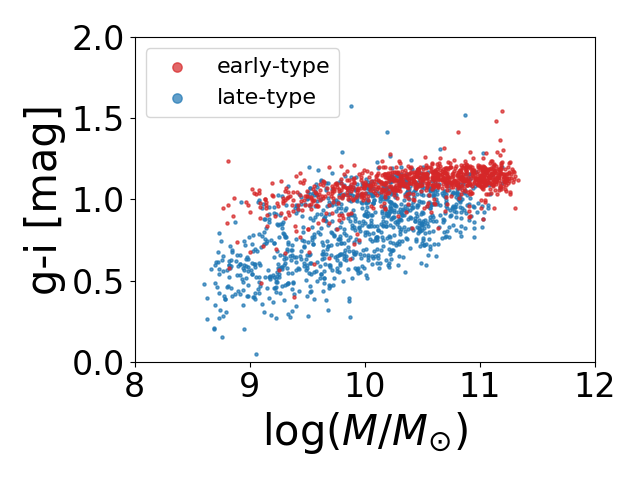}
  \includegraphics[width=.49\linewidth]{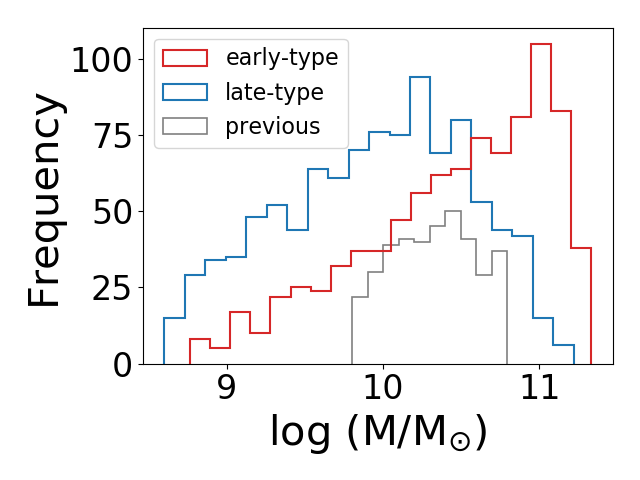}
\caption{Left: Colour-mass diagram for our selection of ETGs (red) and LTGs (blue) after applying S/N and inclination cuts to galaxies from SD15. The masses come from the NASA Sloan Atlas catalogue \citep[NSA,][]{Blanton2005} and the morphologies are based on a Deep Learning Value Added Catalog \citep{Fischer2019}. Right: The stellar mass distributions for the two galaxy types are shown, with the ETG sample from \citetalias{Parikh2019} overlaid in grey.}
\label{fig:sample}
\end{figure*}

\begin{figure}
\centering
  \includegraphics[width=.8\linewidth]{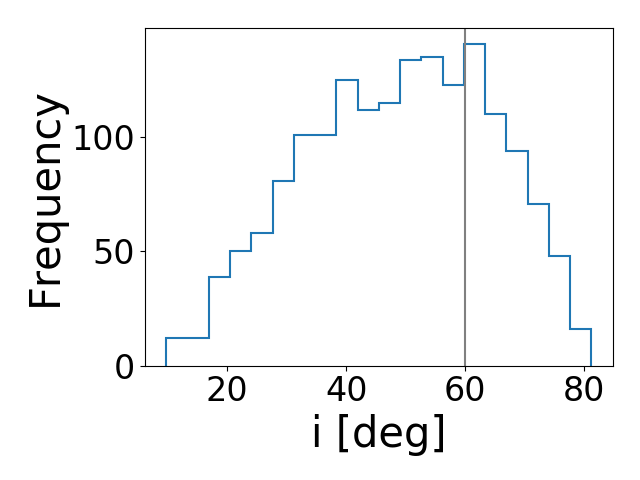}
\caption{Distribution of inclinations for the spiral galaxies. We exclude galaxies with inclinations greater than 60 degrees in order to select mostly face-on objects, since these have a smaller ISM path-length and are affected by contamination to a lesser extent.}
\label{fig:inc}
\end{figure}

\section{Data and analysis tools}
\label{sec:data}
MaNGA is an ongoing project obtaining spatially resolved spectroscopy for 10,000 nearby galaxies at a spectral resolution of $R\sim 2000$ in the wavelength range $3,600-10,300\;$\AA. Independent fibre-bundles provide 17 simultaneous observations of galaxies \citep{Drory2015}, which are fed into the BOSS spectrographs \citep{Smee2013} on the Sloan $2.5\;$m telescope \citep{Gunn2006}. MaNGA targets are chosen from the NASA Sloan Atlas catalogue \citep[NSA,][]{Blanton2005} such that there is a uniform distribution in log stellar mass \citep{Wake2017}. We make use of Elliptical Petrosian quantities for the $r$-band isophotal ellipticity, position angle, half-light radius, and galaxy mass \citep[based on a Chabrier IMF,][]{Chabrier2003} from this catalogue.

Optical fibre bundles of different sizes are chosen to ensure all galaxies are covered out to at least $1.5 R_\mathrm{e}$ for the `Primary' and `Color-enhanced' samples, together known as `Primary$+$', and to $2.5 R_\mathrm{e}$ for the `Secondary' sample \citep{Wake2017}. The Color-enhanced sample supplements colour space that is otherwise under-represented relative to the overall galaxy population. The spatial resolution is $1 - 2\;$kpc at the median redshift of the survey ($z\sim 0.03$), and the $r$-band S/N is $4-8\;$\AA$^{-1}$, for each 2\arcsec\ fibre, at the outskirts of MaNGA galaxies. For more detail on the survey we refer the reader to \citet{Law2015} for MaNGA's observing strategy, to \citet{Yan2016a} for the spectrophotometry calibration, to \citet{Wake2017} for the survey design, and to \citet{Yan2016b} for the initial performance.

Using the general stacking technique described in \citetalias{Parikh2018}, we proceed to bin ETGs and LTGs from the latest SDSS data release out to 1.5 R$_e$. There are several improvements we have made to the sample selection and binning method, which are described below. 

\begin{table}
	\centering
	\caption{We split our sample of ETGs (top panel) into six stellar mass bins and LTGs (bottom panel) into seven bins. For each bin, the number of galaxies and median velocity dispersion, effective radius and redshift are given.}
	\label{tab:new_bins}
	\begin{tabular}{ccccc} 
		\hline
		Mass range ($\log M/M_{\odot}$) & Number & $\sigma$ (km/s) & $R_\mathrm{e}$ (kpc) & z\\
		\hline
		8.8 - 9.8 & 148 & 72 & 3.74 & 0.023\\
		9.8 - 10.3 & 150 & 144 & 3.89 & 0.027\\
		10.3 - 10.6 & 149 & 188 & 4.99 & 0.031\\
		10.6 - 10.8 & 149 & 227 & 6.02 & 0.040\\
		10.8 - 11.0 & 150 & 250 & 6.27 & 0.063\\
		11.0 - 11.3 & 149 & 272 & 5.95 & 0.099\\
		\hline
		8.6 - 9.2 & 143 & 55 & 4.98 & 0.020\\
		9.2 - 9.6 & 143 & 65 & 5.80 & 0.024\\
		9.6 - 9.9 & 144 & 66 & 6.37 & 0.026\\
		9.9 - 10.1 & 144 & 92 & 6.29 & 0.027\\
		10.1 - 10.3 & 144 & 117 & 7.84 & 0.029\\
		10.3 - 10.6 & 143 & 139 & 7.75 & 0.033\\
		10.6 - 11.2 & 144 & 177 & 7.80 & 0.057\\
		\hline
	\end{tabular}
\end{table}

\subsection{Major changes}
\begin{enumerate}
\item The latest SDSS-IV data release with new MaNGA data, DR15 \citep{Aguado2019}, contains 4672 datacubes, allowing us to considerably increase our sample size and parameter space.

\begin{figure*}
\centering
  \includegraphics[width=.4\linewidth]{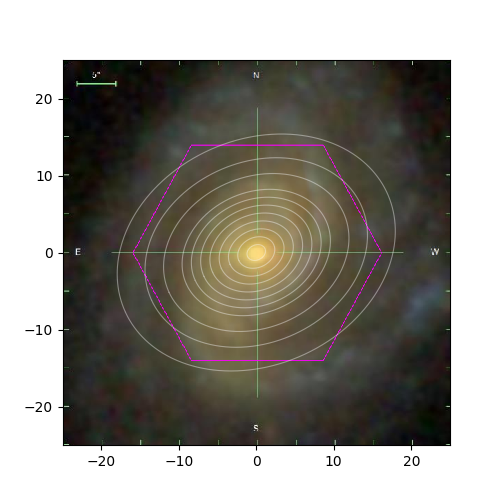}
  \includegraphics[width=.4\linewidth]{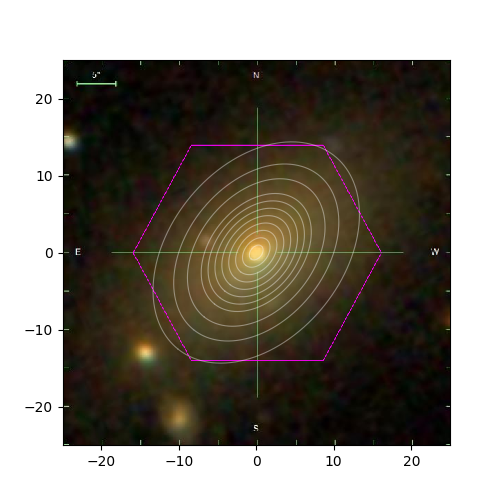}
\caption{Example images of a LTG, MaNGA-ID 1-591917, (left) and an ETG, MaNGA-ID 1-314719 (right) from our sample. The magenta hexagon shows MaNGA's IFU bundle. Overlaid in white are the radial bins calculated from the position angle and ellipticity of the galaxy, out to 1.5 R$_e$. The bins are in steps of 0.1R$_e$ in the centre, with increased widths in the outermost regions in order to compensate for the decreasing surface brightness.}
\label{fig:image}
\end{figure*}

\item We stack spectra of individual spaxels, but make use of kinematics determined for Voronoi-binned spectra to a S/N of 10 from MaNGA's data analysis pipeline \citep[DAP][]{Westfall2019}. Hence, individual spectra belonging to the same Voronoi bin would have the same stellar velocity and velocity dispersion. The benefit of adopting this approach is to avoid using unreliable kinematics for lower S/N spectra. We impose a S/N \textgreater \ 5\; pixel$^{-1}$ threshold for individual spectra. In \citetalias{Parikh2018}, since we used kinematics for individual spectra, we had to impose a higher S/N threshold of 7\; pixel$^{-1}$ to ensure reliable kinematics, leading to more data loss. In the present approach, lowering our S/N criteria allows a large sample that does not affect the accuracy of the kinematics. Stellar velocities are used along with the galaxy redshift to bring individual spectra to the rest frame, and dispersions are propagated to obtain profiles for the final stacked spectra. Emission line fitting and subtraction, and correcting for velocity dispersion broadening, are carried out after stacking. See \citetalias{Parikh2018} for a detailed description of these procedures.
\item For morphological classifications, we make use of the Deep Learning-based catalog \citep{Fischer2019}, which provides the T-type for each galaxy such that T-type \textgreater 0 gives late types, while T-type \textless 0 gives early types, including ellipticals and S0s. We choose this classification since \citet{Fischer2019} show that galaxies with a high probability of being ETGs from Galaxy Zoo can be significantly contaminated by LTGs. The left hand panel of \autoref{fig:sample} shows our final sample in a colour-mass plot, resulting in 895 early type, and 1005 late type galaxies, ranging in stellar mass from $8.60 - 11.33\;\log M/M_{\odot}$. The two types occupy different regions of this parameter space, with early types having redder g - i colour and extending to higher masses. The late type galaxies on the other hand display a greater spread in g - i. The right hand panel shows the mass histograms, with the limited sample of ETGs from \citetalias{Parikh2018} overlaid. The increased mass range that we now study is a substantial improvement. These final distributions are affected by the various cuts, but generally late types are less massive than early types.

\item We split these into $6$ ETG mass bins and $7$ LTG mass bins, with the boundaries calculated such that there are an equal number of galaxies ($\sim 150$) in each bin. These details as well as the median velocity dispersion, effective radius, and redshift for each mass bin are given in \autoref{tab:new_bins}.
\item For LTGs, we apply an inclination cut to select mostly face on galaxies. This minimises the ISM path-length, which is known to affect resonant lines such as NaD at 5900 \AA. \autoref{fig:inc} shows a histogram of the inclinations of LTGs, after making the S/N cut. The distribution peaks near 60 degrees, and this is also the maximum inclination of galaxies that we include in our sample. The galaxies plotted in \autoref{fig:sample} already included this cut in inclination.

\item We decided to modify the radial binning scheme used in \citetalias{Parikh2018} to compensate for the loss in S/N near the half-light radius. Due to lower surface brightness at larger radii, we have increased the widths of the radial bins in order to have more spectra to stack together in these regions, compared to linear bins of equal widths of 0.1~R$_e$. Images of a spiral and an elliptical galaxy are shown in \autoref{fig:image} with the new elliptical annuli bins overlaid. The MaNGA coverage is out to 1.5 R$_e$, and our final three radial bins are twice or thrice as wide as the inner ones.
\item \citetalias{Parikh2018} calculated the error spectrum by taking the standard deviation of the radially stacked spectra from each galaxy, which would go into the stacked spectrum for a particular mass bin. However this method did not give weight to the number of spectra that each galaxy had originally contributed, leading to errors being overestimated. To rectify this, we now take the standard deviation of all raw spectra contributing to a stack. We discuss the effect on the errors in \autoref{sec:results}.
\end{enumerate}

\begin{figure*}
\centering
  \includegraphics[width=.9\linewidth]{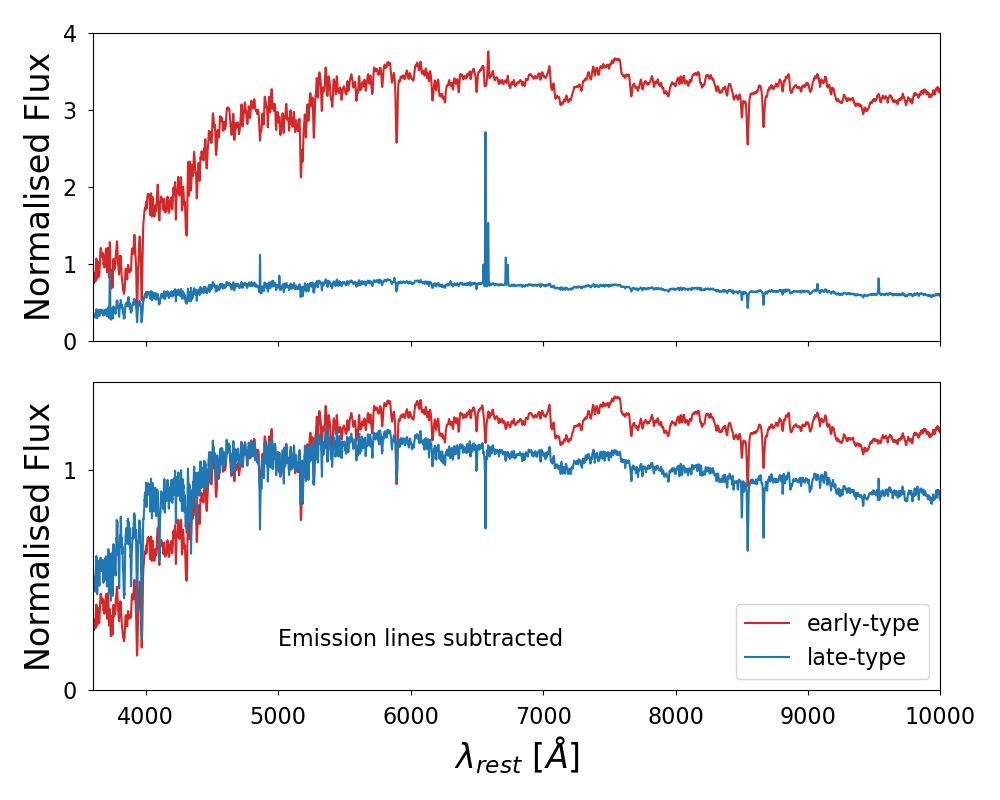}
\caption{Stacked spectra of $\sim$150 LTGs (blue) and ETGs (red) from the central 0.1~R$_e$. The bottom panel shows the normalised spectra after fitting for and subtracting emission lines using pPXF.}
\label{fig:spectra}
\end{figure*}

Stacked spectra from the central 0.1~R$_e$ of the two galaxy types for an intermediate mass bin, containing $\sim$150 galaxies, are shown in \autoref{fig:spectra}. The differences in the spectra of the two galaxy types are evident i.e. LTGs have strong emission lines and the continuum flux level is much lower. The bottom panel shows the same spectra after fitting for and subtracting emission lines using the penalised pixel-fitting algorithm, pPXF \citep{Cappellari2004}. The first fit provides the stellar continuum and stellar kinematics, and the second fit to the spectrum after subtracting the continuum provides a fit for 21 emission lines \citep{NIST_ASD}.  After subtracting emission lines and normalising the spectra to the same flux scale, we can see the same absorption features also present in the LTG spectra, and that ETGs are more flux-enhanced at redder wavelengths compared to LTGs. The MaNGA DAP provides emission line fits for individual spaxels however we fit for and subtract emission lines using pPXF ourselves after stacking spectra to take advantage of the higher S/N ratio.

\begin{figure*}
\centering
    \includegraphics[width=.9\linewidth]{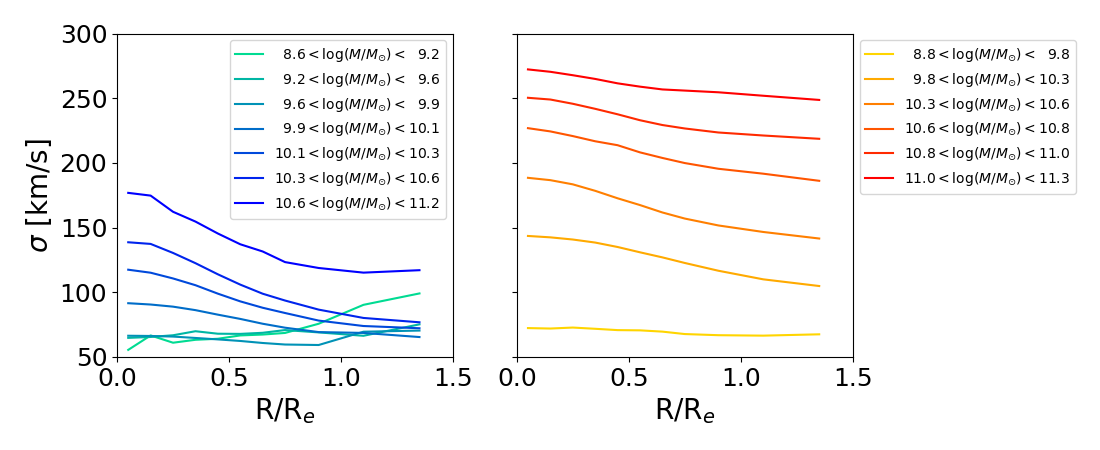}
  \caption{Velocity dispersion profiles for ETGs and LTGs are shown. Different colours represent different mass bins, as denoted in the legends.}
\label{fig:sigma_profiles}
\end{figure*}

\begin{figure}
\centering
  \includegraphics[width=\linewidth]{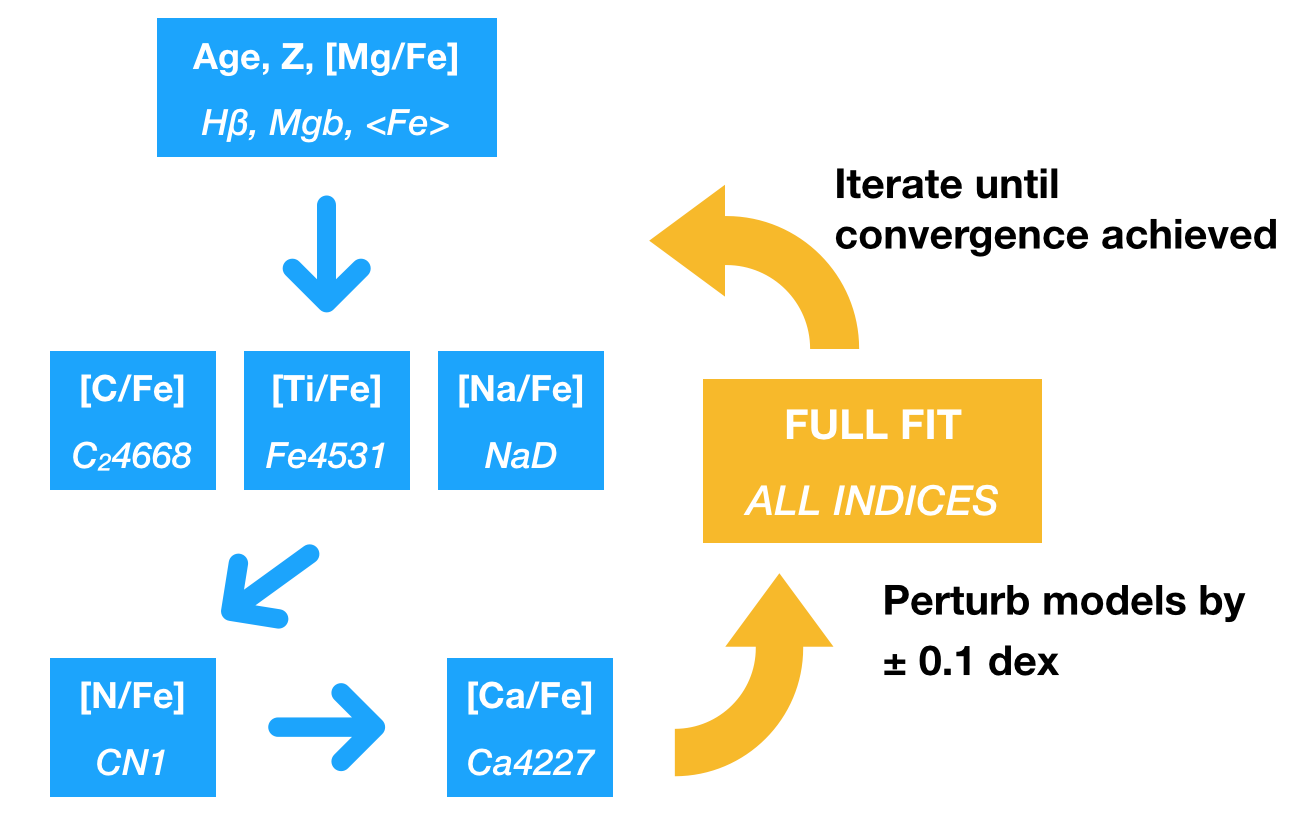}
\caption{Flowchart showing the method adopted to derive chemical abundances. An initial fit is obtained using H$\beta$, Mg$b$, and <Fe> to derive age, metallicity, and [Mg/Fe]. Other abundances are subsequently derived in steps using absorption features which are most sensitive to changes in those parameters. The models are then perturbed by $\pm$0.1~dex and a full fit of all indices is carried out. This process is repeated until there is no significant change in the model parameters.}
\label{fig:flowchart}
\end{figure}

\subsection{Local velocity dispersion}
\label{sec:localsigma}
The velocity dispersion profiles of our binned galaxy sample are shown in \autoref{fig:sigma_profiles}. These values are calculated as the median of the velocity dispersions from the DAP of each spectrum in a particular bin. We use these values as the local $\sigma$ in our analysis to look at correlations between stellar population parameters and velocity dispersions. These differ from the velocity dispersions measured on spectra after stacking at large radii, due to velocity errors becoming convolved as an effective dispersion \citepalias[][Section 2.3]{Parikh2018}. The profiles show expected decreasing trends, except for the lowest mass LTGs, which rise beyond 0.5~R$_e$. We do not include this mass bin in our analysis because of large uncertainties in H$\beta$ absorption measurements, as shown in \autoref{sec:datamodels}.

These velocity dispersion are a combination of bulge and disk contributions. We do not attempt to separate the components in this work but include a discussion of other results in \autoref{sec:bulgedisk}. The true physical meaning of sigma needs to be investigated in an in-depth analysis, however such a study goes well beyond the scope of the present paper.

\subsection{Method}
Key absorption features are measured and corrected for velocity-dispersion broadening, and are modelled using \citet[][TMJ]{ThomasD2011b} line index models. These models are an update and extension of the earlier models by \citet{Thomas2003,Thomas2004} and based on the \citet{Maraston1998,Maraston2005} evolutionary synthesis code, using new empirical calibrations by \citet{Johansson2010} based on the MILES stellar library \citep{Sanchez-Blazquez2006} and element response functions from \citet{Korn2005}. The models are carefully calibrated with galactic globular clusters and reproduce the observations well for the spectral featured used in this study \citep{Thomas2011a}.

\begin{figure}
\centering
  \includegraphics[width=.9\linewidth]{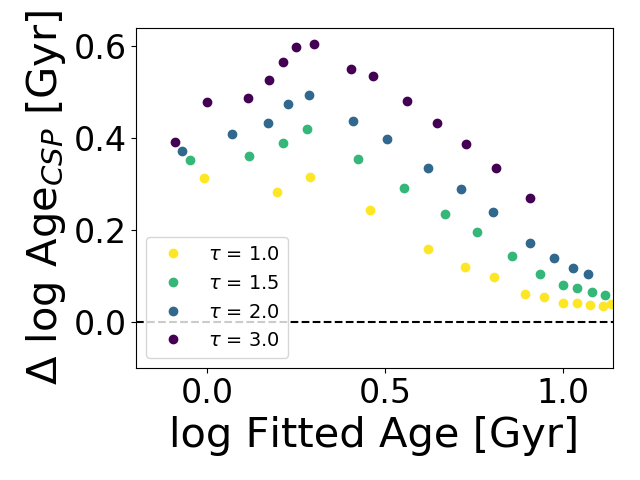}
\caption{CSP ages plotted against derived age from fitting optical Lick indices with SSPs. Different colours show different $\tau$ models with an exponentially declining SFR. The age is always underestimated, and the effect is larger for greater $\tau$ models.}
\label{fig:mocks_cspage}
\end{figure}

The models are available for different ages, metallicities, variable element abundance ratios, and a Salpeter IMF, at MILES resolution.  Element variation are calculated at constant total metallicity, hence the \citetalias{ThomasD2011b} models enhance the $\alpha$-elements and suppress the Fe-peak elements according to Equations 1-3 in \citet{Thomas2003}. The ages span from 0.1 - 15 Gyrs (in steps of 0.2 Gyrs up to 1Gyr, and then in steps of 1 Gyr), metallicities of -2.25, -1.35, -0.33, 0.0, +0.35, +0.67, [Mg/Fe] values of -0.3, 0.0, +0.3, +0.5 dex, and other abundances ratios [X/Mg] of -0.3, 0.0, +0.3 dex. When fitting the models, we set an upper limit on the age corresponding to the age of the universe, 13.7 Gyrs, so that an age older than this is not allowed. Parameters are derived using chi-squared minimisation by interpolating the SSP model grids in the n-dimensional parameter space to fit the data.

To compare measured absorption features to the \citetalias{ThomasD2011b} models, we correct the equivalent widths to MILES resolution using correction factors. The procedure to derive these is described in detail in \citetalias{Parikh2018}. Briefly, we measure absorption features on \citet{Maraston2011} models at MILES resolution and at the resolution of the data, and use the ratio between these measurements as the correction factors.

We follow our previous approach from \citetalias{Parikh2019} of modelling selected features in steps in order to break degeneracies between parameters. We make some modifications to follow the method in \citetalias{Johansson2012} more closely, by iterating using perturbed models until convergence is reached. This is summarised in a flowchart in \autoref{fig:flowchart}. An initial fit is carried out using H$\beta$, Mg$b$, Fe5270, Fe5335 to derive the parameters of age, metallicity, and [Mg/Fe]. We do not make use of higher order Balmer lines since these are much more sensitive to [$\alpha$/Fe] than H$\beta$ \citep{Thomas2004}. Individual features reacting to different element abundances are then modelled in steps as before, C$_2$4668 for [C/Fe], Fe4531 for [Ti/Fe], and NaD for [Na/Fe], followed by CN1 for [N/Fe], and finally Ca4227 for [Ca/Fe]. The latter two are done in steps because they are affected by some of the previously derived abundances. IMF sensitive indices (Mg1, Mg2, TiO1, TiO2) are excluded from the analysis \citep[see also][]{Vazdekis2001, Maraston2003, Conroy2012a}. Next, we perturb the age, metallicity and [Mg/Fe] by $\pm$ 0.1 dex in steps of 0.02 (for age and metallicity) or 0.05 (for element abundances). We then find the minimum chi-squared solution using the entire combination of indices to re-derive these parameters. Finally, we repeat this entire process, except for the initial fit, until the residuals improve by less than 1\%. This usually requires three iterations to achieve.

\begin{figure*}
\centering
  \includegraphics[width=\linewidth]{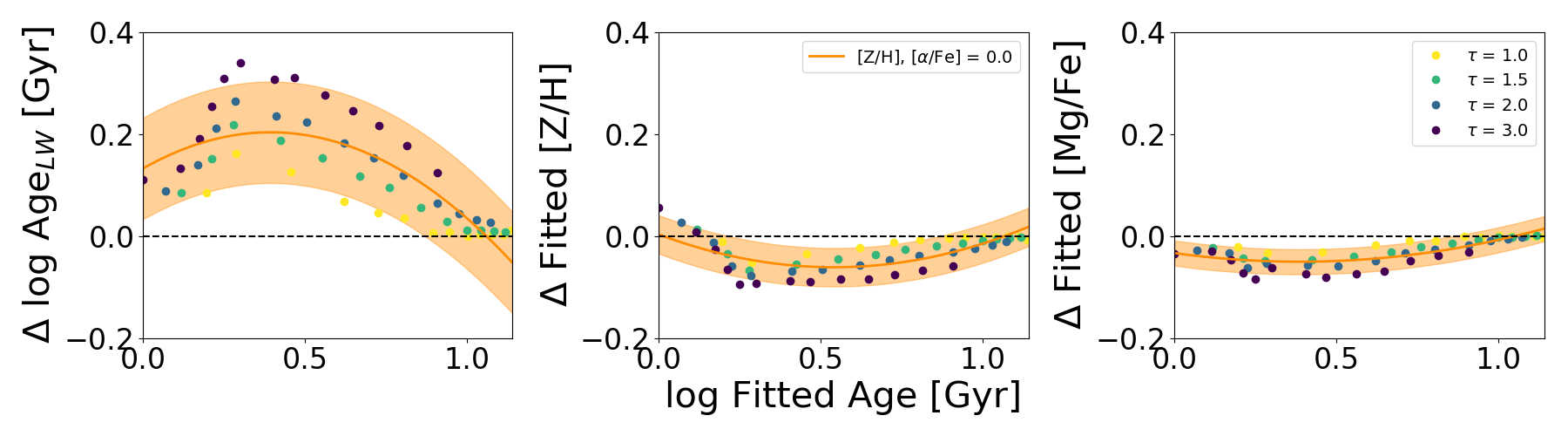}
\caption{LW Age, fitted metallicity and fitted [Mg/Fe] as a function of fitted age for a solar metallicity and solar abundance composite population. The fitted ages correspond better to LW ages of a composite population rather than the oldest age present. At young ages, the metallicity is overestimated by up to +0.2 dex. The fitted [Mg/Fe] appears to be offset from the expected solar value by +0.15 dex at young ages. This decreases for older populations. The orange curves are polynomial fits to the CSP results for all values of $\tau$ and the shaded orange regions represent the 2-$\sigma$ uncertainty in the polynomial coefficients. This captures the scatter due to the different $\tau$ values.}
\label{fig:mocks_taueffect}
\end{figure*}

\begin{figure}
\centering
\flushleft
    \includegraphics[width=.8\linewidth]{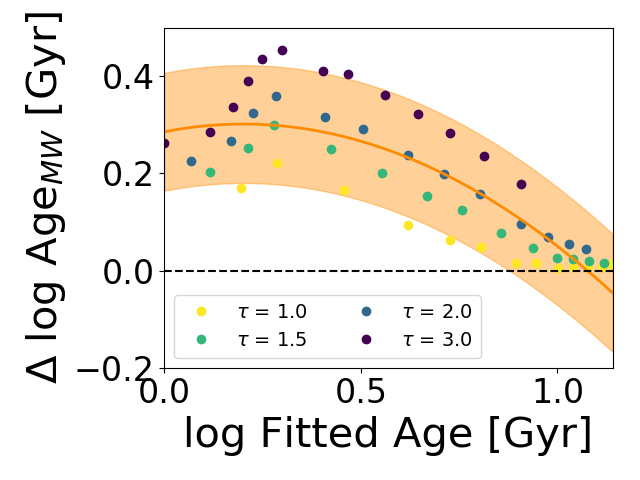}
\caption{Same as the left-hand panel of \autoref{fig:mocks_taueffect} for MW ages.}
\label{fig:mocks_mwage}
\end{figure}

A number of studies have used absorption line indices to infer element abundances \citep{Sanchez-Blazquez2003, Gallazzi2006, Sanchez-Blazquez2006, Clemens2006, Kelson2006, Graves2007, Gallazzi2008, Graves2008, Smith2009, Thomas2010, Price2011, Johansson2012, Worthey2014}. While the methods and models vary greatly between these works, including our work in \citetalias{Parikh2019}, qualitatively the trends agree well. Differences arise for certain elements, possibly due to the treatment of enhancing and depressing various groups in the models \citepalias[see][for a comprehensive discussion]{Parikh2019}. To this end, the greatly increased sample size in this work will shed further light on the robustness of stellar population gradients in different galaxy types.

\subsection{Method validation using mock CSPs}
\label{sec:mocks}
As mentioned in \autoref{sec:intro}, LTGs have complicated star formation histories, and modelling these with SSPs can lead to biases. We use CSP models to test our method and investigate this. These CSP models are constructed by integrating line indices over all the ages contributing to the composite population for different star formation rates. We assume an exponentially declining SFR, $\propto exp(-t/\tau)$, and $\tau$ is the characteristic decay time. Such $\tau$ models were introduced by \citet{Bruzual1983} and are used routinely for representing spiral galaxies \citep[e.g.][]{Bell2001}. We use values of $\tau$ = 1, 1.5, 2, 3 Gyr, and times t = 2 to 15 Gyr in steps of 1 Gyr. These $\tau$ values cover the expected range for galaxies and the age range covers our results of fitted ages. The absorption index strength, I, is calculated as follows:

\begin{equation}
I(t, \tau)_{CSP} = \frac{\int_{0}^{t} \exp^{-\frac{(t-t')}{\tau}} \frac{1}{M/L(t')} I_{SSP} dt'}{\int_{0}^{t} \exp^{-\frac{(t-t')}{\tau}} \frac{1}{M/L(t')} dt'},
\end{equation}

where the SFR is at time t-t', the time at which the population formed, and M/L(t') is the mass-to-light ratio for the stellar population age t'. The coordinate t traces time since the beginning of star formation, corresponding to the age of the oldest SSP contributing to the composite population. We refer to this as the CSP age. More relevant quantities are LW and MW ages. In particular, analyses made using spectral indices correspond to LW quantities. These quantities are calculated by replacing $I_{SSP}$ with age t' and further removing the M/L(t') factors for the MW age.

We apply the initial fit from \autoref{fig:flowchart} to the four optical CSP indices and derive the age, metallicity, and [Mg/Fe]; these quantities derived using our methodology are referred to in this section as the \textit{fitted} quantities. Shown in \autoref{fig:mocks_cspage} is the difference between the CSP age and fitted age as a function of the fitted age for a solar metallicity and solar abundance CSP. All ages are plotted in log Gyrs and the deviations from the dashed line indicate the effect of the extended star formation histories. These results show that fitting with SSPs leads to ages being underestimated, and by a larger value for models with larger $\tau$ (longer SF decay time) because these have a larger fraction of young stars. The bias is also a function of age and improves for older populations.

The left-hand panel of \autoref{fig:mocks_taueffect} shows the difference between the LW age and the fitted age. The differences from the \textit{true} age are much smaller with a maximum of 0.3~dex. Again, ages are underestimated for larger $\tau$ values and for younger fitted ages. Also shown are the fitted metallicity (centre) and fitted Mg abundance (right) as a function of the fitted age.Fitted ages are shown for \textgreater 1 Gyr, the youngest fitted age for the galaxies in \autoref{sec:results}. 

The effect on metallicity and abundance is much smaller in comparison. The metallicity is likely to be overestimated by up to 0.1 dex for intermediate ages, while at very young, and older ages, there is negligible difference. [Mg/Fe] is overestimated by even smaller values of up to 0.05 dex. Additionally, repeating this exercise for different metallicities suggests that these biases have a complicated relation with the true underlying parameters (see Appendix~\ref{sec:app_mocks}), while for different abundances the corrections remain the same as for the solar case i.e. negligible. Hence, we do not attempt to correct these parameters. Furthermore, \citep{Vazdekis2015} show that abundance ratios from indices are contributed mostly by old stars.

\begin{figure*}
\centering
  \includegraphics[width=\linewidth]{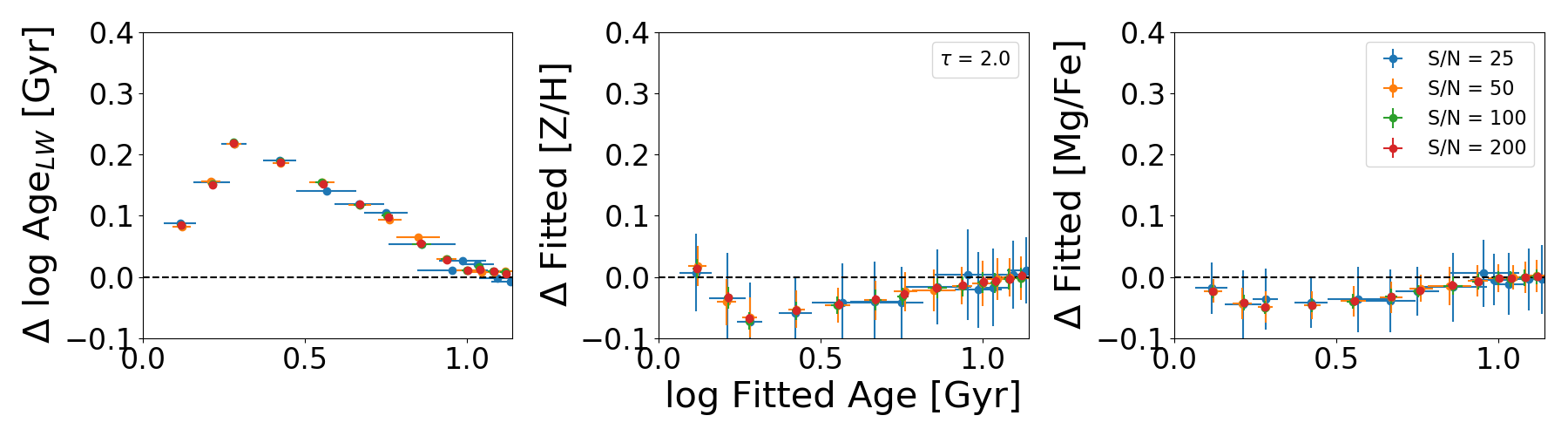}
\caption{For $\tau$ = 1 Gyr, parameters are derived from mock CSPs with S/N values of 50, 100, and 200. The symbols show the ages, metallicities and [Mg/Fe] derived from the average index measurements on 100 realisations of perturbing the mocks, and the errors are propagated from the standard deviation of the measured indices. The S/N does not affect the value of the derived parameter significantly, and at S/N = 200 (green), the errors are small.}
\label{fig:mocks_sn}
\end{figure*}

\autoref{fig:mocks_mwage} shows the equivalent correction for MW ages. In this case, fitted ages must be corrected by a larger value, and the correction peaks at the youngest ages. The LW age is equal to $-0.459 x^2 + 0.360 x + 0.133$, and the mass weighted age is equal to $-0.394 x^2 + 0.158 x + 0.285$, where x is the fitted age between 1 - 15 Gyrs, with all ages in units of log Gyr. We apply these corrections to convert our final derived ages to LW and MW ages.

We also study the effect of S/N on the fitted parameters. To this end we perturb each index with a Gaussian with standard deviation equal to the index value divided by S/N. This is repeated 100 times and the mean and standard deviation of the measured values are used as the index measurement and error in the fitting routine. We choose S/N values of 25, 50, 100, and 200, since these span the S/N of our LTG indices, with low mass galaxies having lower S/N. \autoref{fig:mocks_sn} shows the results for $\tau$ = 2 Gyr. The different colours represent different S/N. It is clear that the S/N does not affect the value of the parameter significantly, but affects the error bars. The errors for S/N = 200 are very small. Typical errors on the fitted parameters for S/N = 25 are $\pm0.1$ log Gyrs for age, $\pm0.1$ dex for metallicity, and $\pm0.05$ dex for [Mg/Fe].

In conclusion, we find that SSP-equivalent ages for populations with a complex SFH are underestimated compared to the true LW and MW ages. We construct functions using polynomial fits to CSPs to convert our derived ages into these parameters. In comparison, SSP-equivalent [Z/H] and [Mg/Fe] are good tracers of the true metallicity and abundance.


\section{Results}
\label{sec:results}
This section presents the fits to the absorption features, and the derived stellar population parameters for ETGs and LTGs. The parameters are first presented as a function of radius, and then as functions of velocity dispersion and galaxy stellar mass. All absorption index measurements and stellar population parameters are provided in Appendix~\ref{sec:app_tables}.

\subsection{Index measurements and model fits}
\label{sec:datamodels}
\begin{figure*}
\centering
  \includegraphics[width=.32\linewidth]{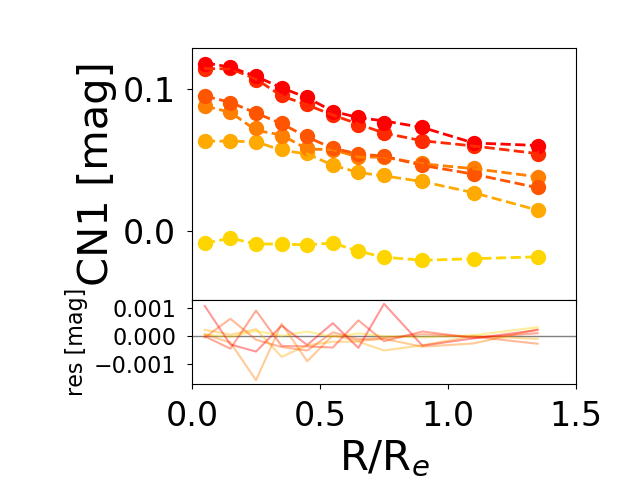}
  \includegraphics[width=.32\linewidth]{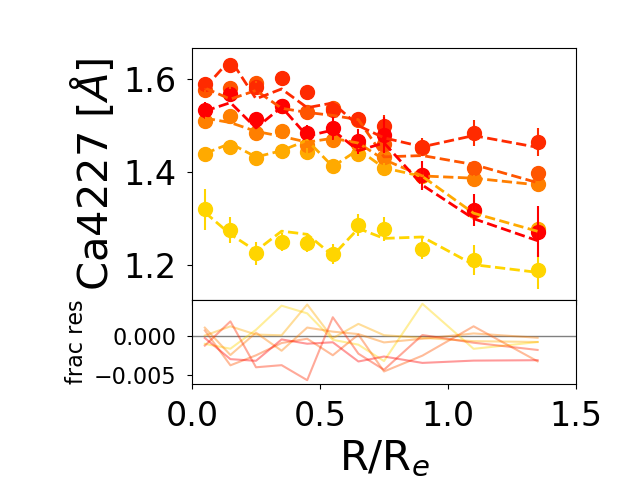}
    \includegraphics[width=.32\linewidth]{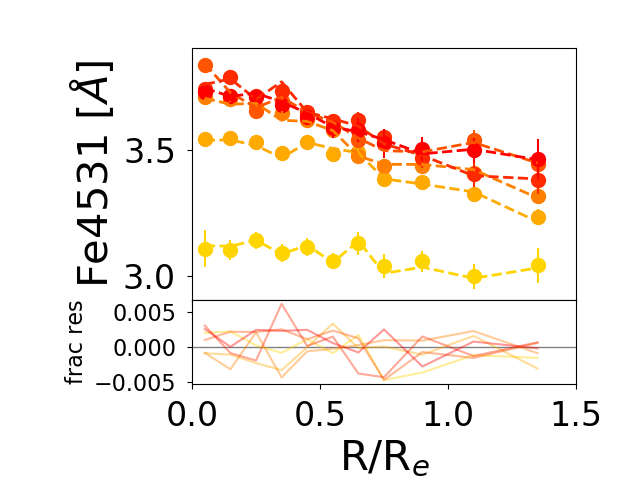}
    \includegraphics[width=.32\linewidth]{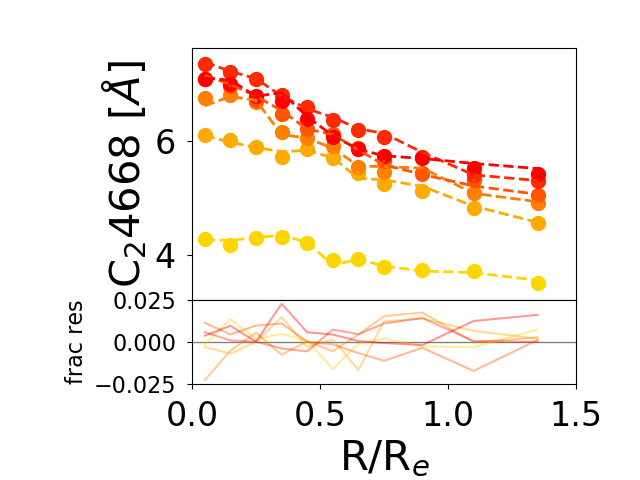}
    \includegraphics[width=.32\linewidth]{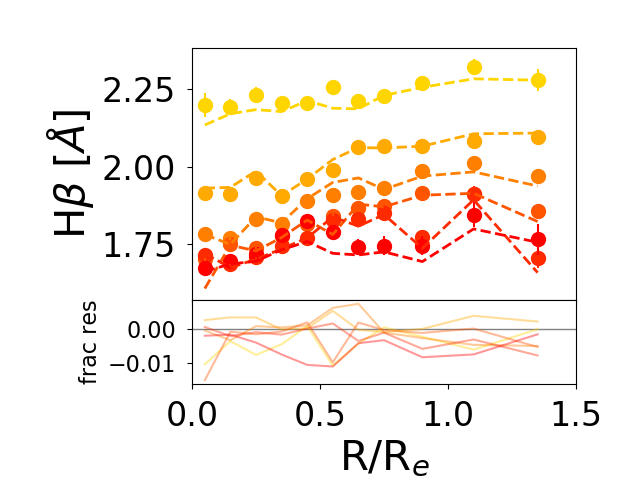}
    \includegraphics[width=.32\linewidth]{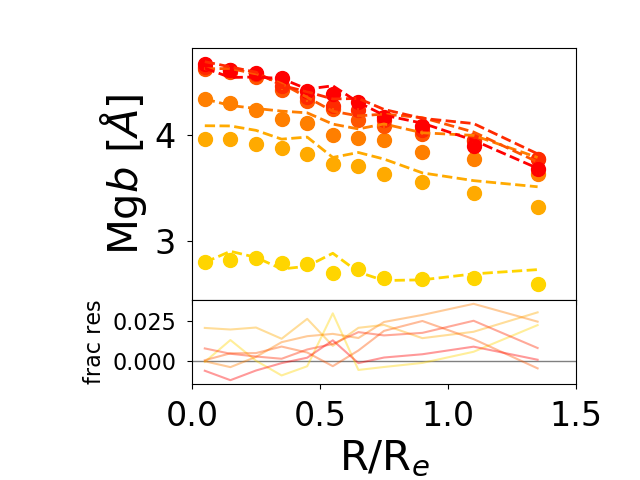}
    \includegraphics[width=.32\linewidth]{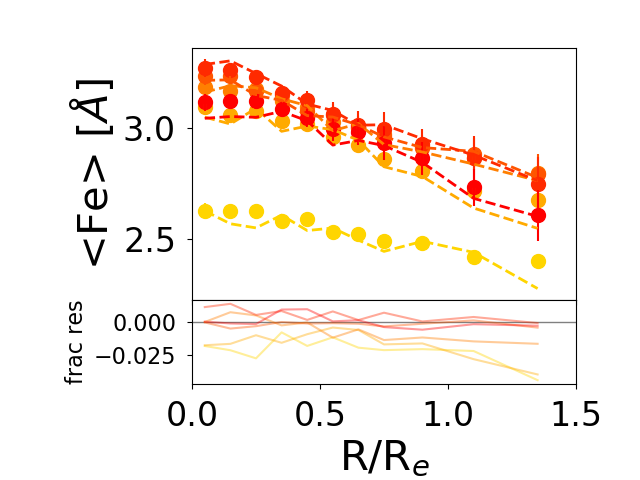}
    \includegraphics[width=.32\linewidth]{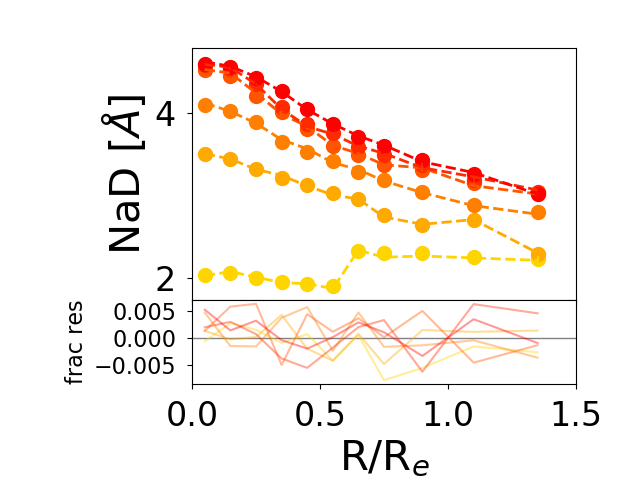}
    \includegraphics[width=.32\linewidth]{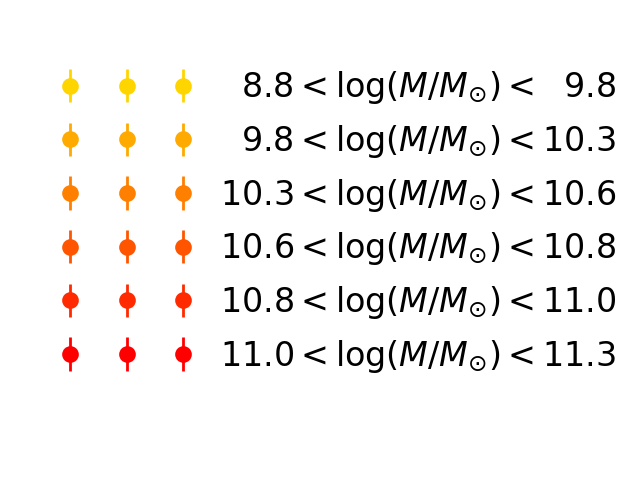}
\caption{Absorption index measurements from stacked spectra of ETGs are shown as a function of radius as coloured circles, with Monte-Carlo based 1-$\sigma$ errors. Yellow to red represents increasing galaxy mass in six bins. Best fitting stellar population models obtained at each point are shown as dashed lines. The bottom panels show the residuals between the data and the models divided by the dynamical range of each index, except for CN1 which is measured in magnitude units.}
\label{fig:fit_etg}
\end{figure*}

The absorption index measurements for ETGs are shown as symbols in \autoref{fig:fit_etg}. There are 6 mass bins with low mass in yellow going to high mass in red.  All indices display negative radial gradients, with more massive galaxies showing stronger absorption and steeper profiles. An exception to this is H$\beta$, which increases with radius and decreasing mass, in line with previous literature \citep[trends with mass,][]{Bender1992, Trager1998} \citep[trends with radius,][]{Carollo1993, Davies1993, Mehlert2000}.

The lowest mass bin has significantly weaker absorption compared to the others (stronger for H$\beta$), this could be due to the large spread in galaxy mass for this particular bin. We carried out a test to split this further into two mass bins with equal number of galaxies, and find that the upper half $9.5 - 9.8\;\log M/M_{\odot}$ is still offset from the other masses, and has the same radial trends as the lower half $8.8 - 9.5\;\log M/M_{\odot}$. This suggests a physical difference in absorption strengths for the lowest mass ETGs. Note also that NaD absorption shows a jump for the lowest mass bin at $\sim 0.6$ R$_e$, while the other mass bins show a smooth decline in strength.

Following the iterative procedure outlined above, we obtain best-fitting models to the measured absorption indices with small residuals. The model predictions for all the indices are shown as dashed lines; these are the final fits derived after constraining the stellar population age, metallicity, and individual element abundances. The bottom panel shows the residuals between the data and models, divided by the dynamical range of each index. For CN1 which is measured in magnitude units, the residual difference is shown. These small residuals indicate that the indices are reproduced well by the models.

The equivalent figure for late types is presented in \autoref{fig:fit_ltg} with 7 low to high mass bins going from green to blue. Generally these galaxies display weaker absorption than their ETG counterparts. Again, H$\beta$ increases with radius, and a clear mass-dependent radial trend is seen, unlike ETGs, such that high mass galaxies have steep radial gradients which flatten for low mass galaxies. Ongoing star formation in spiral galaxies leads to larger H$\beta$ values, since this index anti-correlates with age. Most of the features display very smooth profiles, with striking radial gradients and a clear distinction between the different mass bins. The transition from low to high masses is smoother for LTGs, than for ETGs. The residuals between the models and data are larger compared to ETGs. While the residuals scatter around zero in most cases, H$\beta$ and $<$Fe$>$ are always slightly under-predicted by the models.

The lowest mass bin shows the strongest H$\beta$ absorption but also with the largest uncertainties on this measurement, because this spectral region containing emission lines has a large associated error. Since all our parameters depend on the H$\beta$ equivalent width, we exclude this mass bin from the rest of our analysis.

As mentioned before, we have updated our error estimation so that we take the standard deviation of all individual spectra from each galaxy contributing to a mass bin, rather than the standard deviation of the radially stacked spectra from galaxies. It can be seen in \autoref{fig:fit_etg} and \ref{fig:fit_ltg} that the biggest effect on the index errors is that they are smaller than our previous errors close to 1R$_e$, while errors in galaxy centres continue to remain negligible. Since the number of spectra contributing to the stack increases as a function of radius, our improved method makes a difference to the errors at large radii. In the centre, the radial bins cover small areas and hence the number of galaxies is similar to the total number of spectra.

\subsection{Stellar population parameters}
\begin{figure*}
\centering
  \flushleft
  \includegraphics[width=.32\linewidth]{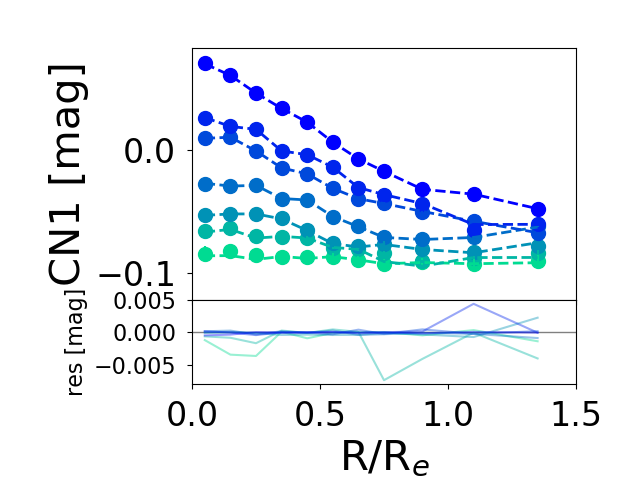}
  \includegraphics[width=.32\linewidth]{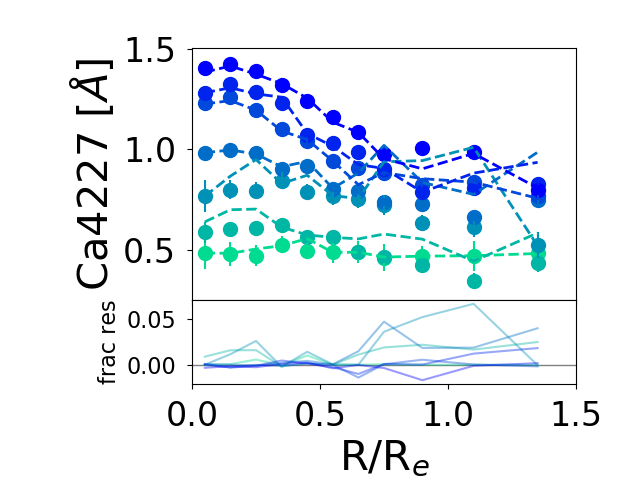}
    \includegraphics[width=.32\linewidth]{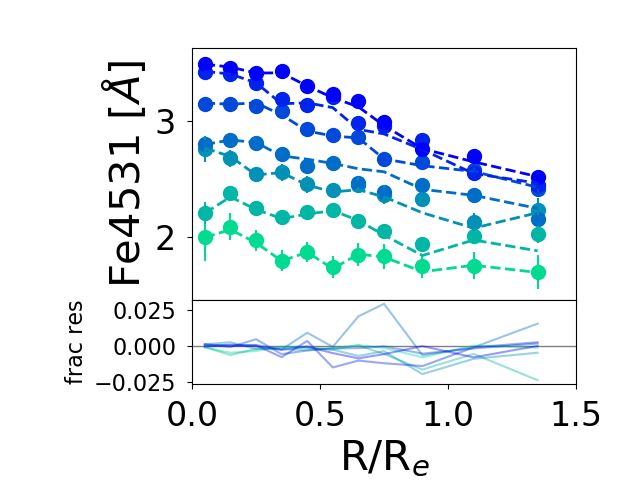}
    \includegraphics[width=.32\linewidth]{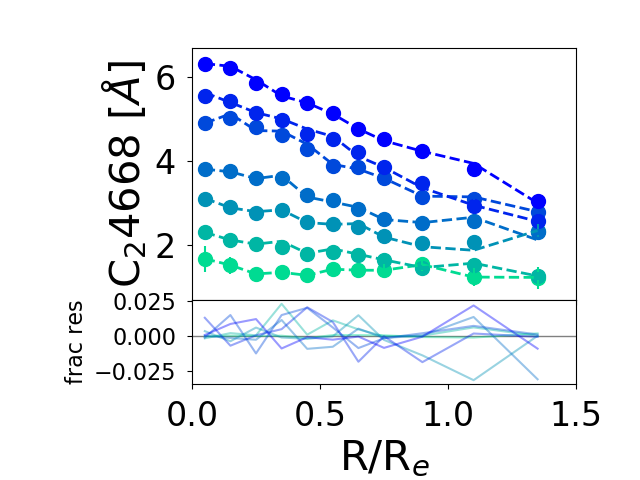}
    \includegraphics[width=.32\linewidth]{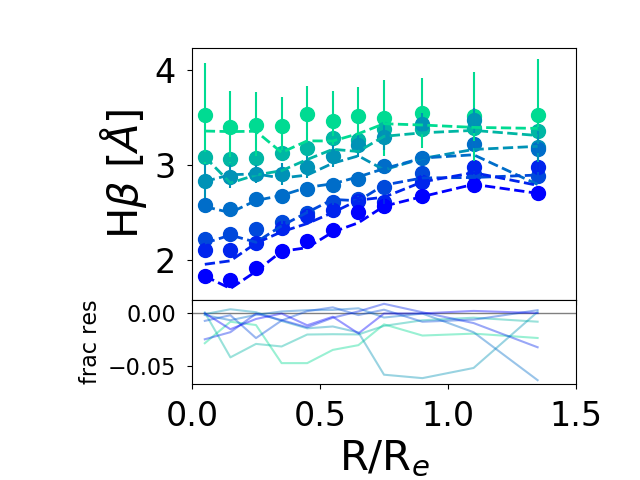}
    \includegraphics[width=.32\linewidth]{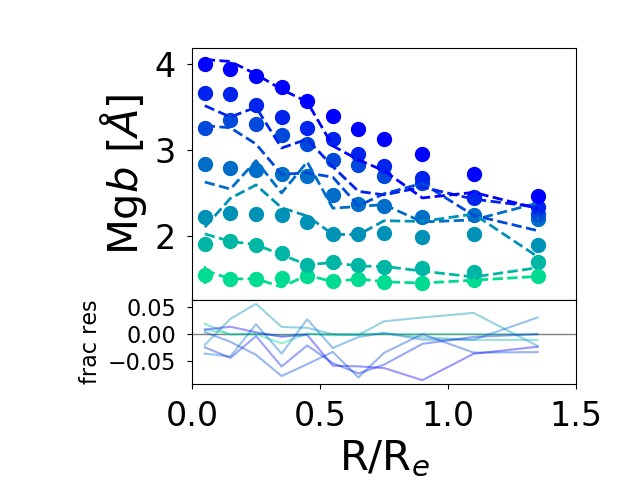}
    \includegraphics[width=.32\linewidth]{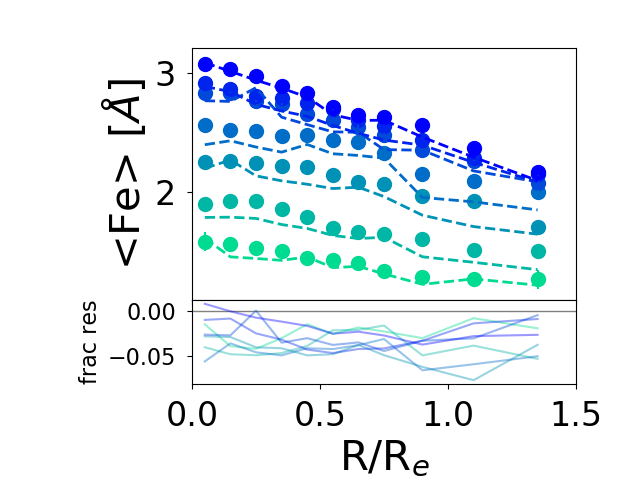}
    \includegraphics[width=.32\linewidth]{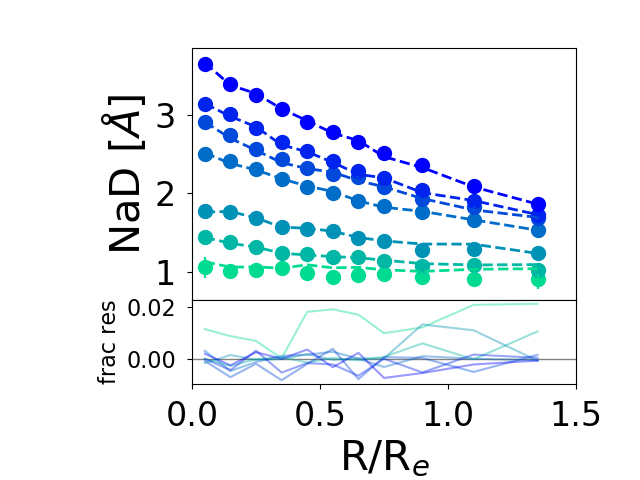}
    \includegraphics[width=.32\linewidth]{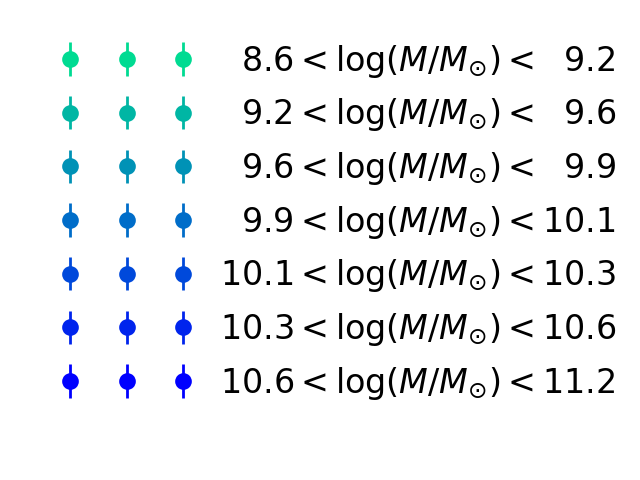}
\caption{Same as \autoref{fig:fit_etg} for late types. There are seven galaxy mass bins, with green shades representing lower mass and blue shades representing higher mass galaxies.}
\label{fig:fit_ltg}
\end{figure*}

In the next series of plots, we present the parameters corresponding to the best-fit models shown in \autoref{fig:fit_etg} and \ref{fig:fit_ltg}. ETGs are shown on the left and LTGs on the right. The scale on the y-axis is kept the same for each parameter to allow a direct visual comparison.

The symbols show the derived parameter for each stacked spectrum with 1-$\sigma$ error-bars. These errors are calculated using a Monte Carlo approach, such that index measurements are randomly perturbed by their errors 100 times and stellar population parameters are derived each time by fitting models. The standard deviation between the parameters derived during each realisation is used to determine the final error on the parameter. Hence these are statistical errors based on the S/N of the stacked spectra. Errors are generally smaller than symbol sizes. The typical error in age is 0.5 Gyr, increasing to 1 Gyr for the lowest mass LTGs, and the typical error in total metallicity and abundances is 0.03 dex. We note that further uncertainty might be introduced for the LTGs due to fitting with SSP models (\autoref{sec:mocks}), depending on the star formation histories of these galaxies.

The radial gradient and its error on each parameter for the different galaxy masses are given in Tables~\ref{tab:new_etg_gradients} \& \ref{tab:new_ltg_gradients}, and a comparison of the radial gradients with galaxy mass and type, as well as with \citetalias{Parikh2019}, follows after the results for all parameters are presented. Radial gradients are calculated from a linear fit (in log-linear scale) to the points and errors on the gradients are calculated based on the deviation of points from the fit. Since the scatter is greater than the errors on the parameters, the former dominates. We provide the gradient within 1 R$_e$ and between 1-1.5 R$_e$ separately, because we notice differences at larger radii.

None of the calculated parameters of age, metallicity, and [Mg/Fe] hit the edges of the model grids except for three regions in the highest mass galaxies for which we derive the maximum age of 13.7 Gyrs. For the other abundances [X/Mg], the lower limit of -0.3 dex is reached for LTGs in some instances: low mass galaxies for Na, Ca, Ti, and most galaxies for N. Thus when we perturb the models by $\pm$ 0.1 dex, we allow them to extend beyond the grids by extrapolation, to obtain a better fit.

\subsubsection{Radial gradients in age and total metallicity}
\begin{figure*}
\centering
  \flushleft
  \includegraphics[width=.33\linewidth]{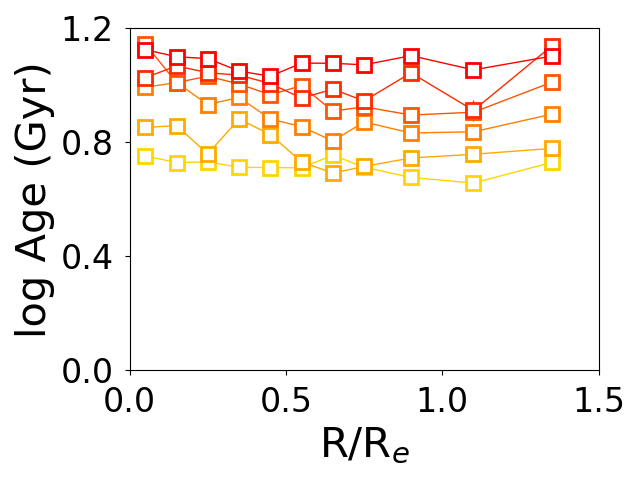}
  \includegraphics[width=.33\linewidth]{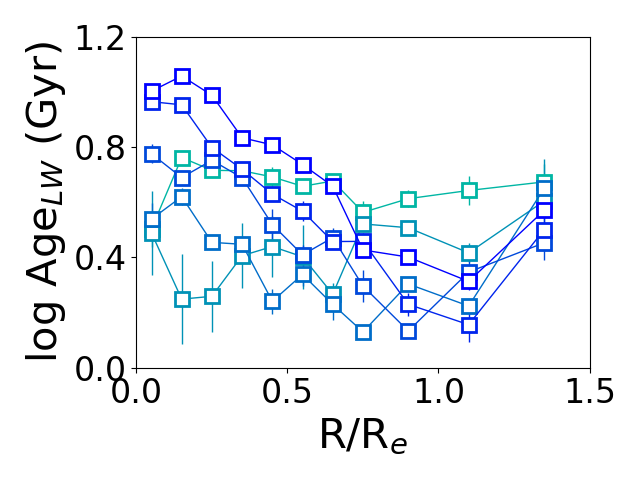}
    \includegraphics[width=.33\linewidth]{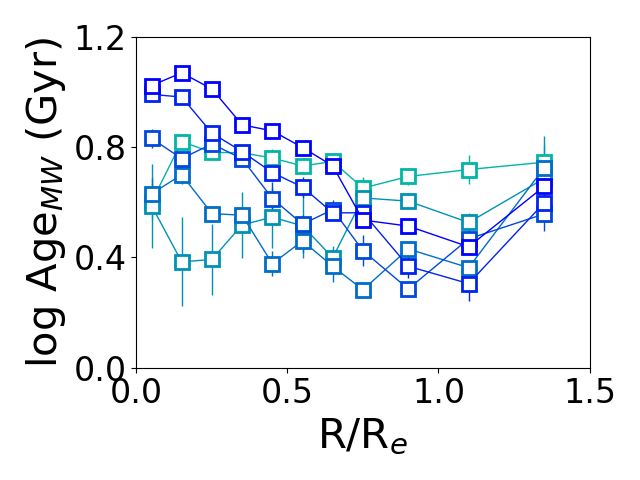}
  \includegraphics[width=.33\linewidth]{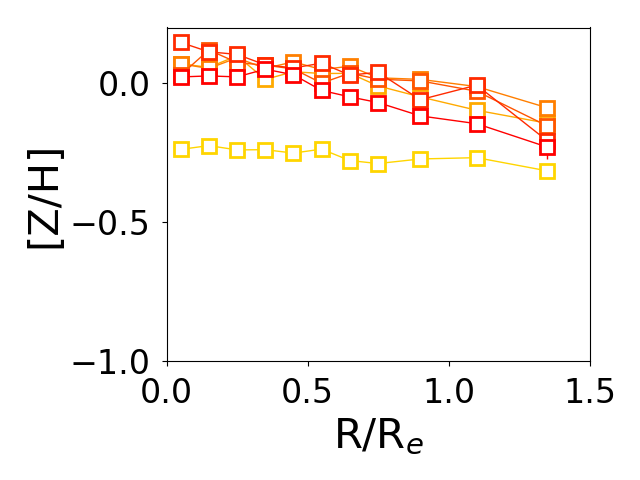}
  \includegraphics[width=.33\linewidth]{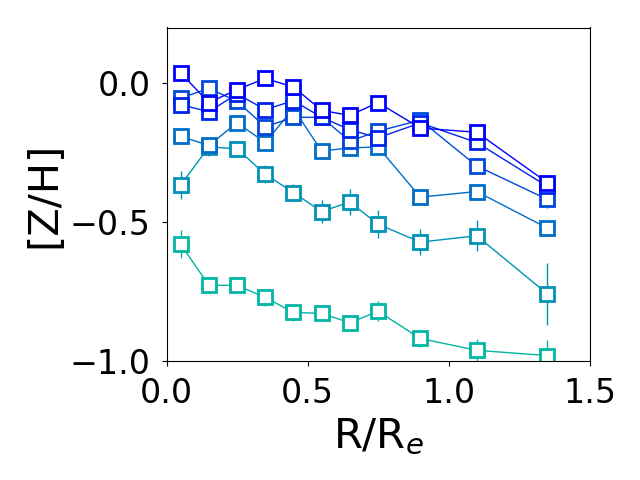}
  \caption{The ages and metallicities as a function of radius are shown for ETGs (left) and LTGs (centre and right). For LTGs, both LW and MW ages are shown. The colours represent different mass bins, as before. The symbols are the derived parameters for each stacked spectrum and are connected by lines. 1-$\sigma$ error bars on the parameters, based on 100 realisations using the index errors, are shown.}
\label{fig:parameters_1}
\end{figure*}

\autoref{fig:parameters_1} shows the the stellar population age (top panel) and metallicity (bottom panel), with the colour scheme as before indicating galaxy type, and within these types, different mass bins. The top left panel shows that more massive ellipticals and lenticulars above $11\;\log M/M_{\odot}$ are older, at $\sim$ 13 Gyr, and the age steadily decreases as galaxies become less massive, reaching $\sim$ 5 Gyr at the lowest masses, $8.8 - 9.8\;\log M/M_{\odot}$. The gradients in age are negligibly small. These galaxies have negative metallicity gradients with slightly super-solar metallicities in centres, and sub-solar metallicities at large radii. The metallicity gradients, shown in the bottom left panel are slightly shallow for low mass galaxies, $-0.09\pm0.03$, and steepen, $-0.23\pm0.03$~dex/R$_e$, for the more massive galaxies. Our results of flat age gradients and negative metallicity gradients in ETGs are qualitatively in agreement with previous results \citep{Kobayashi1999, Mehlert2000, Mehlert2003, Kuntschner2010, Spolaor2010, Greene2015, GonzalezDelgado2015, Goddard2017, Zibetti2019}. Comparing the gradients in detail, our age gradients for intermediate-mass galaxies are slightly steeper at $-0.29 \pm 0.05$ compared to $-0.03\pm0.03$~dex/R$_e$ from \citet{Goddard2017}. However our low and high mass galaxies are consistent with their results of negligible gradients. Our metallicity gradients are also slightly steeper than their reported  $-0.12\pm0.03$~dex/R$_e$ for the highest mass galaxies.

For spiral galaxies we show both LW and MW ages. These ages have been derived by correcting for biases due to modelling with SSPs as described in \autoref{sec:mocks}. Note that these corrections have been derived for solar metallicity and abundance CSPs. Spiral galaxies display younger ages between 2 - 6 Gyrs, and only the innermost regions of galaxies more massive than $10.6\;\log M/M_{\odot}$ are as old as ETGs with ages of $\sim$ 10 Gyrs. These central ages are slightly older than previously reported for bulges \citep[e.g.][]{Proctor2002, Thomas2006}, between 1.3 - 6 Gyrs.

LTGs have negative age gradients, as expected from the H$\beta$ profiles seen in \autoref{fig:fit_ltg}. These gradients clearly represent the transition from bulge to disk regions. Galaxies with masses \textgreater $10.1\;\log M/M_{\odot}$ have very steep age gradients of $-0.60\pm0.04$ on average within the half-light radius. \citet{Goddard2017} also find negative radial gradients for age and metallicity in LTGs however interestingly, they do not find age gradients steepening with galaxy mass like us. Their study uses full spectrum fitting with linear combinations of SSPs to derive stellar population parameters, hence we expect to find differences due to the vastly varied methodology, but it is encouraging to see some agreement.

In the outer regions the age gradients reverse to $0.49\pm0.19$ on average for the massive galaxies. This can also be seen for the intermediate and low mass galaxies. This reversal in age gradients beyond 1 R$_e$ is very interesting and could be due to radial migration. \citet{Zheng2017} derived luminosity-weighted ages for MaNGA galaxies using full spectrum fitting that showed a minimum near the half-light radius and increased with radius beyond this.

The correction for MW ages is larger than for LW ages (from \autoref{sec:mocks}), hence these ages are older. They range from 3.5 - 7.5 Gyrs, with the central regions of massive galaxies reaching 12.5 Gyrs. The gradients in MW age are negative but shallower and display a weaker reversal beyond the half-light radius than the LW ages. \citet{Boardman2020} find similar negative age gradients in their sample of Milky Way-like galaxies from the MaNGA survey. \citet{Goddard2017} also find older MW ages but flat radial gradients, suggesting that the inside-out formation signature is weak and the overall mass budget is little affected. We note that although our methodology is not expected to provide MW quantities, we use the correction derived from \autoref{sec:mocks} to calculate these ages, which provide an indication of how LW and MW parameters might differ. In all following plots we show LW ages for LTGs.

LTGs are more metal poor than ETGs. The lowest mass spirals are very metal poor with values between -0.5 and -1.0 dex (a tenth of solar metallicity). All galaxies have negative metallicity gradients, with the average across all masses being $-0.18\pm0.02$~dex/R$_e$. \citet{Goddard2017} find gradients ranging between $0.06\pm0.01$ to $-0.32\pm0.06$~dex/R$_e$. Negative radial gradients have also been obtained in the Milky Way \citep{Carollo2007, Hayden2015}, and for disk galaxies from the CALIFA survey \citep{Sanchez-Blazquez2014, GonzalezDelgado2015}.

\citet{Peletier2007, Ganda2007} measure 2 dimensional absorption features (H$\beta$, Fe5015, Mgb) for a small sample of spiral galaxies and translate these to age, metallicity, and [Mg/Fe]. They find younger ages, lower metallicities and abundances close to solar for spirals compared to ETGs. \citet{Scott2017} report the same using the SAMI survey.

\subsubsection{Radial gradients in element abundances}
We now present the derived element abundances for C, N, Na, Mg, Ca, and Ti. See \citetalias{Parikh2019} and references therein for a detailed comparison with the literature and implications of element abundances. Here, we will focus on describing the extension in mass and radius for ETGs, any differences compared to \citetalias{Parikh2019}, and the results for LTGs. Abundances of individual elements in these galaxy types have been studied in a very limited sense in literature before. However, there are some studies which look at a single [$\alpha$/Fe] parameter, which we compare with in \autoref{sec:discussion}.

\begin{figure}
\centering
  \includegraphics[width=.49\linewidth]{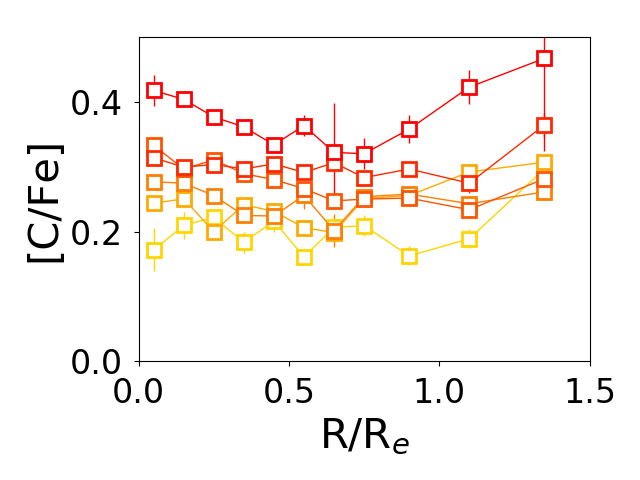}
  \includegraphics[width=.49\linewidth]{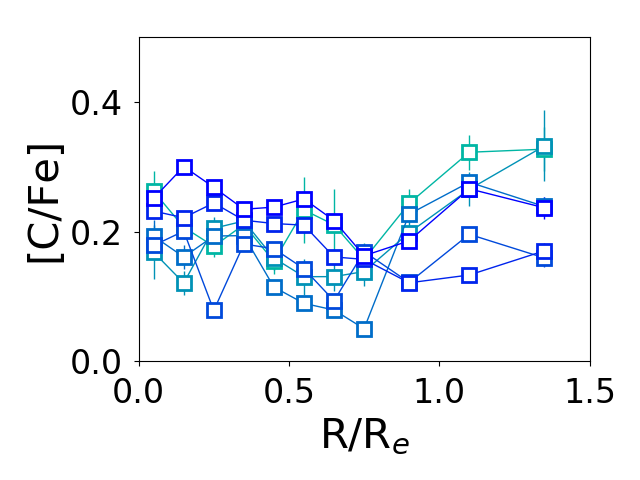}
   \includegraphics[width=.49\linewidth]{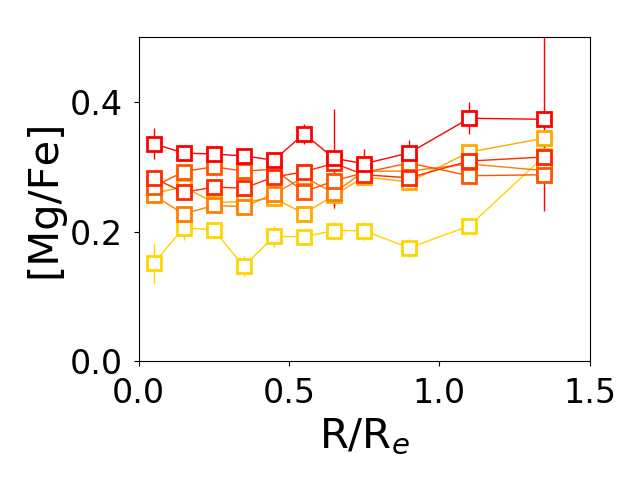}
  \includegraphics[width=.49\linewidth]{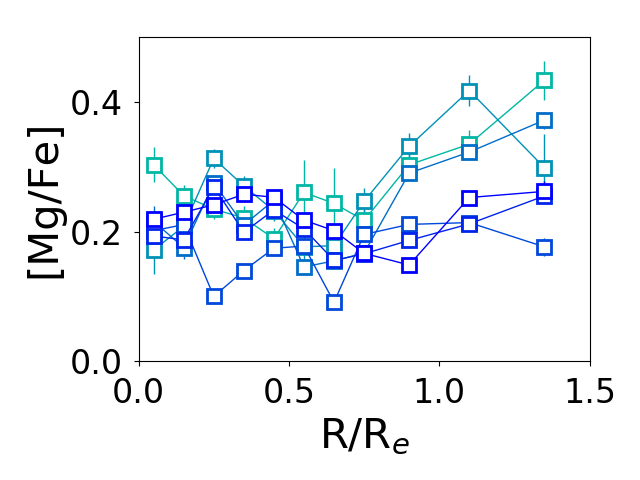}
  \caption{Same as \autoref{fig:parameters_1} for C and Mg abundances.}
\label{fig:parameters_2}
\end{figure}

C and Mg abundances are shown in \autoref{fig:parameters_2}. [C/Fe] values for ETGs range between 0.2 to 0.4 dex for different masses. There is a clear trend of [C/Fe] enhancement with increasing galaxy mass. For spiral galaxies, these abundances range between 0.1 to 0.3~dex, ignoring values with very large errors. For these galaxies, the trend with mass is not as evident and there is more scatter. Both types display negligible gradients with radius, and the values inside and outside 1 R$_e$ are generally consistent.

\begin{figure}
\centering
  \includegraphics[width=.49\linewidth]{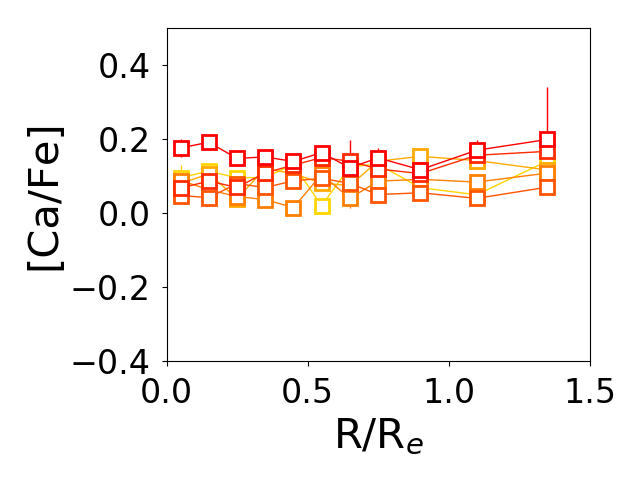}
  \includegraphics[width=.49\linewidth]{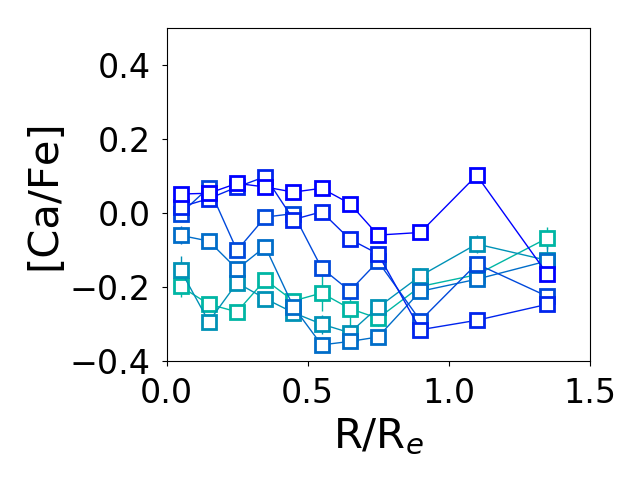}
    \includegraphics[width=.49\linewidth]{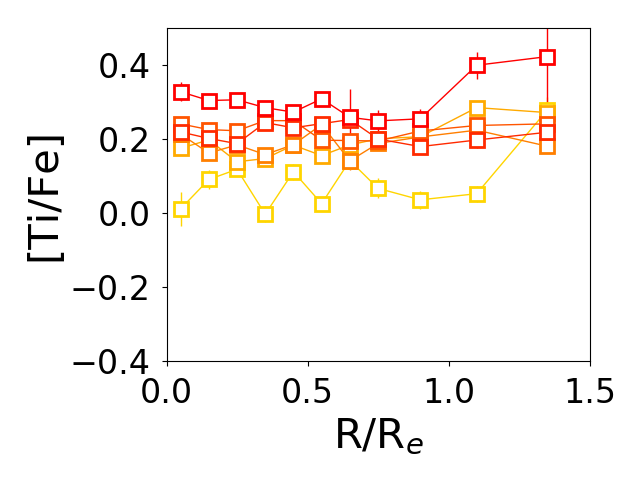}
  \includegraphics[width=.49\linewidth]{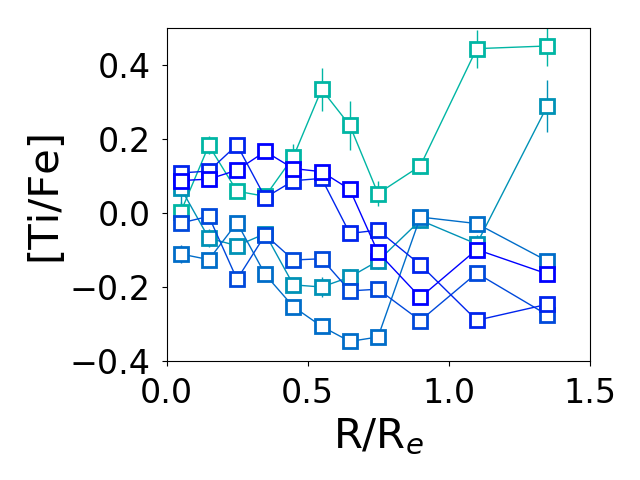}
  \caption{Same as \autoref{fig:parameters_1} for Ca and Ti abundances.}
\label{fig:parameters_3}
\end{figure}

The behaviour of [Mg/Fe] for ETGs and LTGs is very similar to C: it has similar abundances and little variation with radius. This is consistent with previous studies for ETGs \citep{Mehlert2003, Kuntschner2004, Johansson2012, Alton2018}. However, beyond 1 R$_e$, the gradients become more positive for all galaxy types and masses. This effect is stronger for lower masses and hence, for LTGs this causes a reversed trend with mass such that more massive galaxies are less enhanced. This is potentially interesting in determining accretion histories. \citet{Greene2013} find radially constant large [Mg/Fe] ratios out to the haloes of elliptical galaxies, and rising ratios for lower dispersion galaxies. \citet{Thomas1999} suggest that large [Mg/Fe] ratios at large radii, in relatively metal-poor regions, show evidence for a fast clumpy collapse model for the formation of massive ellipticals, rather than a merging spirals scenario which would result in solar [Mg/Fe] values in the very outer regions \citep{Coccato2010}.

Next we move on to the heavier $\alpha$ elements, Ca, and Ti, shown in \autoref{fig:parameters_3}. As found before in \citetalias{Parikh2019} for ETGs, Ca does not follow Mg, and instead is under-enhanced, with values between 0 and 0.2 dex, and no variation with radius. Ca under-abundance in ETGs, also previously reported by \citep{Saglia2002, Cenarro2003, Thomas2003b, Graves2007, Smith2009, Price2011}, was interpreted as contribution to Ca from delayed Type Ia supernovae. Spiral galaxies are even further depleted in Ca, with sub-solar abundances between -0.4 and 0 dex. Low mass spirals are expected to be enriched in both Ca and Fe through their extended star formation history hence these low [Ca/Fe] ratios suggest that Ca under-abundance cannot be a result of short star formation timescales. LTGs interestingly display negative radial gradients within the half-light radius, as steep as $-0.36\pm0.10$ for some of the intermediate mass spirals. The regions with low [Ca/Fe] ratios are very metal poor hence metallicity could be causing Ca under-abundance in LTGs.

\begin{figure}
\centering
    \includegraphics[width=.49\linewidth]{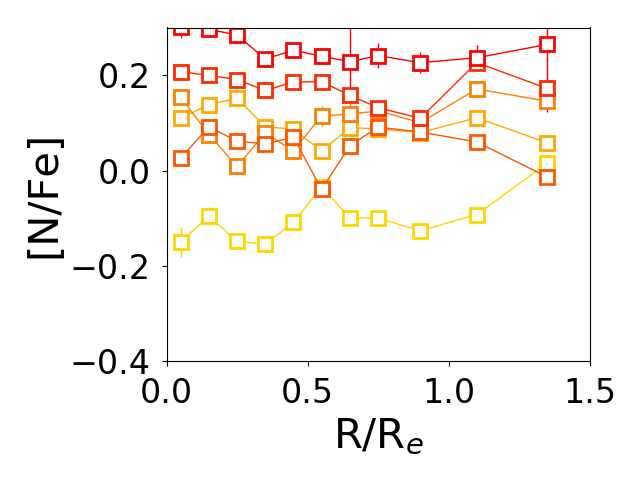}
  \includegraphics[width=.49\linewidth]{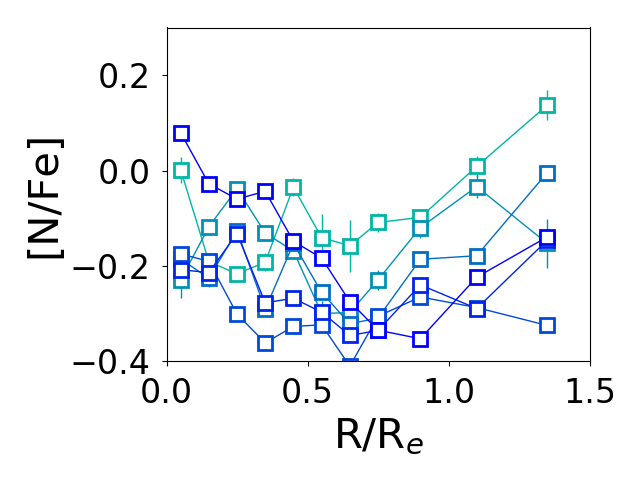}
    \includegraphics[width=.49\linewidth]{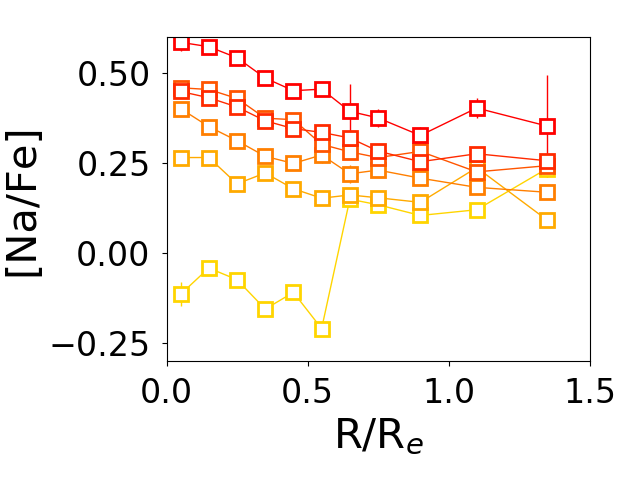}
  \includegraphics[width=.49\linewidth]{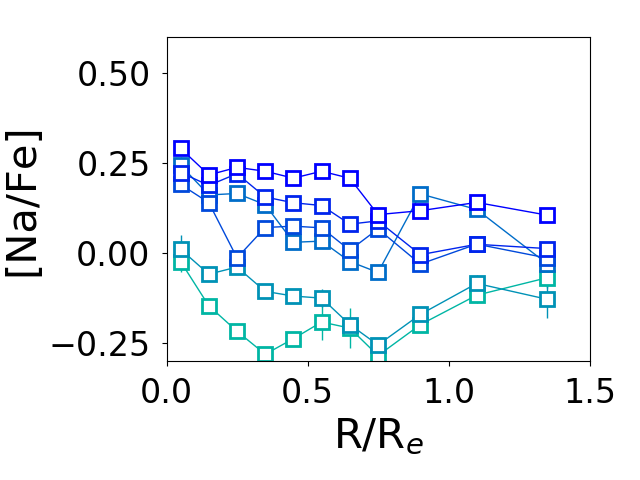}
\caption{Same as \autoref{fig:parameters_1} for N and Na abundances.}
\label{fig:parameters_4}
\end{figure}

Ti values lie somewhere between Ca and Mg, with values between 0 and 0.3 dex for ETGs, similar to \citepalias{Johansson2012, Parikh2019}, and between -0.3 and 0.2 dex for LTGs, neglecting outliers with large errors. Interestingly, for LTGs, Ti abundances not as low as Ca, which fits in to the picture that low Ca cannot be caused by delayed Type Ia enrichment. The radial gradients are also similar to Ca, with ETGs showing no variation, and LTGs displaying negative radial gradients. LTGs also show varying gradients in the inner and outer regions, without any clear trends, making the results unreliable and difficult to interpret.

Lastly, \autoref{fig:parameters_4} shows N and Na abundances, the two elements for which we detected negative radial gradients in \citetalias{Parikh2019}. Looking at [N/Fe], we find abundance values ranging from -0.15 to 0.3 dex from low to high mass ETGs. We now see shallow radial gradients compared to \citetalias{Parikh2019}, based on a more robust analysis out to larger radii. For the same mass range, we previously derived $-0.25\pm0.05$ and $-0.18\pm0.05$~dex/R$_e$, and now find $0.05\pm0.06$, and $-0.07\pm0.03$~dex/R$_e$, which are inconsistent to 2-3$\sigma$. The present results are more trustworthy and cause us to revise our conclusion regarding [N/Fe] gradients. Spiral galaxies have a scattered trend of [N/Fe] with mass and display negligible radial gradients, except for the highest mass bin.

As before, we find high [Na/Fe] in ETGs, with the more massive ones reaching up to 0.6 dex. In \citetalias{Parikh2019}, the highest mass bin showed 0.5 dex [Na/Fe] in the centre, and in this work we have two mass bins higher than this. We see strong radial gradients, which are steeper for more massive galaxies. The gradients range from $-0.15\pm0.03$ to $-0.32\pm0.02$~dex/R$_e$. These gradients within the effective radius are consistent with the gradients from \citetalias{Parikh2019} of $-0.26\pm0.03$ and $-0.29\pm0.02$. Beyond 1 R$_e$ [Na/Fe] seems to flatten. [N/Fe] also flattens (or even rises for LTGs), similar to [Mg/Fe].

We note that there is a sharp jump in Na abundance for the lowest mass bin which cannot be physical. This is caused by the sudden change in NaD absorption (see \autoref{fig:fit_etg}), and the fact that [Na/Fe] is highly sensitive to changes in this feature. Hence we will not consider this mass bin in plots when measuring the [Na/Fe] gradient with radius or velocity dispersion. Low mass spiral galaxies are less enhanced in Na, with values between -0.25 and 0.25 dex, and high mass spirals are enhanced to twice solar Na abundances in the centres. These galaxies also show negative radial gradients in Na. It is very interesting to find the same radial behaviour in Na for LTGs.

\begin{table*}
	\centering
	\caption{Radial gradients of stellar population parameters for ETGs in dex/$R_e$; age is in log Gyr/$R_e$. The top rows for each parameter provide the gradient within 1~R$_e$, and the bottom rows provide the gradient between 1 and 1.5~R$_e$.}
	\label{tab:new_etg_gradients}
	\resizebox{\linewidth}{!}{%
	\begin{tabular}{cccccccccc}
		\hline
		Mass bin 		& 	Age 			   & [Z/H] 			 & [C/Fe] 		       & [N/Fe] 		     & [Na/Fe] 		  & [Mg/Fe] 	       	  & [Ca/Fe] 	     & [Ti/Fe]\\
		\hline
		$8.8 - 9.8$	&	$-0.05 \pm 0.03$ & $-0.06 \pm 0.02$ & $-0.02 \pm 0.03$ & $0.05 \pm 0.05$ & $0.29 \pm 0.14$ & $0.02 \pm 0.03$ & $-0.01 \pm 0.05$ & $0.01 \pm 0.07$\\
					&      $0.12 \pm 0.11$ & $-0.10 \pm 0.06$ & $0.30 \pm 0.09$ & $0.32 \pm 0.07$ & $0.30 \pm 0.11$ & $0.32 \pm 0.07$ & $0.17 \pm 0.13$ & $0.55 \pm 0.23$\\
		\hline
		$9.8 - 10.3$	&	$-0.18 \pm 0.06$ & $-0.13 \pm 0.03$ & $0.00 \pm 0.03$ & $-0.07 \pm 0.03$ & $-0.15 \pm 0.03$ & $0.02 \pm 0.02$ & $0.07 \pm 0.04$ & $0.04 \pm 0.03$\\
					&      $0.07 \pm 0.0$ & $-0.21 \pm 0.02$ & $0.11 \pm 0.03$ & $-0.05 \pm 0.10$ & $-0.13 \pm 0.30$ & $0.15 \pm 0.04$ & $-0.08 \pm 0.01$ & $0.14 \pm 0.13$\\
		\hline
		$10.3 - 10.6$	&	$-0.23 \pm 0.04$ & $-0.06 \pm 0.02$ & $-0.03 \pm 0.03$ & $0.03 \pm 0.06$ & $-0.21 \pm 0.03$ & $0.07 \pm 0.02$ & $0.02 \pm 0.04$ & $0.01 \pm 0.04$\\
					&	$0.15 \pm 0.06$ & $-0.23 \pm 0.05$ & $0.01 \pm 0.04$ & $0.09 \pm 0.13$ & $-0.09 \pm 0.02$ & $-0.00 \pm 0.03$ & $0.04 \pm 0.04$ & $-0.06 \pm 0.08$\\
		\hline
		$10.6 - 10.8$	&	$-0.24 \pm 0.05$ & $-0.09 \pm 0.04$ & $-0.10 \pm 0.01$ & $0.02 \pm 0.05$ & $-0.26 \pm 0.03$ & $0.01 \pm 0.02$ & $0.01 \pm 0.03$ & $-0.04 \pm 0.02$\\
					&	$0.26 \pm 0.10$ & $-0.36 \pm 0.09$ & $0.07 \pm 0.08$ & $-0.21 \pm 0.05$ & $-0.08 \pm 0.10$ & $-0.04 \pm 0.03$ & $0.04 \pm 0.06$ & $0.04 \pm 0.02$\\
		\hline
		$10.8 - 11.0$	&	$-0.07 \pm 0.05$ & $-0.19 \pm 0.03$ & $-0.02 \pm 0.01$ & $-0.10 \pm 0.02$ & $-0.23 \pm 0.01$ & $0.03 \pm 0.02$ & $0.07 \pm 0.03$ & $-0.01 \pm 0.04$\\
					&	$0.24 \pm 0.44$ & $-0.35 \pm 0.3$ & $0.16 \pm 0.13$ & $0.12 \pm 0.23$ & $0.00 \pm 0.05$ & $0.07 \pm 0.03$ & $0.13 \pm 0.06$ & $0.09 \pm 0.00$\\
		\hline
		$11.0 - 11.3$	&	$-0.03 \pm 0.04$ & $-0.17 \pm 0.04$ & $-0.09 \pm 0.03$ & $-0.09 \pm 0.02$ & $-0.32 \pm 0.02$ & $-0.01 \pm 0.02$ & $-0.06 \pm 0.02$ & $-0.08 \pm 0.02$\\
					&	$0.00 \pm 0.13$ & $-0.25 \pm 0.06$ & $0.24 \pm 0.04$ & $0.09 \pm 0.02$ & $0.05 \pm 0.17$ & $0.11 \pm 0.08$ & $0.18 \pm 0.04$ & $0.36 \pm 0.18$\\
		\hline
	\end{tabular}}
\end{table*}

\begin{table*}
	\centering
	\caption{Same as \autoref{tab:new_etg_gradients} for LTGs.}
	\label{tab:new_ltg_gradients}
	\resizebox{\linewidth}{!}{%
	\begin{tabular}{cccccccccc}
		\hline
		Mass bin 		& 	Age 			   & [Z/H] 			 & [C/Fe] 		       & [N/Fe] 		     & [Na/Fe] 		  & [Mg/Fe] 	       	  & [Ca/Fe] 	     & [Ti/Fe]\\
		\hline
		$9.2 - 9.6$	&	$-0.05 \pm 0.11$ & $-0.32 \pm 0.05$ & $-0.02 \pm 0.05$ & $0.00 \pm 0.10$ & $-0.16 \pm 0.09$ & $-0.00 \pm 0.05$ & $-0.02 \pm 0.05$ & $0.12 \pm 0.13$\\
					&	$0.13 \pm 0.01$ & $-0.13 \pm 0.04$ & $0.18 \pm 0.11$ & $0.52 \pm 0.01$ & $0.29 \pm 0.06$ & $0.29 \pm 0.07$ & $0.29 \pm 0.07$ & $0.69 \pm 0.44$\\
		\hline
		$9.6 - 9.9$	&	$0.15 \pm 0.13$ & $-0.35 \pm 0.08$ & $-0.01 \pm 0.05$ & $-0.09 \pm 0.11$ & $-0.26 \pm 0.05$ & $0.07 \pm 0.07$ & $-0.04 \pm 0.08$ & $-0.13 \pm 0.11$\\
					&	$0.24 \pm 0.34$ & $-0.43 \pm 0.27$ & $0.3 \pm 0.02$ & $-0.09 \pm 0.26$ & $0.08 \pm 0.17$ & $-0.09 \pm 0.26$ & $0.08 \pm 0.17$ & $0.72 \pm 0.51$\\
		\hline
		$9.9 - 10.1$	&	$-0.47 \pm 0.12$ & $-0.19 \pm 0.09$ & $-0.08 \pm 0.08$ & $-0.09 \pm 0.09$ & $-0.22 \pm 0.11$ & $0.00 \pm 0.07$ & $-0.32 \pm 0.11$ & $-0.14 \pm 0.16$\\
					&	$0.81 \pm 0.6$ & $-0.26 \pm 0.18$ & $0.02 \pm 0.11$ & $0.42 \pm 0.19$ & $-0.44 \pm 0.11$ & $0.18 \pm 0.01$ & $0.18 \pm 0.01$ & $-0.27 \pm 0.09$\\
		\hline
		$10.1 - 10.3$	&	$-0.75 \pm 0.09$ & $-0.16 \pm 0.05$ & $-0.05 \pm 0.05$ & $-0.14 \pm 0.09$ & $-0.17 \pm 0.07$ & $0.01 \pm 0.06$ & $-0.34 \pm 0.08$ & $-0.29 \pm 0.06$\\
					&	$0.7 \pm 0.18$ & $-0.63 \pm 0.10$ & $0.07 \pm 0.15$ & $-0.13 \pm 0.01$ & $0.03 \pm 0.12$ & $-0.08 \pm 0.05$ & $0.13 \pm 0.32$ & $0.01 \pm 0.31$\\
		\hline
		$10.3 - 10.6$	&	$-0.85 \pm 0.05$ & $-0.13 \pm 0.05$ & $-0.13 \pm 0.02$ & $-0.14 \pm 0.07$ & $-0.25 \pm 0.03$ & $-0.05 \pm 0.04$ & $-0.36 \pm 0.10$ & $-0.31 \pm 0.07$\\
					&	$0.63 \pm 0.5$ & $-0.51 \pm 0.08$ & $0.11 \pm 0.03$ & $0.22 \pm 0.23$ & $0.03 \pm 0.05$ & $0.15 \pm 0.01$ & $0.15 \pm 0.01$ & $-0.22 \pm 0.26$\\
		\hline
		$10.6 - 11.2$	&	$-0.82 \pm 0.08$ & $-0.18 \pm 0.05$ & $-0.12 \pm 0.03$ & $-0.52 \pm 0.04$ & $-0.18 \pm 0.04$ & $-0.10 \pm 0.03$ & $-0.14 \pm 0.04$ & $-0.33 \pm 0.11$\\
					&	$0.4 \pm 0.42$ & $-0.45 \pm 0.18$ & $0.10 \pm 0.15$ & $0.47 \pm 0.09$ & $-0.03 \pm 0.07$ & $0.24 \pm 0.14$ & $-0.28 \pm 0.52$ & $0.12 \pm 0.26$\\
		\hline
	\end{tabular}}
\end{table*}

\subsubsection{Radial gradients as a function of galaxy mass and type}
\begin{figure*}
\centering
\flushleft
  \includegraphics[width=.33\linewidth]{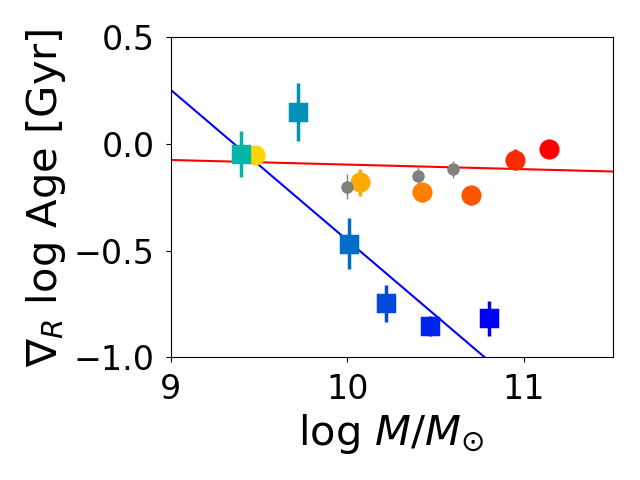}
  \includegraphics[width=.33\linewidth]{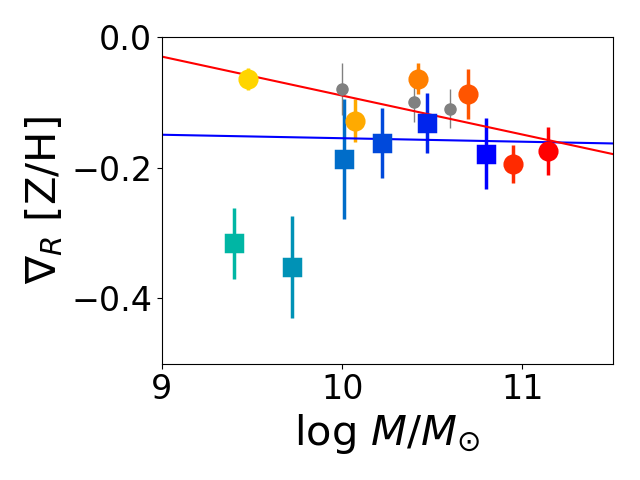}
       \includegraphics[width=.25\linewidth]{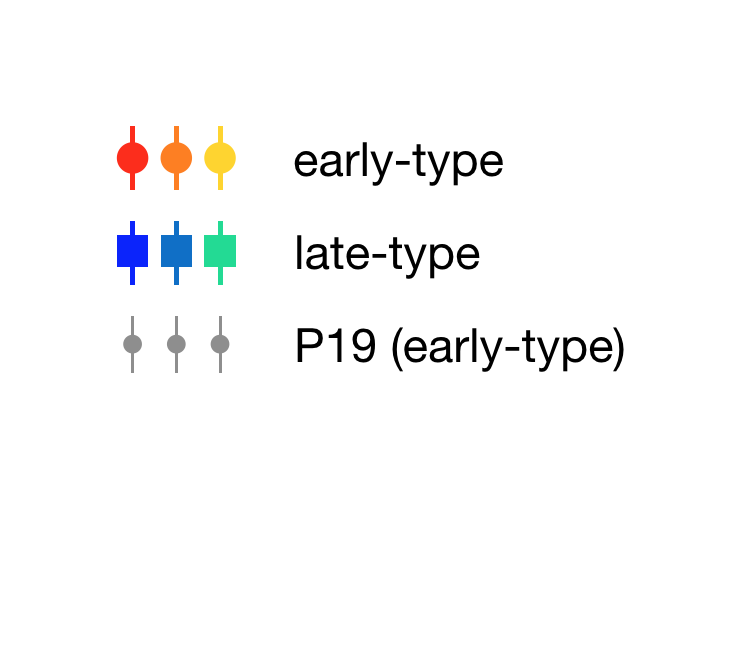}
  \includegraphics[width=.33\linewidth]{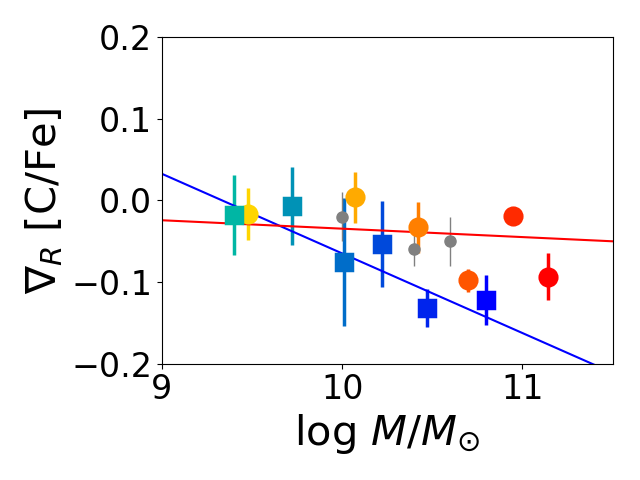}
    \includegraphics[width=.33\linewidth]{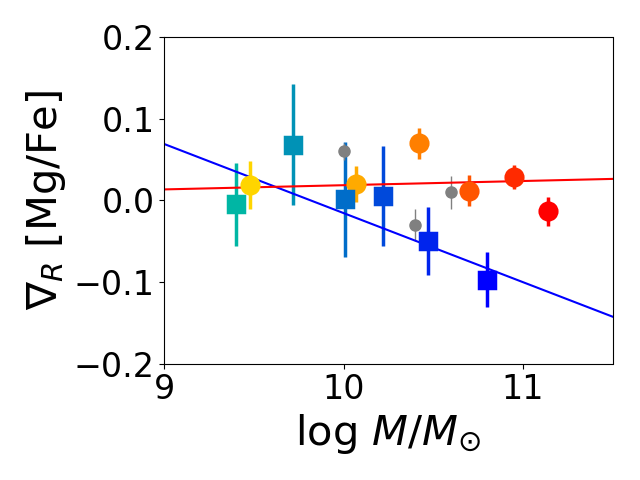}
      \includegraphics[width=.33\linewidth]{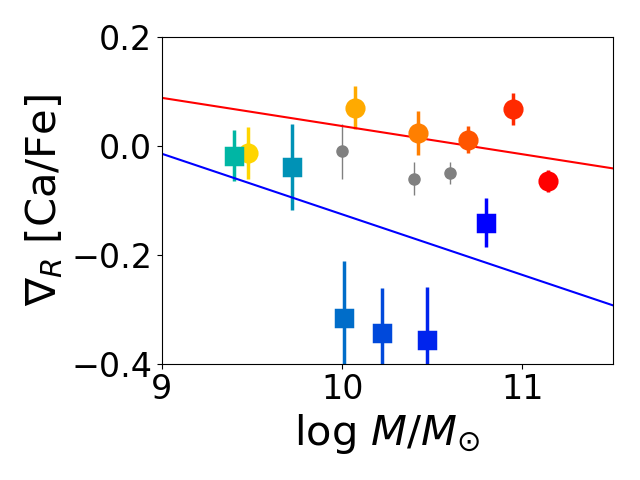}
  \includegraphics[width=.33\linewidth]{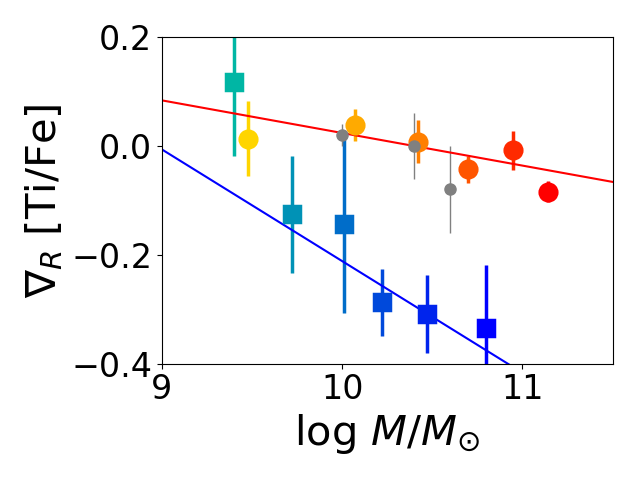}
  \includegraphics[width=.33\linewidth]{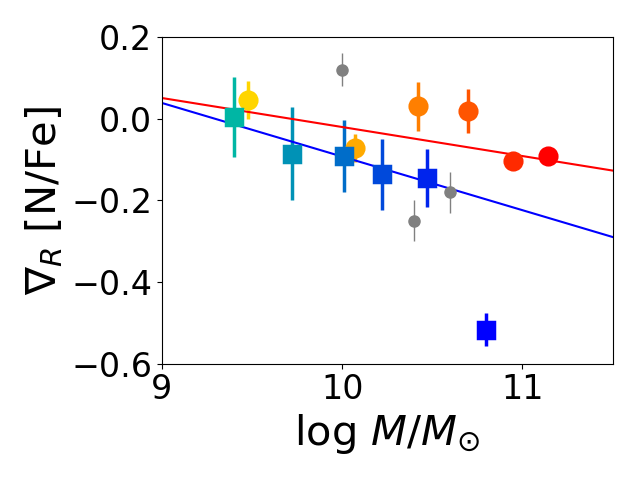}
  \includegraphics[width=.33\linewidth]{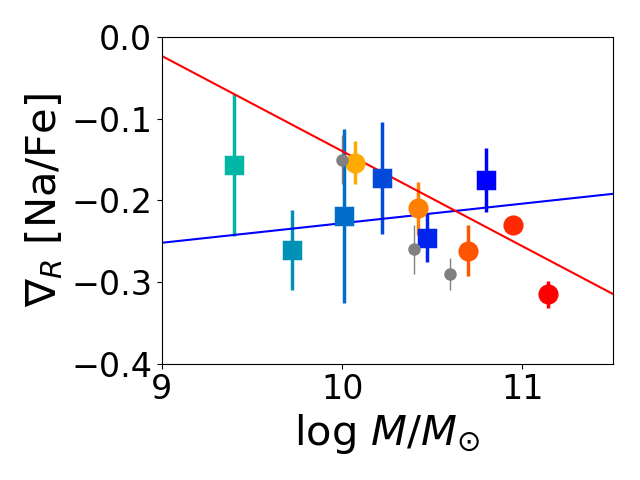}
\caption{For each parameter, the radial gradient and error within 1~R$_e$ is shown as a function of galaxy stellar mass. Square symbols show LTGs and circles show ETGs, with the colours represent different galaxy mass bins. $\sigma$-clipped linear fits to both galaxy types are shown. The grey symbols are the results for ETGs from \citetalias{Parikh2019}.}
\label{fig:rad_comparison}
\end{figure*}

\autoref{fig:rad_comparison} provides a comparison of the radial gradients within 1 R$_e$ in each parameter, for the different galaxy types. The value of the radial gradient and its error is shown as a function of mass, where the mass is the average of all galaxies in that mass bin. ETGs are shown as circles and LTGs are shown as squares; each mass bin is a different colour. The overlap in mass allows us to compare trends for different galaxy types at the same mass to determine whether stellar populations vary depending on morphology. Linear fits, with $\sigma$-clipping to exclude outliers, are shown for both galaxy types, and the corresponding slopes are given in \autoref{tab:mass_gradient_relations}. Also shown here are our results from \citetalias{Parikh2019}, as grey circles with error bars. These are consistent with our present ETG results for all parameters except [N/Fe].

For galaxies with log M/M$_{\odot}$ \textless 10, stellar population age gradients are small and do not depend on the type of galaxy. More massive galaxies however show a very clear divergence: for increasing galaxy mass, ETGs show negligible gradients, while  LTG gradients are negative. This is clearly shown by the flat red linear fit and the steep blue linear fit, which cross over at low masses. For metallicity, while ETG gradients become increasingly negative for more massive galaxies, LTGs have the same negative gradient regardless of mass. Metallicity gradients steepening for more massive ETGs is consistent with other studies from the MaNGA survey, \citet{Oyarzun2019} (after noting that they probe larger radii) and \citet{Lacerna2020}

For C and Mg, at low masses the radial gradients are slightly positive, and they become shallower or slightly negative as galaxy mass increases. The gradients are independent of the type of galaxy. For the heavier elements, ETGs display similar gradients at all masses for [Ca/Fe], and become slightly negative as mass increases for [Ti/Fe]. High mass spirals show larger variation in gradients with mass. The overall trend for Ca is unclear, but for Ti, the gradients become more negative as galaxy mass increases, like ETGs.

N gradients go from slightly positive or shallow at low masses, to negative at high masses for both types. Finally, for Na, ETGs display a slight steepening in the gradient for the most massive galaxies while gradients are consistent for both types at intermediate masses. This suggests that the mass or some other property drives Na abundance variation rather than galaxy type.

\begin{table*}
	\centering
	\caption{Slopes of the mass-gradient relations for the galaxy types i.e. gradients of the linear fits from \autoref{fig:rad_comparison}}
	\label{tab:mass_gradient_relations}
	\resizebox{\linewidth}{!}{%
	\begin{tabular}{cccccccccc}
		\hline
		Galaxy type 		& 	Age 			   & [Z/H] 			 & [C/Fe] 		       & [N/Fe] 		     & [Na/Fe] 		  & [Mg/Fe] 	       	  & [Ca/Fe] 	     & [Ti/Fe]\\
		\hline
		ETGs			&	$-0.02 \pm 0.06$ & $-0.06 \pm 0.03$ & $-0.01 \pm 0.05$ & $-0.07 \pm 0.03$ & $-0.12 \pm 0.05$ & $0.01 \pm 0.01$ & $-0.05 \pm 0.05$ & $-0.06 \pm 0.03$\\
		LTGs			&	$-0.70 \pm 0.18$ & $-0.01 \pm 0.06$ & $-0.10 \pm 0.03$ & $-0.13 \pm 0.02$ & $0.02 \pm 0.04$ & $-0.08 \pm 0.03$ & $-0.11 \pm 0.09$ & $-0.21 \pm 0.06$\\
		\hline
	\end{tabular}}
\end{table*}

Briefly, we find several interesting features while studying these radial gradients in these parameters for different galaxy types. The most striking difference is in the age gradients. While ETGs display relatively flat age gradients, at all masses, LTGs show negative age gradients, which become steeper with mass. Metallicity behaves exactly the opposite way, with negative gradients steepening with mass for ETGs and negative gradients overall for LTGs. These properties could provide clues into how these different galaxy types and their stellar populations assembled and evolved. C and Mg abundances and radial gradients are remarkably similar for both types, suggesting that star formation timescales are independent of galaxy type but rather depend on some other physical property i.e. galaxy mass or velocity dispersion. Na behaves in the same way as metallicity, showing negative gradients for LTGs, and negative gradients that steepen with mass for ETGs. We explore relations with velocity dispersion in the following section.

\subsubsection{Local vs global relationships}
Along with radial gradients, it is informative to look at trends with velocity dispersion. We plot the spatially-resolved parameters for both galaxy types, in different mass bins, so that we can identify local deviations (radial) from the global relation with velocity dispersion. The increased sample size and inclusion of different morphologies considerably expands our parameter space compared to \citetalias{Parikh2019}, Section 3.3.

\begin{figure*}
\centering
  \includegraphics[width=.33\linewidth]{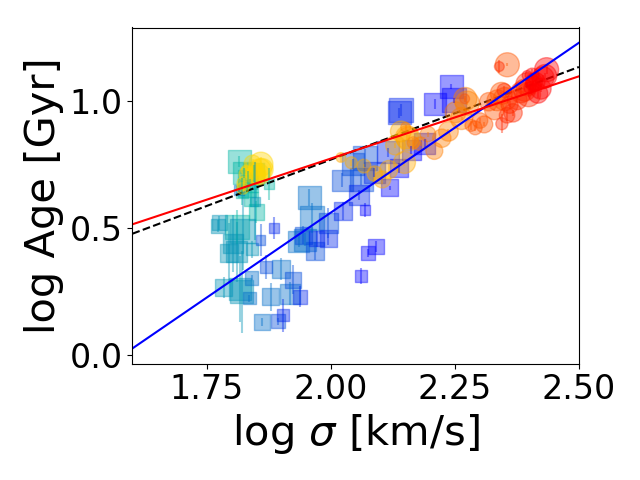}
  \includegraphics[width=.33\linewidth]{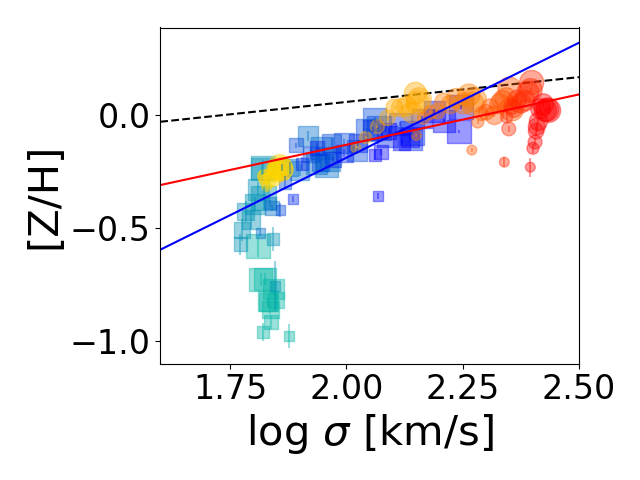}
       \includegraphics[width=.33\linewidth]{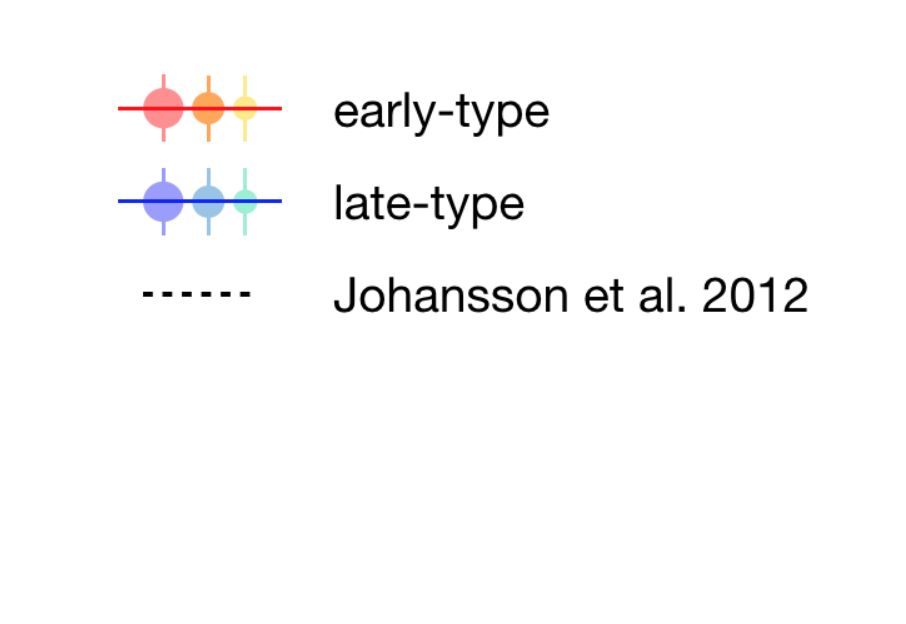}
  \includegraphics[width=.33\linewidth]{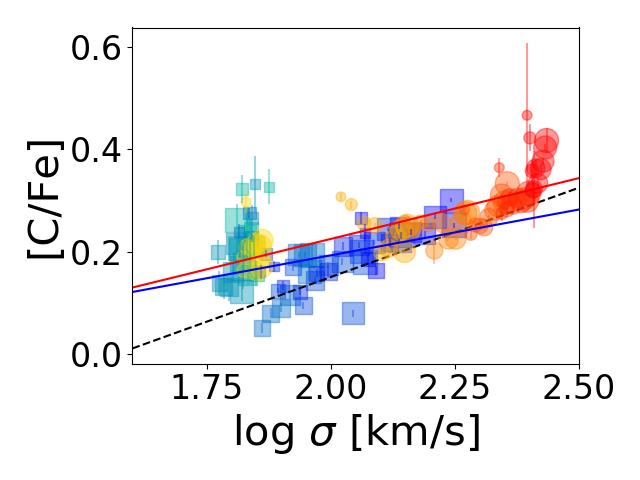}
    \includegraphics[width=.33\linewidth]{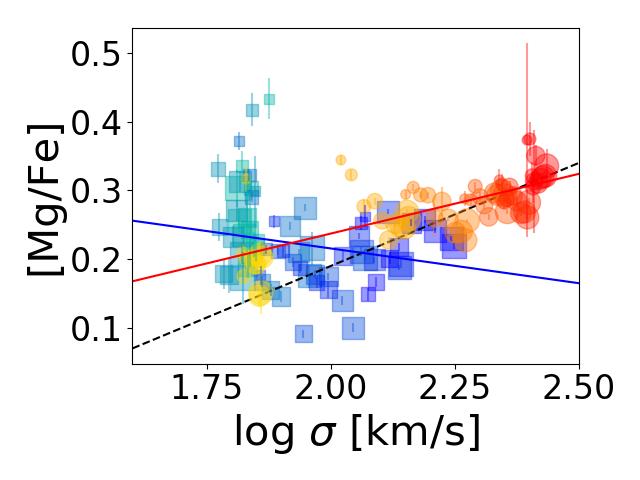}
      \includegraphics[width=.33\linewidth]{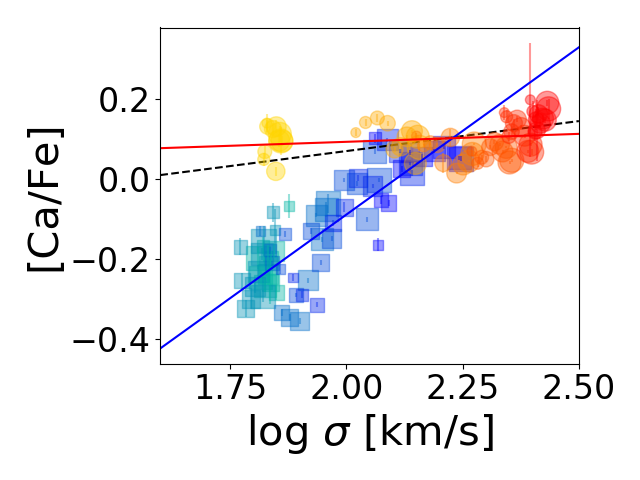}
  \includegraphics[width=.33\linewidth]{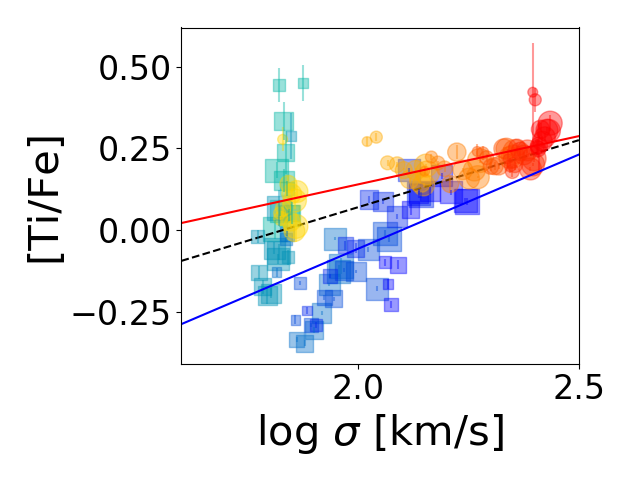}
  \includegraphics[width=.33\linewidth]{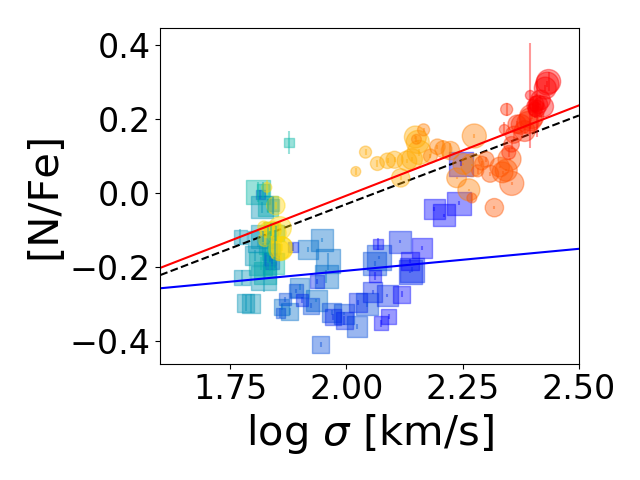}
  \includegraphics[width=.33\linewidth]{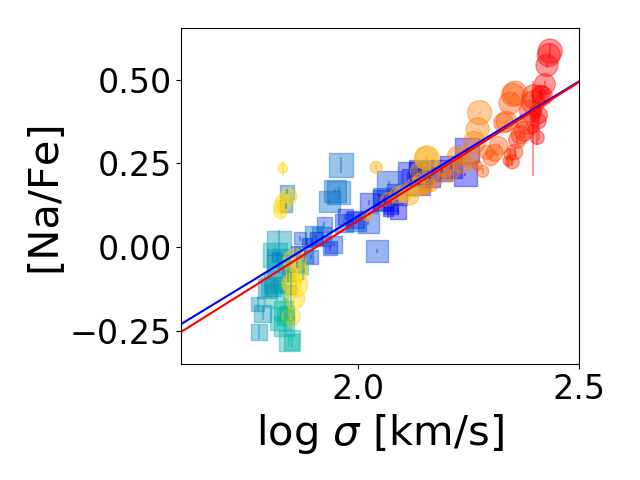}
\caption{Stellar population parameters are shown against the local velocity dispersion, for ETGs (circles) and LTGs (squares). The colours represent different galaxy mass bins, and decreasing symbol size represents increasing radius from the central radial bin, 0-0.1~R$_e$, to the outermost radial bin, 1.3-1.5~R$_e$. Linear fits (in log-log space) to all ETGs and LTGs are shown as red and blue solid lines, respectively. The lowest mass LTG bin is always offset, skewing the relation, and hence is not included in the fit. The relation from \citetalias{Johansson2012} is shown as a dashed black line for reference.}
\label{fig:comparison}
\end{figure*}

\autoref{fig:comparison} shows all our derived stellar population parameters against the velocity dispersion. These velocity dispersion profiles shown in \autoref{fig:sigma_profiles} have been calculated by combining the dispersions from the individual spectra belonging to each bin. The symbols represent 3 key pieces of information: i) ETGs are shown as circles and LTGs are shown as squares, ii) different colours within each type represent different mass bins, and iii) each symbol is one radial bin with decreasing symbol size represents increasing radius from the centre, 0-0.1~R$_e$, to the outermost bin, 1.3-1.5~R$_e$.

We also plot linear fits to all ETG results in red, and to LTGs in blue. Deviations from this line suggest internal gradients within galaxies are different from the overall global relation with $\sigma$. For LTGs, the lowest mass bin in this plot is always offset from the other points, causing any fit to be misleading, hence we do not include these points in the linear fit. Lastly, the relation from \citetalias{Johansson2012} is shown as a dashed black line for comparison. 

Starting with stellar population age, ETGs are older with increasing velocity dispersion. This effect is owing to higher mass galaxies being older and not due to radial gradients, since the points within each mass bin lie flat or parallel to the global relation (solid red line). LTGs show a steeper increase than ETGs, and also have steep local age gradients since the data points within galaxies do not lie along the global relation (solid blue line). At high $\sigma$ ETGs and LTGs have similar ages. This resembles the colour-magnitude diagram for galaxies in terms of a blue cloud and red sequence. For the metallicity, LTGs again show a steeper increase with velocity dispersion globally. High mass ETGs have steep local metallicity gradients, which are very evident in this figure, suggesting that internal processes within these galaxies are responsible for these steep gradients. These age and metallicity trends with velocity dispersion for ETGs and LTGs are consistent with \citet{Li2018}.

[C/Fe] and [Mg/Fe] both increase at a similar rate with $\sigma$, independent of galaxy type although [Mg/Fe]-$\sigma$ shows some scatter. The relations within galaxies deviates only for the highest mass ETGs which have a steep local [C/Fe] gradient. Since [Mg/Fe] is linked with formation timescales, this suggests that the formation time is the same regardless of galaxy type or as a function of radius but only depends on the velocity dispersion. This was also hinted at by \autoref{fig:rad_comparison}. \citep{Greene2019} find [$\alpha$/Fe] for ETGs to be correlated with velocity dispersion and stellar mass, and saturating at higher mass. 

\begin{figure*}
\centering
\flushleft
  \includegraphics[width=.33\linewidth]{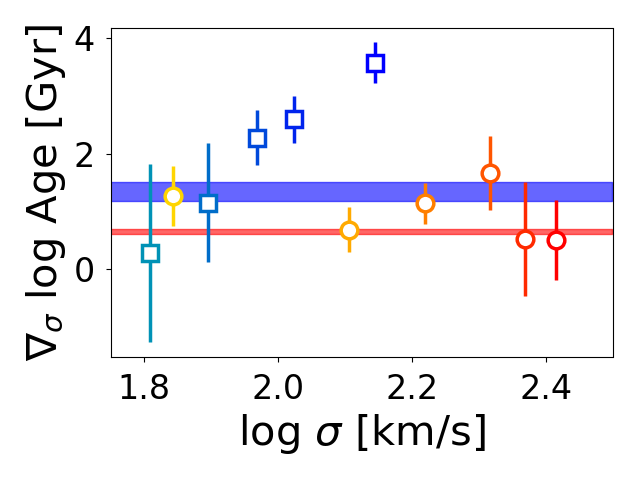}
  \includegraphics[width=.33\linewidth]{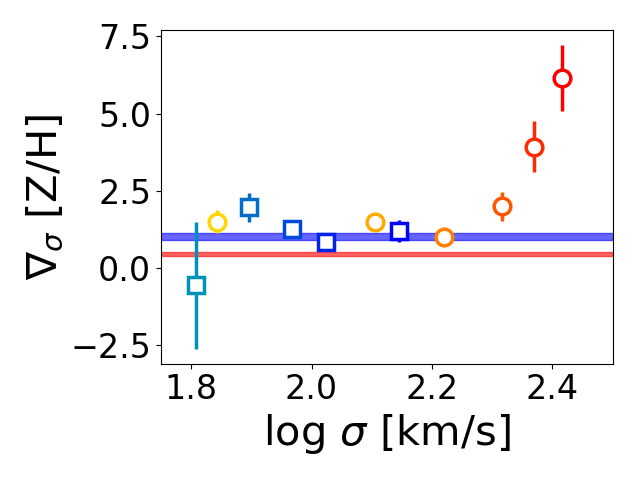}
      \includegraphics[width=.25\linewidth]{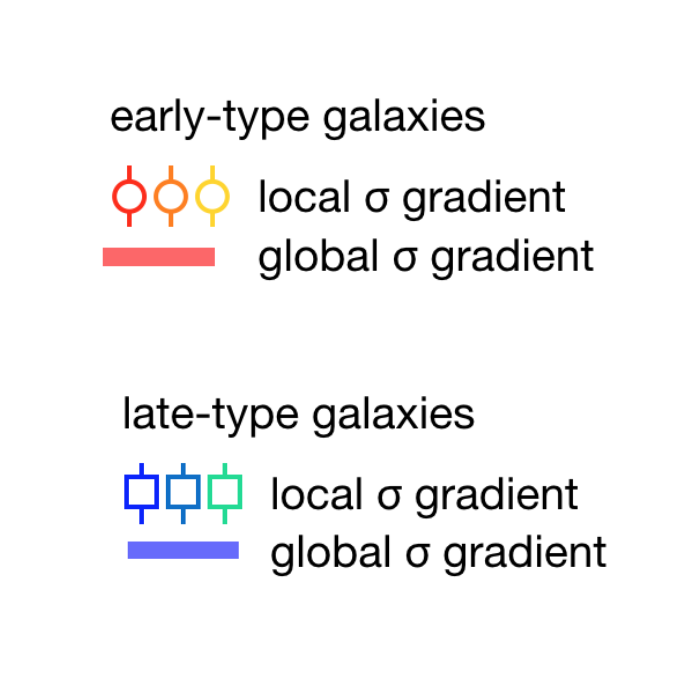}
  \includegraphics[width=.33\linewidth]{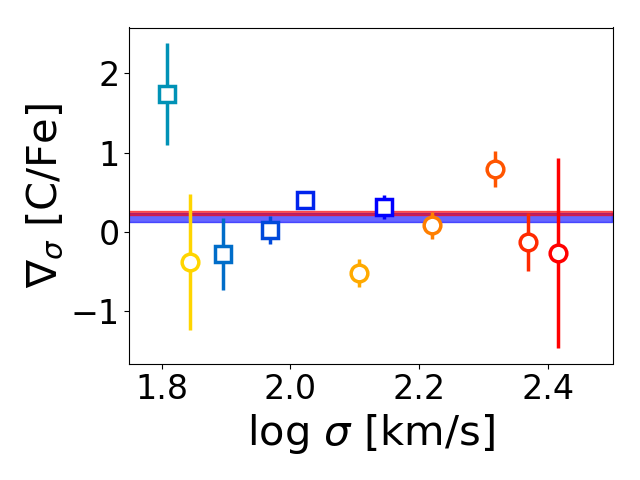}
   \includegraphics[width=.33\linewidth]{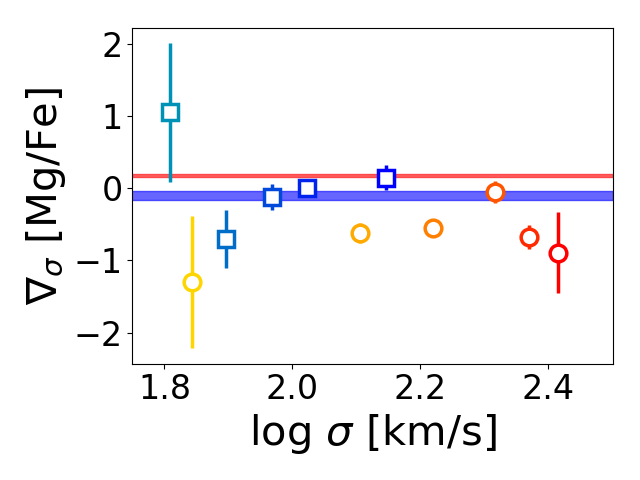}
   \includegraphics[width=.33\linewidth]{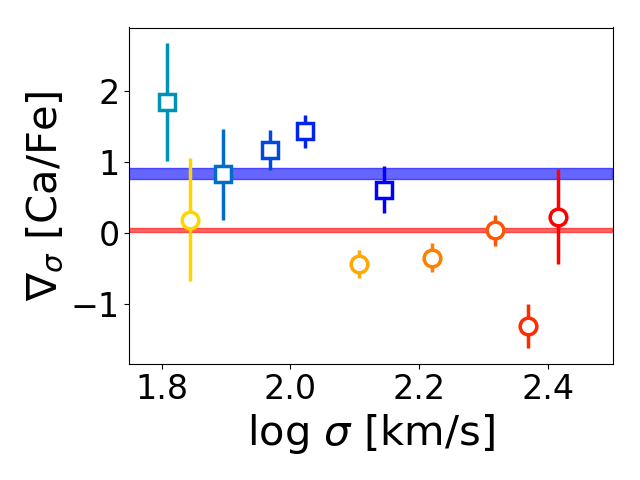}
  \includegraphics[width=.33\linewidth]{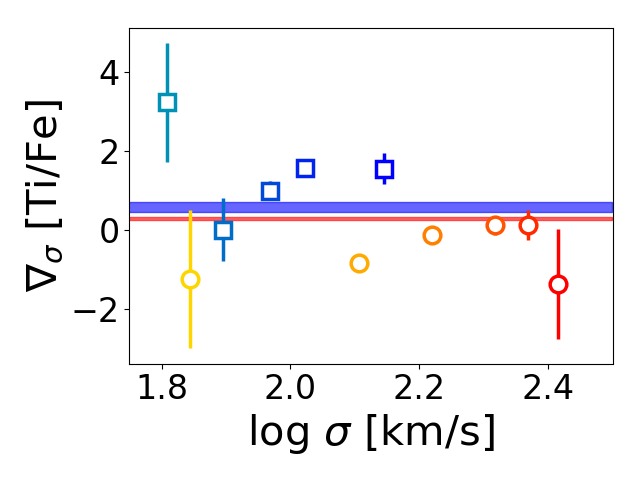}
  \includegraphics[width=.33\linewidth]{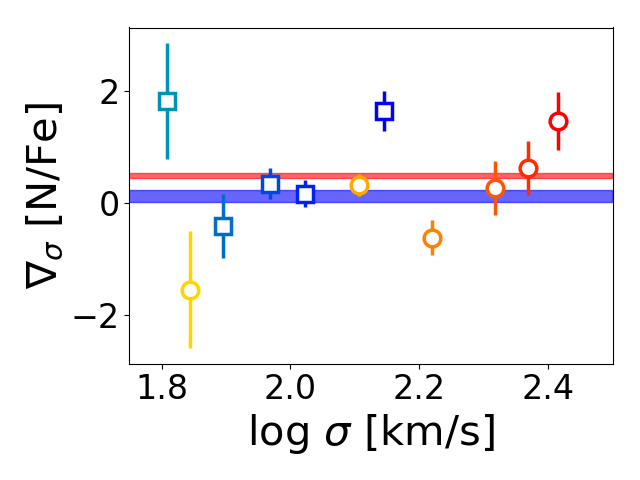}
  \includegraphics[width=.33\linewidth]{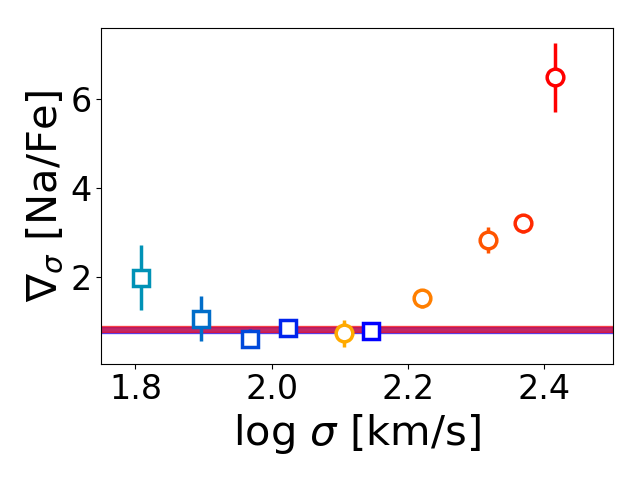}
\caption{Gradients of stellar population parameters with velocity dispersion are plotted as a function of the velocity dispersion. The global gradients across all galaxy masses are shown as shaded regions for ETGs (red) and LTGs (blue). The local gradients within galaxies for each mass bin are shown as circles for ETGs and squares for LTGs. Error-bars and width of the shaded region represent $\pm$ 1-$\sigma$ errors.}
\label{fig:gradientcomparison}
\end{figure*}

\citet{Thomas2006} find that the relations of age, metallicity, and [$\alpha$/Fe] with $\sigma$ are the same for ETGs and LTGs. It is interesting that with the added spatial information, we find differences between the ages and metallicities in these galaxy types. Due to the negative age and metallicity gradients within LTGs, both age and metallicity show steeper relations with $\sigma$ than ETGs. A recent result from the SAMI survey \citep{Croom2012} obtained ETG ages, metallicities, [Mg/Fe], and [C/Fe], showing that the velocity dispersion is the dominating driver behind chemical gradients \citep{Ferreras2019}. They find, consistent with our results, that [Mg/Fe] gradients are less steep for more massive galaxies.

We can see that for both [Ca/Fe] and [Ti/Fe], ETGs display a clear flat or shallow increase with $\sigma$. LTGs are a bit more complicated and appear to have steep relations with the velocity dispersion such that lower $\sigma$ galaxies have much lower [X/Fe] values. 

There is some local variation in the [N/Fe]-$\sigma$ relation for ETGs, and LTGs show some scatter with $\sigma$. [Na/Fe] increases very steeply with velocity dispersion. The different types seamlessly blend together. It is very evident for the high mass ETGs that the local change in Na abundance is much steeper than the overall trend, and not at all for LTGs, very similar to the trends with metallicity. Due to the steep [Na/Fe]-$\sigma$ relation, high mass spirals have Na abundances similar to their C and Mg abundances, while high mass ETGs have much larger Na abundances compared to the other elements.

In all cases, the agreement with \citetalias{Johansson2012} is remarkably good, with a significant difference seen only for metallicity. Their metallicities are consistent with our results for the innermost radial bins, while the metallicity gradients within galaxies, which lead to lower values at large radii, cause these radial bins to be offset.

To illustrate the trends seen in \autoref{fig:comparison}, \autoref{fig:gradientcomparison} shows gradients of stellar population parameters with velocity dispersion, as a function of the velocity dispersion. This is similar to \citetalias[][Figure 5]{Parikh2019}, with additions in mass and type. For ETGs, the red shaded region is the gradient of each parameter with velocity dispersion for all galaxies i.e. the slope of the solid red line from each panel in \autoref{fig:comparison}. The circles are the gradients with velocity dispersion within galaxies, with each galaxy mass bin represented by a different colour as before. These have been placed at the average $\sigma$ for each bin and are shown with error bars. The blue shaded regions and square symbols are the same for LTGs. A positive gradient with $\sigma$ means a negative radial gradient. Similar to the radial gradients, the errors on the gradients of parameters with the velocity dispersion come from the deviation of points around the linear fit. The two lowest mass bins of LTGs are not shown since one has large errors due to H$\beta$ and the other has not been included in the linear fits in \autoref{fig:comparison}.

The extension in mass compared to \citetalias{Parikh2019} confirms that for ETGs, the local age gradients (circles) are consistent with the global age gradient (red shaded region, 0.8 log Gyr/log km/s), and the local metallicity gradients are steeper than the global relation (0.3 dex/log km/s). The local metallicity gradients become steeper with increasing mass, with $\nabla_\sigma$ [Z/H] $\sim5$~dex/log km/s for the highest mass galaxies. Very interestingly, making the same comparison for LTGs shows the converse relations. For these galaxies, the local age gradients (between 2 - 3 log Gyr/ log km/s) are steeper than the global relation (1 log Gyr/log km/s), and higher mass spirals have very steep age gradients with stars becoming increasingly younger moving outwards from the galaxy. Meanwhile the local metallicity gradients for these galaxies are generally consistent with the global gradient (1 dex/log km/s). Hence internal processes regulate metallicity in ETGs but not age, while the converse is true for LTGs. Metallicity gradients with respect to the velocity dispersion show a much stronger dependence on mass than the radial gradients seen in \autoref{fig:rad_comparison}.

For both C and Mg, independent of galaxy type, the global gradients with velocity dispersion are similar. The same is true locally within galaxies, independent of mass and type since the local $\sigma$ gradients (symbols) are consistent with the global relations (shaded regions). These results are consistent with \citetalias{Parikh2019} for ETGs and we now show that the same is true for LTGs. For ETGs, [Mg/Fe] increases slightly in the outer regions, causing the local $\sigma$ gradients to be negative and systematically offset from the global relation.

The global gradient of Ca with velocity dispersion is steeper for LTGs than for ETGs. Very low [Ca/Fe] ratios in LTGs at large radii lead to the steeper relation. The relations within galaxies are generally consistent with the global trend for each type. For Ti, the global gradients of the galaxy types are consistent with each other.

It appears that the most massive galaxies have slightly steeper local relations in N than the global relation for their respective types, as also found in \citetalias{Parikh2019} for ETGs, but the effect is weaker here and disappears at low masses, particularly for LTGs. For Na, we expand upon our earlier picture and show that higher mass galaxies continue to have increasingly steep local abundance gradients, reaching 6 dex/log km/s. LTGs display shallower $\nabla_\sigma$ [Na/Fe] seem to fit into this picture because of their velocity dispersion, and hence it seems that galaxy type does not drive this relation. There is a significant difference between the global and local relations only for the galaxies with the largest velocity dispersions ($>$230km/s), hence only for ETGs. Both types have the same overall global gradient with velocity dispersion.

In summary, the age, total metallicity, and [Na/Fe] are the most striking results, with interesting differences between the galaxy types. We discuss the interpretations of our results in the next Section, including how these results help answer some of the questions raised by \citetalias{Parikh2019}.

\section{Discussion} \label{sec:discussion}
We have derived ages, metallicities, and individual chemical abundances of 6 elements as a function of galaxy mass, velocity dispersion, radius, and type. In this Section, we discuss what new insights these results offer and the implications of our results, particularly for the different galaxy types.

There were several changes to the sample selection and stacking method compared to \citetalias{Parikh2019}, including a different morphological selection, binned kinematic data, and a lower S/N criterion for individual spectra. It is therefore encouraging to find that our results for ETGs are generally consistent with what we found in \citetalias{Parikh2019}. The addition of lower mass and higher mass ETGs has allowed us to extend and confirm trends we spotted with our previously limited sample in mass. The increased S/N ratio at large radii, due to stacking more galaxies in each mass bin and increasing the bin width, makes our radial gradients in this work more robust.

\begin{figure*}
\centering
    \includegraphics[width=.8\linewidth]{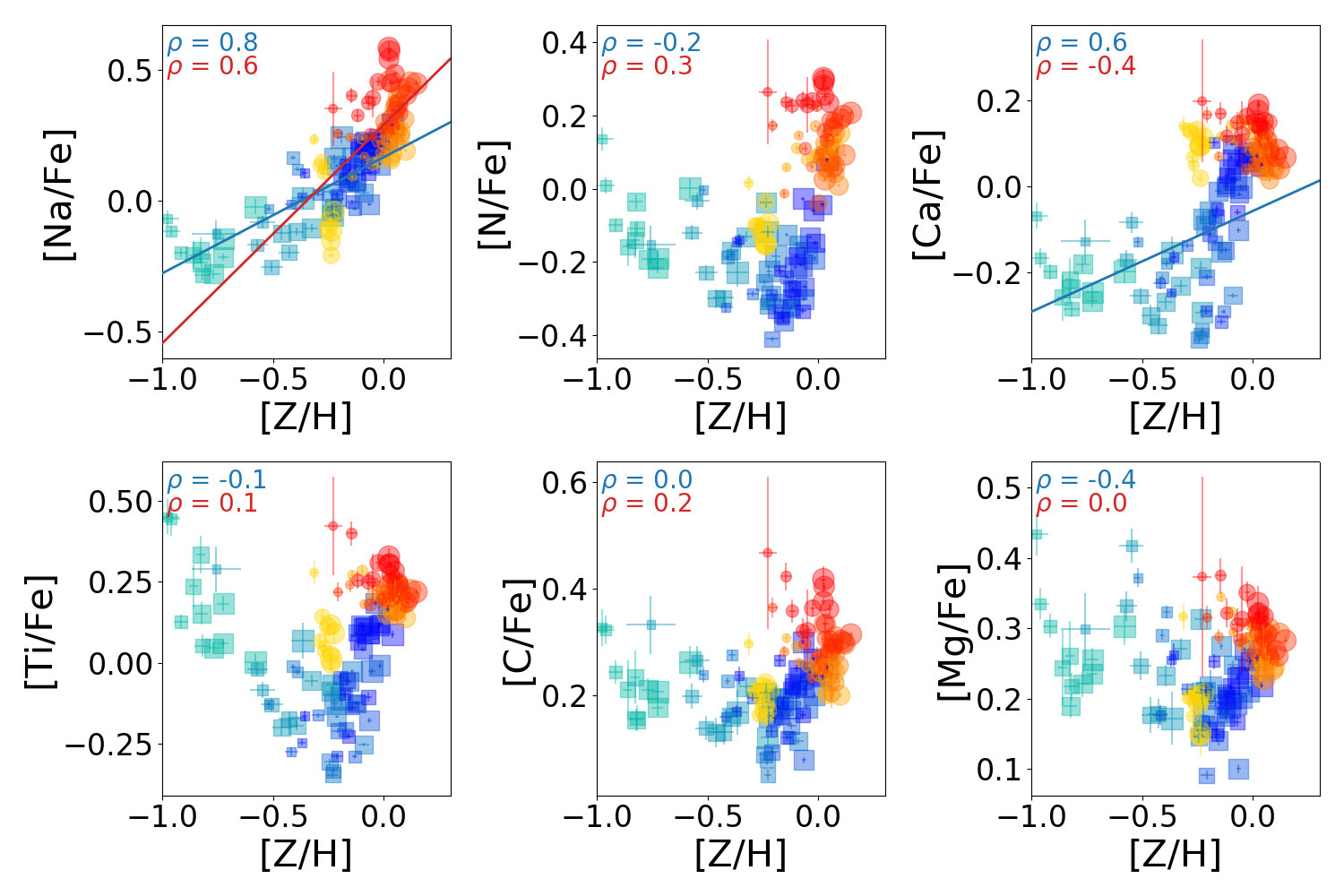}
  \caption{Correlations of element abundances with total metallicity. ETGs are shown as circles and LTGs are shown as squares. Different colours represent different mass bins, and symbol size decreases with increasing radius. The Spearman correlation coefficient, $\rho$, is shown in each panel for ETGs and LTGs separately. Linear fits are plotted for correlations \textgreater 0.5.}
\label{fig:z_correlation}
\end{figure*}

\begin{figure*}
\centering
    \includegraphics[width=.8\linewidth]{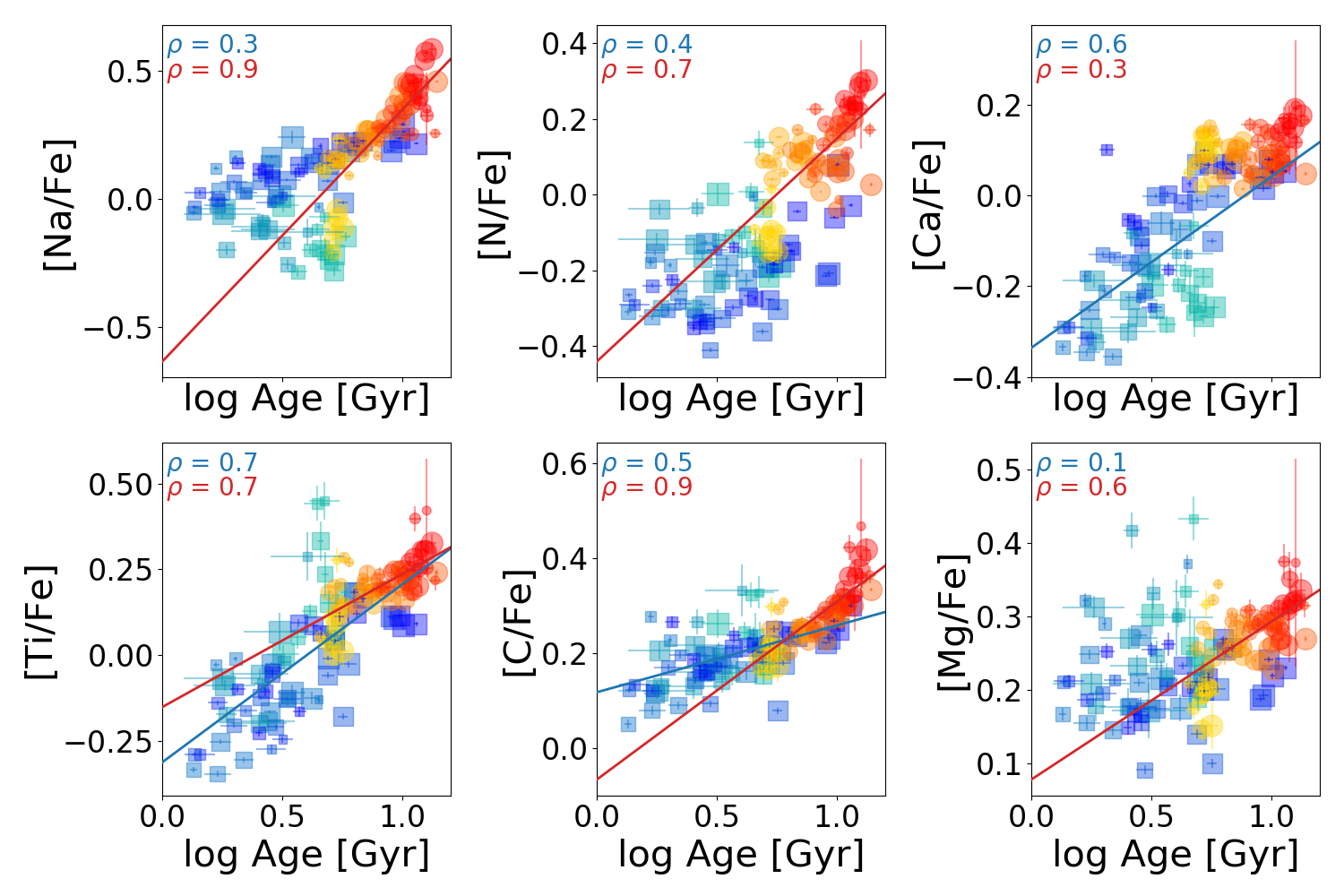}
  \caption{Same as \autoref{fig:z_correlation} for correlations with stellar population age.}
\label{fig:age_correlation}
\end{figure*}

\subsection{Correlations between stellar population parameters}
We study the correlations of all element abundances with metallicity in \autoref{fig:z_correlation}. The derived abundances are plotted as circles for ETGs and squares for LTGs with error-bars, with colours representing different mass bins. Each symbol is a different radial bin and decreasing circle size showing increasing radius. The Spearman correlation coefficient, $\rho$, is shown in each panel for ETGs and LTGs separately, and a linear fit is plotted when $|\rho| \textgreater 0.5$.

With a larger parameter space in [X/Fe] and [Z/H] compared to \citetalias{Parikh2019}, we confirm the correlation between [Na/Fe] and metallicity. LTGs continue this relation into the low metallicity regime, showing further evidence for metallicity-dependent Na enrichment from supernovae. There are internal trends with galactic radius and an overall global trend of [Na/Fe] with [Z/H]. [N/Fe] is expected to also be strongly correlated with metallicity due to secondary N production, however we do not find correlations for either galaxy types. [Ca/Fe] is correlated with metallicity for LTGs, and this could be the reason for extreme Ca under-abundance observed in these galaxies. The remaining elements do not show clear correlations.

We explore correlations with age in \autoref{fig:age_correlation}, since the spirals are mostly different from ETGs through their younger ages. Interestingly, all elements appear to show some correlation with age. [Na/Fe] shows an age dependence on top of the metallicity dependence seen in \autoref{fig:z_correlation}, likely driven by enhanced Na enrichment at high metallicity.

\citet{Smith2009} explored two-parameter relations for abundance patterns with the velocity dispersion and metallicity, and found that these greatly reduced the scatter. This is consistent with at least our [Na/Fe] results appearing to be driven by two physical processes.

\subsection{New insights and constraints on nucleosynthetic yields}
Our ETG abundances from \citetalias{Parikh2019} left some open questions. Here we explore whether the new work sheds further light on these. We find that LTGs have steep negative age gradients, and negative metallicity gradients. These gradients support inside-out formation of disk galaxies \citep[see discussion in][]{Goddard2017}. In spiral galaxies with masses $> 9.6 \log M/M_{\odot}$, we see that the age flattens out, and even increases in some cases, at large radii. This could be due to the radial migration of slightly older stars moving outwards through the disk. \citet{Cook2016} measure stellar population gradients for ETGs from the Illustris simulation and find that galaxies with large accreted fractions have shallower metallicity gradients and no effect on the age gradients in the halo. They estimate 2 - 4 R$_e$ for the galaxy halo. In the future, it would be interesting to use the secondary sample from MaNGA, which reaches out to 2.5~R$_e$, in order to determine ages and abundances in the inner halo.

Solar [C/Mg] ratios places a lower limit on star formation timescales \citepalias{Johansson2012}. Both [C/Fe] and [Mg/Fe] increase with velocity dispersion, independent of galaxy type and mass, and with shallow or negligible radial gradients. Hence, galaxies with higher velocity dispersions formed their stars earliest, and these timescales do not vary with radius. We now confirm the same across a larger parameter space in galaxy mass, and LTGs also generally follow this trend. Although there are some interesting differences at large radii. Spirals appear to show an increase in C and Mg abundances beyond 0.8 R$_e$, with lower mass galaxies showing a greater increase such that at large radii, abundances are anti-correlated with mass. This is potentially very interesting and could be a result of complicated accretion histories.

We again find low [Ca/Fe] ratios, close to zero, with negligible radial gradients for ETGs of all masses. The same is seen for spiral galaxies, however the low mass galaxies have extremely sub-solar Ca abundances of -0.5 dex. However, Ca4227 is the weakest among the absorption features used in this work, hence it not be strong enough for accurate measurements for the low mass galaxies. Note also that this index is sensitive to both [C/Fe] and [N/Fe], bringing additional sources of error.

Since Ti was found to behave like the lighter elements C and Mg, we attributed this to the production of Ti in Type II supernovae. We find this to be the case again for all but the lowest mass ETGs. Puzzlingly, for LTGs we find that high mass galaxies have slightly sub-solar Ti abundances. It remains to be seen whether the strange Ca and Ti trends in spiral galaxies point to the difficulty in obtaining these parameters.

We previously reported radial gradients for [N/Fe], but the trend with mass was unclear. Low mass galaxies had a positive gradient and higher mass galaxies showed negative gradients. With our improved study, we show that this weak signal disappears and we find flat gradients. As a result, N under-abundance is restricted to low mass galaxies, while high-mass galaxies have super-solar N abundances. Finally, for Na we find that the new sample extends the relations with velocity dispersion and metallicity, such that as $\sigma$ and [Z/H] decrease, Na abundances reach solar and sub-solar values, matching Mg abundances and eventually being less enhanced than Mg. This provides further support for metallicity-dependent Na enrichment \citep{Kobayashi2006}.

\subsection{Bulge-Disk Decomposition}
\label{sec:bulgedisk}
As mentioned in \autoref{sec:localsigma}, we do not attempt to separate the contributions from the bulge and the disk, of which the local $\sigma$ is a combination of. Other works provide interesting clues regarding these components, and our correlations with $\sigma$ also motivate further investigations.

\citet{Tabor2019} perform a spectroscopic bulge-disk decomposition of MaNGA ETGs and show that these components represent different stellar populations with distinct kinematics. They find that galaxy bulges have a lower stellar spin parameter, $\lambda_R$ \citep{Emsellem2007}, while disks tend towards higher values. In terms of the stellar populations, both bulge and disk regions are found to have similar ages, while the former are metal-rich than the latter. These stellar population results are consistent with \citet{Fraser-McKelvie2018} who separate bulge and disk regions in S0 galaxies.

Furthermore, \citet{Cortese2016} find using SAMI data that kinematically fast rotator ETGs and LTGs lie on a continuous kinematic plane defined by the spin and Sersic index. Analysing detailed stellar populations while dividing galaxies using kinematic classifications, as well as morphological, would help shed further light on this matter.



\section{Conclusions}
\label{sec:conclusions}
In this work we investigated stellar population gradients in a large galaxy sample of 895 ETGs and 1005 LTGs, ranging in stellar mass from $8.60 - 11.33\;\log M/M_{\odot}$. We used a morphological classification scheme based on a Deep Learning Catalog to separate early and LTGs. We modified our stacking procedure, including the use of wider radial bins in outer regions of galaxies to compensate for the loss in surface brightness, to produce high quality spectra in mass and radial bins out to 1.5 Re. The LTG sample is cut based on inclination in order to select mostly face-on galaxies, to reduce the effect due to ISM contamination of absorption features such as NaD. We measured absorption features and determined best-fit stellar population models to derive ages, metallicities, and individual element abundances for both galaxy types.\\

\noindent Our main results are summarised below:

\begin{itemize}
\item We find stark differences between the ages and metallicities of the two types. As well as recovering results from \citetalias{Parikh2019} for intermediate-mass ETGs, we show across a wider parameter space that ETGs have flat age and negative metallicity gradients where the latter are steeper than the global relation with velocity dispersion. LTGs show negative age and negative metallicity gradients, pointing to inside-out formation scenarios. For these galaxies, it is the local age variation that is steep compared to the general trend. We observe a reversal of the age gradient beyond the half-light radius.

\item For the abundances, we find C and Mg to be similarly enhanced in ETGs and LTGs, increasing as a function of velocity dispersion, independent of radius. This implies constant star-formation timescales at all locations in all galaxy types, which are only driven by the velocity dispersion.

\item We find [Ca/Fe] to be less enhanced also in LTGs, with negative radial gradients. The inference that Ti behaves similar to the lighter $\alpha$ elements and therefore is produced mostly in Type II supernovae is not clear since high-mass LTGs show much lower [Ti/Fe] values than Mg, although not as low as Ca. LTGs show flattening or even increase in slopes of [C/Fe], [Mg/Fe], and [N/Fe] beyond 1R$_e$, suggesting that the stellar populations in the outskirts of LTGs are different.

\item We confirm steep local Na gradients for ETGs, which steepen with increasing galaxy mass, while LTGs also show radial gradients in Na.

\item Looking at trends with the velocity dispersion for both galaxy types, we find that differences arise due to the local age gradients, and very low abundances ratios for [Ca/Fe] and [Ti/Fe] in LTGs. The most striking differences between galaxy types are seen for the age, metallicity, and Na abundance, providing clues regarding the formation scenarios for these galaxies. Constraints on Ca and Ti must be improved through updated stellar population models in order to gain better insights.

\item Strong Na abundance gradients within galaxies suggest internal processes drive Na enrichment. [Na/Fe]-[Z/H] relations show that Na abundances increase with both of these properties, independent of galaxy type and provide support for metallicity-dependent supernova yields for this element. Na, as well as the other elements, also correlate with age suggesting a combination of effects leading to high Na abundances.
\end{itemize}

In the future, we plan to derive the low-mass IMF slope for these galaxies. Of particular interest are LTGs, for which the IMF has been studied to a very limited extent. The spectral features used in this work are not sensitive to the IMF, but the MaNGA wavelength range makes a several such features accessible.

\section*{Acknowledgements}
We thank the referee for comments which have greatly improved the paper. TP acknowledges funding from a University of Portsmouth PhD bursary. The Science, Technology and Facilities Council is acknowledged for support through the Consolidated Grant Cosmology and Astrophysics at Portsmouth, ST/N000668/1. Numerical computations were performed on the Sciama High Performance Computer (HPC) cluster which is supported by the Institute of Cosmology of Gravitation, SEPnet and the University of Portsmouth. This research made use of the \textsc{python} packages \textsc{numpy} \citep{VanDerWalt2011}, \textsc{scipy} \citep{Jones2001}, \textsc{matplotlib} \citep{Hunter2007}, and \textsc{astropy} \citep{Astropy2013}. Galaxy images were obtained using the Marvin web interface \citep{Cherinka2019}.

Funding for the Sloan Digital Sky Survey IV has been provided by the Alfred P. Sloan Foundation, the U.S. Department of Energy Office of Science, and the Participating Institutions. SDSS-IV acknowledges support and resources from the Center for High-Performance Computing at the University of Utah. The SDSS web site is www.sdss.org.

SDSS-IV is managed by the Astrophysical Research Consortium for the Participating Institutions of the SDSS Collaboration including the Brazilian Participation Group, the Carnegie Institution for Science, Carnegie Mellon University, the Chilean Participation Group, the French Participation Group, Harvard-Smithsonian Center for Astrophysics, Instituto de Astrof\'isica de Canarias, The Johns Hopkins University, Kavli Institute for the Physics and Mathematics of the Universe (IPMU) / University of Tokyo, Lawrence Berkeley National Laboratory, Leibniz Institut f\"ur Astrophysik Potsdam (AIP), Max-Planck-Institut f\"ur Astronomie (MPIA Heidelberg), Max-Planck-Institut f\"ur Astrophysik (MPA Garching), Max-Planck-Institut f\"ur Extraterrestrische Physik (MPE), National Astronomical Observatories of China, New Mexico State University, New York University, University of Notre Dame, Observat\'ario Nacional / MCTI, The Ohio State University, Pennsylvania State University, Shanghai Astronomical Observatory, United Kingdom Participation Group, Universidad Nacional Aut\'onoma de M\'exico, University of Arizona, University of Colorado Boulder, University of Oxford, University of Portsmouth, University of Utah, University of Virginia, University of Washington, University of Wisconsin, Vanderbilt University, and Yale University.

\section*{Data Availability}
The data underlying this article are available at https://www.sdss.org/dr15/manga/ for DR15. Additional data generated by the analysis in this work are available in the article.


\bibliographystyle{mnras}
\bibliography{Refs}


\appendix
\section{Effect of different metallicities and alpha abundances}
\label{sec:app_mocks}
\begin{figure*}
\centering
  \includegraphics[width=\linewidth]{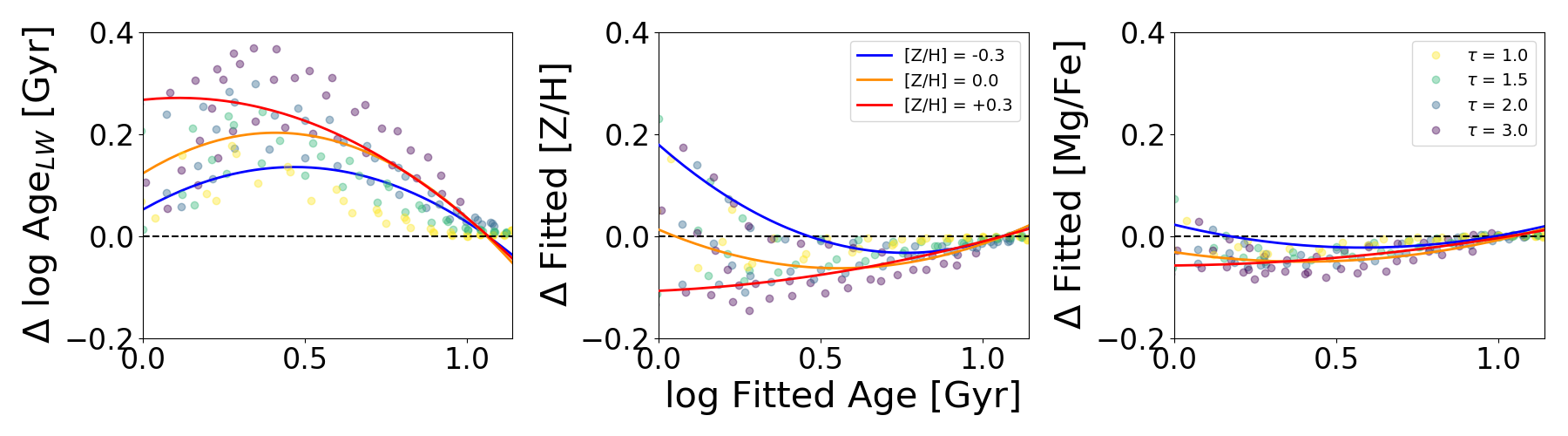}
  \includegraphics[width=\linewidth]{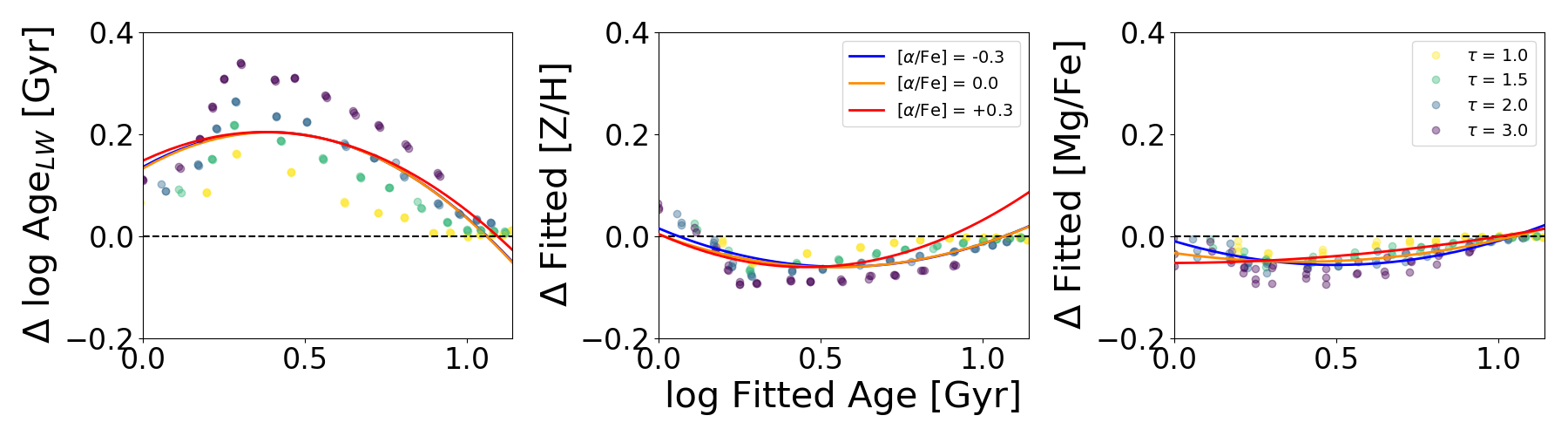}
\caption{LW Age, fitted metallicity and fitted [Mg/Fe] as a function of fitted age for different metallicities (top) and different abundances (bottom). The fitted ages correspond better to LW ages of a composite population. At young ages, the metallicity is overestimated by up to +0.2 dex. The fitted [Mg/Fe] appears to be offset from the expected solar value by +0.15 dex at young ages. This decreases for older populations. Red circles show the same for fitting SSPs. The orange curves are polynomial fits to the CSPs.}
\label{fig:mocks_zalphaeffect}
\end{figure*}

The effect of CSP models at different metallicities and abundances is shown in \autoref{fig:mocks_zalphaeffect}, analogous to \autoref{fig:mocks_taueffect}. The top panel shows [Z/H] = -0.3, 0.0, and +0.3 at fixed [$\alpha$/Fe] = 0.0, while the bottom panel shows [$\alpha$/Fe] = -0.3, 0.0, and +0.3 at fixed [Z/H] = 0.0. Tests are performed as described in \autoref{sec:mocks} for four different $\tau$ values. For each metallicity or abundance, a fit is performed to the results from all the $\tau$ models. The orange lines in both panels correspond to the results for solar metallicity and abundance shown in \autoref{fig:mocks_taueffect}.

Focussing on the effect of different metallicities first, we see for all parameters that biases are similar at larger fitted ages. At younger fitted ages, the LW Age is underestimated more for super-solar models and less for sub-solar models compared to solar metallicity. The difference between these models is however comparable to the spread due to different $\tau$s. Since our LTGs are most affected by the assumption of a singular SFH, and these galaxies are solar or more metal-poor, we can expect the level of correction to be no more than 0.2 dex. The middle panel shows that the metallicity for metal-poor populations could be underestimated for ages younger than 0.4 log Gyrs. Lastly the difference between the true and fitted [Mg/Fe] remains negligible regardless of the metallicity. We conclude that ages must be corrected using solar CSPs, while the effect on metallicities and abundances is small, with the caveat that for very young ages, metallicities might be underestimated by up to 0.2 dex.

The lower panel shows that the biases for all parameters are almost identical to the solar case, regardless of super- or sub-solar element abundances. Hence applying the correction based on the solar model is valid for our LTGs.

\newpage

\section{Index measurements and parameters}
\label{sec:app_tables}
We provide our index measurements for the optical indices CN1, Ca4227, Fe4531, C$_2$4668, H$\beta$, Mg$b$, Fe5270, Fe5335, and NaD for all radial and mass bins in \autoref{tab:ind_measurements_ET} for ETGs and \autoref{tab:ind_measurements_LT} for LTGs, at MILES resolution.

Also provided are the following parameters at each radial bin: age, metallicity, and the individual element abundances for C, N, Na, Mg, Ca, and Ti, derived using the TMJ models, in \autoref{tab:parameters_ET} and \autoref{tab:parameters_LT} for ETGs and LTGs respectively.

\onecolumn
\begin{landscape}

\begin{longtable}{ccccccccccccc}
  \caption{Measured equivalent widths at all radial bins from $0 - 1.5\;R_\mathrm{e}$ for ETGs. The measurements have been corrected to MILES resolution.}
  \label{tab:ind_measurements_ET} 
  \endfirsthead
  \caption[]{(continued)}
  \endhead \\
  \hline
  $10^{8.8 - 9.8}\;M_{\odot}$ & 0.0-0.1 & 0.1-0.2 & 0.2-0.3 & 0.3-0.4 & 0.4-0.5 & 0.5-0.6 & 0.6-0.7 & 0.7-0.8 & 0.8-1.0 & 1.1-1.2 & 1.2-1.5 \\
  \hline
    CN1 & -0.01$\pm$0.0 & -0.0$\pm$0.0 & -0.01$\pm$0.0 & -0.01$\pm$0.0 & -0.01$\pm$0.0 & -0.01$\pm$0.0 & -0.01$\pm$0.0 & -0.02$\pm$0.0 & -0.02$\pm$0.0 & -0.02$\pm$0.0 & -0.02$\pm$0.0  \\
    Ca4227 & 1.32$\pm$0.04 & 1.28$\pm$0.03 & 1.22$\pm$0.02 & 1.25$\pm$0.02 & 1.25$\pm$0.02 & 1.22$\pm$0.02 & 1.28$\pm$0.02 & 1.28$\pm$0.03 & 1.23$\pm$0.02 & 1.21$\pm$0.03 & 1.19$\pm$0.04  \\
    Fe4531 & 3.11$\pm$0.07 & 3.11$\pm$0.04 & 3.14$\pm$0.03 & 3.09$\pm$0.04 & 3.12$\pm$0.04 & 3.06$\pm$0.03 & 3.13$\pm$0.04 & 3.04$\pm$0.05 & 3.06$\pm$0.04 & 3.0$\pm$0.05 & 3.04$\pm$0.07  \\
    C24668 & 4.27$\pm$0.1 & 4.17$\pm$0.06 & 4.29$\pm$0.05 & 4.31$\pm$0.05 & 4.21$\pm$0.05 & 3.91$\pm$0.05 & 3.93$\pm$0.06 & 3.79$\pm$0.08 & 3.73$\pm$0.06 & 3.7$\pm$0.08 & 3.5$\pm$0.1  \\
    Hb & 2.2$\pm$0.04 & 2.19$\pm$0.02 & 2.23$\pm$0.02 & 2.21$\pm$0.02 & 2.21$\pm$0.02 & 2.26$\pm$0.02 & 2.21$\pm$0.02 & 2.23$\pm$0.02 & 2.27$\pm$0.02 & 2.32$\pm$0.02 & 2.28$\pm$0.04  \\
    Mgb & 2.8$\pm$0.04 & 2.82$\pm$0.02 & 2.84$\pm$0.02 & 2.79$\pm$0.02 & 2.78$\pm$0.02 & 2.7$\pm$0.02 & 2.74$\pm$0.02 & 2.65$\pm$0.03 & 2.64$\pm$0.03 & 2.65$\pm$0.04 & 2.59$\pm$0.04  \\
    Fe5270 & 2.74$\pm$0.04 & 2.71$\pm$0.02 & 2.72$\pm$0.02 & 2.65$\pm$0.02 & 2.66$\pm$0.02 & 2.6$\pm$0.03 & 2.63$\pm$0.03 & 2.58$\pm$0.03 & 2.61$\pm$0.03 & 2.56$\pm$0.03 & 2.56$\pm$0.04  \\
    Fe5335 & 2.52$\pm$0.05 & 2.55$\pm$0.03 & 2.53$\pm$0.03 & 2.51$\pm$0.03 & 2.52$\pm$0.02 & 2.47$\pm$0.03 & 2.42$\pm$0.03 & 2.41$\pm$0.03 & 2.36$\pm$0.03 & 2.28$\pm$0.04 & 2.25$\pm$0.04  \\
    NaD & 2.03$\pm$0.04 & 2.06$\pm$0.02 & 2.0$\pm$0.02 & 1.95$\pm$0.02 & 1.92$\pm$0.02 & 1.89$\pm$0.02 & 2.33$\pm$0.03 & 2.3$\pm$0.03 & 2.3$\pm$0.03 & 2.25$\pm$0.03 & 2.23$\pm$0.04  \\
  \hline
  $10^{9.8 - 10.3}\;M_{\odot}$ & 0.0-0.1 & 0.1-0.2 & 0.2-0.3 & 0.3-0.4 & 0.4-0.5 & 0.5-0.6 & 0.6-0.7 & 0.7-0.8 & 0.8-1.0 & 1.1-1.2 & 1.2-1.5 \\
  \hline
    CN1 & 0.06$\pm$0.0 & 0.06$\pm$0.0 & 0.06$\pm$0.0 & 0.06$\pm$0.0 & 0.05$\pm$0.0 & 0.05$\pm$0.0 & 0.04$\pm$0.0 & 0.04$\pm$0.0 & 0.04$\pm$0.0 & 0.03$\pm$0.0 & 0.02$\pm$0.0  \\
    Ca4227 & 1.44$\pm$0.01 & 1.45$\pm$0.01 & 1.43$\pm$0.01 & 1.44$\pm$0.01 & 1.44$\pm$0.01 & 1.41$\pm$0.01 & 1.44$\pm$0.01 & 1.41$\pm$0.01 & 1.39$\pm$0.01 & 1.31$\pm$0.01 & 1.28$\pm$0.02  \\
    Fe4531 & 3.54$\pm$0.03 & 3.55$\pm$0.02 & 3.53$\pm$0.01 & 3.49$\pm$0.01 & 3.53$\pm$0.01 & 3.48$\pm$0.02 & 3.49$\pm$0.02 & 3.38$\pm$0.02 & 3.37$\pm$0.02 & 3.32$\pm$0.03 & 3.24$\pm$0.03  \\
    C24668 & 6.12$\pm$0.03 & 6.02$\pm$0.02 & 5.89$\pm$0.02 & 5.72$\pm$0.02 & 5.86$\pm$0.02 & 5.7$\pm$0.02 & 5.45$\pm$0.03 & 5.24$\pm$0.03 & 5.12$\pm$0.03 & 4.82$\pm$0.03 & 4.56$\pm$0.04  \\
    Hb & 1.91$\pm$0.01 & 1.91$\pm$0.01 & 1.96$\pm$0.01 & 1.91$\pm$0.01 & 1.96$\pm$0.01 & 1.99$\pm$0.01 & 2.06$\pm$0.01 & 2.07$\pm$0.01 & 2.07$\pm$0.01 & 2.08$\pm$0.01 & 2.09$\pm$0.02  \\
    Mgb & 3.96$\pm$0.01 & 3.96$\pm$0.01 & 3.92$\pm$0.01 & 3.88$\pm$0.01 & 3.82$\pm$0.01 & 3.73$\pm$0.01 & 3.71$\pm$0.01 & 3.63$\pm$0.01 & 3.55$\pm$0.01 & 3.46$\pm$0.01 & 3.33$\pm$0.02  \\
    Fe5270 & 3.2$\pm$0.01 & 3.17$\pm$0.01 & 3.18$\pm$0.01 & 3.12$\pm$0.01 & 3.1$\pm$0.01 & 3.04$\pm$0.01 & 3.01$\pm$0.01 & 2.97$\pm$0.01 & 2.92$\pm$0.01 & 2.85$\pm$0.01 & 2.82$\pm$0.02  \\
    Fe5335 & 2.99$\pm$0.02 & 2.94$\pm$0.01 & 2.97$\pm$0.01 & 2.94$\pm$0.01 & 2.93$\pm$0.01 & 2.87$\pm$0.01 & 2.84$\pm$0.02 & 2.76$\pm$0.02 & 2.7$\pm$0.01 & 2.58$\pm$0.02 & 2.54$\pm$0.03  \\
    NaD & 3.5$\pm$0.02 & 3.44$\pm$0.01 & 3.31$\pm$0.01 & 3.21$\pm$0.01 & 3.12$\pm$0.01 & 3.03$\pm$0.01 & 2.95$\pm$0.01 & 2.77$\pm$0.02 & 2.64$\pm$0.01 & 2.7$\pm$0.02 & 2.29$\pm$0.02  \\
  \hline
  $10^{10.3 - 10.6}\;M_{\odot}$ & 0.0-0.1 & 0.1-0.2 & 0.2-0.3 & 0.3-0.4 & 0.4-0.5 & 0.5-0.6 & 0.6-0.7 & 0.7-0.8 & 0.8-1.0 & 1.1-1.2 & 1.2-1.5 \\
  \hline
    CN1 & 0.09$\pm$0.0 & 0.08$\pm$0.0 & 0.07$\pm$0.0 & 0.07$\pm$0.0 & 0.06$\pm$0.0 & 0.06$\pm$0.0 & 0.05$\pm$0.0 & 0.05$\pm$0.0 & 0.05$\pm$0.0 & 0.04$\pm$0.0 & 0.04$\pm$0.0  \\
    Ca4227 & 1.51$\pm$0.01 & 1.52$\pm$0.0 & 1.48$\pm$0.0 & 1.49$\pm$0.0 & 1.46$\pm$0.01 & 1.47$\pm$0.01 & 1.46$\pm$0.01 & 1.43$\pm$0.01 & 1.39$\pm$0.01 & 1.38$\pm$0.01 & 1.37$\pm$0.01  \\
    Fe4531 & 3.71$\pm$0.02 & 3.7$\pm$0.01 & 3.67$\pm$0.01 & 3.64$\pm$0.01 & 3.62$\pm$0.01 & 3.58$\pm$0.01 & 3.47$\pm$0.02 & 3.44$\pm$0.02 & 3.44$\pm$0.02 & 3.41$\pm$0.02 & 3.32$\pm$0.02  \\
    C24668 & 6.76$\pm$0.02 & 6.82$\pm$0.01 & 6.69$\pm$0.01 & 6.16$\pm$0.01 & 6.05$\pm$0.02 & 5.91$\pm$0.01 & 5.54$\pm$0.02 & 5.46$\pm$0.02 & 5.42$\pm$0.02 & 5.08$\pm$0.03 & 4.92$\pm$0.03  \\
    Hb & 1.78$\pm$0.01 & 1.77$\pm$0.0 & 1.83$\pm$0.0 & 1.82$\pm$0.0 & 1.89$\pm$0.01 & 1.91$\pm$0.01 & 1.92$\pm$0.01 & 1.93$\pm$0.01 & 1.99$\pm$0.01 & 2.01$\pm$0.01 & 1.97$\pm$0.01  \\
    Mgb & 4.34$\pm$0.01 & 4.3$\pm$0.0 & 4.23$\pm$0.0 & 4.15$\pm$0.0 & 4.11$\pm$0.01 & 4.0$\pm$0.01 & 3.97$\pm$0.01 & 3.95$\pm$0.01 & 3.84$\pm$0.01 & 3.78$\pm$0.01 & 3.64$\pm$0.01  \\
    Fe5270 & 3.22$\pm$0.01 & 3.19$\pm$0.0 & 3.19$\pm$0.0 & 3.18$\pm$0.0 & 3.13$\pm$0.0 & 3.11$\pm$0.01 & 3.08$\pm$0.01 & 3.05$\pm$0.01 & 3.0$\pm$0.01 & 2.96$\pm$0.01 & 2.89$\pm$0.01  \\
    Fe5335 & 3.15$\pm$0.01 & 3.14$\pm$0.01 & 3.16$\pm$0.01 & 3.06$\pm$0.01 & 3.05$\pm$0.01 & 2.95$\pm$0.01 & 2.93$\pm$0.01 & 2.9$\pm$0.01 & 2.84$\pm$0.01 & 2.77$\pm$0.01 & 2.7$\pm$0.02  \\
    NaD & 4.09$\pm$0.01 & 4.02$\pm$0.0 & 3.88$\pm$0.0 & 3.65$\pm$0.0 & 3.53$\pm$0.0 & 3.41$\pm$0.01 & 3.28$\pm$0.01 & 3.18$\pm$0.01 & 3.04$\pm$0.01 & 2.88$\pm$0.01 & 2.79$\pm$0.01  \\
  \hline
    $10^{10.6 - 10.8}\;M_{\odot}$ & 0.0-0.1 & 0.1-0.2 & 0.2-0.3 & 0.3-0.4 & 0.4-0.5 & 0.5-0.6 & 0.6-0.7 & 0.7-0.8 & 0.8-1.0 & 1.1-1.2 & 1.2-1.5 \\
  \hline
    CN1 & 0.09$\pm$0.0 & 0.09$\pm$0.0 & 0.08$\pm$0.0 & 0.08$\pm$0.0 & 0.07$\pm$0.0 & 0.06$\pm$0.0 & 0.05$\pm$0.0 & 0.05$\pm$0.0 & 0.05$\pm$0.0 & 0.04$\pm$0.0 & 0.03$\pm$0.0  \\
    Ca4227 & 1.58$\pm$0.01 & 1.58$\pm$0.0 & 1.59$\pm$0.01 & 1.54$\pm$0.01 & 1.53$\pm$0.01 & 1.54$\pm$0.01 & 1.51$\pm$0.01 & 1.46$\pm$0.01 & 1.45$\pm$0.01 & 1.41$\pm$0.01 & 1.4$\pm$0.02  \\
    Fe4531 & 3.83$\pm$0.02 & 3.71$\pm$0.01 & 3.65$\pm$0.01 & 3.69$\pm$0.01 & 3.64$\pm$0.02 & 3.59$\pm$0.03 & 3.54$\pm$0.03 & 3.53$\pm$0.03 & 3.5$\pm$0.03 & 3.54$\pm$0.04 & 3.44$\pm$0.05  \\
    C24668 & 7.09$\pm$0.02 & 6.97$\pm$0.01 & 6.73$\pm$0.01 & 6.47$\pm$0.01 & 6.21$\pm$0.01 & 6.13$\pm$0.02 & 5.88$\pm$0.02 & 5.65$\pm$0.02 & 5.43$\pm$0.02 & 5.32$\pm$0.03 & 5.06$\pm$0.04  \\
    Hb & 1.7$\pm$0.01 & 1.75$\pm$0.0 & 1.74$\pm$0.0 & 1.78$\pm$0.0 & 1.82$\pm$0.01 & 1.84$\pm$0.01 & 1.87$\pm$0.01 & 1.87$\pm$0.01 & 1.91$\pm$0.01 & 1.91$\pm$0.01 & 1.86$\pm$0.01  \\
    Mgb & 4.62$\pm$0.01 & 4.6$\pm$0.0 & 4.55$\pm$0.0 & 4.43$\pm$0.01 & 4.32$\pm$0.01 & 4.24$\pm$0.01 & 4.14$\pm$0.01 & 4.08$\pm$0.01 & 4.01$\pm$0.01 & 3.94$\pm$0.02 & 3.77$\pm$0.01  \\
    Fe5270 & 3.28$\pm$0.01 & 3.3$\pm$0.01 & 3.23$\pm$0.0 & 3.18$\pm$0.01 & 3.12$\pm$0.01 & 3.1$\pm$0.01 & 3.06$\pm$0.01 & 3.01$\pm$0.01 & 2.96$\pm$0.01 & 2.92$\pm$0.01 & 2.83$\pm$0.02  \\
    Fe5335 & 3.18$\pm$0.07 & 3.16$\pm$0.05 & 3.1$\pm$0.04 & 3.11$\pm$0.05 & 3.03$\pm$0.07 & 3.01$\pm$0.1 & 2.93$\pm$0.11 & 2.91$\pm$0.12 & 2.86$\pm$0.11 & 2.85$\pm$0.14 & 2.76$\pm$0.18  \\
    NaD & 4.52$\pm$0.01 & 4.44$\pm$0.01 & 4.2$\pm$0.01 & 4.01$\pm$0.01 & 3.81$\pm$0.01 & 3.6$\pm$0.01 & 3.48$\pm$0.01 & 3.37$\pm$0.01 & 3.3$\pm$0.01 & 3.14$\pm$0.02 & 3.02$\pm$0.02  \\
  \hline
  \newpage
  \hline
  $10^{10.8 - 11.0}\;M_{\odot}$ & 0.0-0.1 & 0.1-0.2 & 0.2-0.3 & 0.3-0.4 & 0.4-0.5 & 0.5-0.6 & 0.6-0.7 & 0.7-0.8 & 0.8-1.0 & 1.1-1.2 & 1.2-1.5 \\
  \hline
    CN1 & 0.11$\pm$0.0 & 0.11$\pm$0.0 & 0.11$\pm$0.0 & 0.1$\pm$0.0 & 0.09$\pm$0.0 & 0.08$\pm$0.0 & 0.07$\pm$0.0 & 0.07$\pm$0.0 & 0.06$\pm$0.0 & 0.06$\pm$0.0 & 0.05$\pm$0.0  \\
    Ca4227 & 1.59$\pm$0.01 & 1.63$\pm$0.01 & 1.58$\pm$0.01 & 1.6$\pm$0.01 & 1.57$\pm$0.01 & 1.53$\pm$0.01 & 1.51$\pm$0.02 & 1.5$\pm$0.02 & 1.45$\pm$0.02 & 1.48$\pm$0.03 & 1.46$\pm$0.03  \\
    Fe4531 & 3.74$\pm$0.02 & 3.78$\pm$0.01 & 3.71$\pm$0.02 & 3.73$\pm$0.02 & 3.65$\pm$0.02 & 3.61$\pm$0.02 & 3.62$\pm$0.03 & 3.55$\pm$0.03 & 3.47$\pm$0.04 & 3.41$\pm$0.05 & 3.38$\pm$0.06  \\
    C24668 & 7.36$\pm$0.03 & 7.23$\pm$0.02 & 7.1$\pm$0.02 & 6.82$\pm$0.02 & 6.61$\pm$0.03 & 6.38$\pm$0.04 & 6.19$\pm$0.05 & 6.07$\pm$0.06 & 5.72$\pm$0.05 & 5.4$\pm$0.07 & 5.3$\pm$0.08  \\
    Hb & 1.72$\pm$0.01 & 1.69$\pm$0.01 & 1.71$\pm$0.01 & 1.75$\pm$0.01 & 1.77$\pm$0.01 & 1.83$\pm$0.01 & 1.83$\pm$0.02 & 1.85$\pm$0.02 & 1.77$\pm$0.02 & 1.91$\pm$0.03 & 1.71$\pm$0.03  \\
    Mgb & 4.64$\pm$0.01 & 4.61$\pm$0.01 & 4.57$\pm$0.01 & 4.46$\pm$0.01 & 4.36$\pm$0.01 & 4.27$\pm$0.02 & 4.23$\pm$0.02 & 4.14$\pm$0.02 & 4.05$\pm$0.02 & 3.95$\pm$0.03 & 3.77$\pm$0.03  \\
    Fe5270 & 3.28$\pm$0.08 & 3.29$\pm$0.06 & 3.27$\pm$0.05 & 3.21$\pm$0.06 & 3.17$\pm$0.08 & 3.1$\pm$0.11 & 3.07$\pm$0.13 & 3.03$\pm$0.15 & 3.0$\pm$0.14 & 2.91$\pm$0.19 & 2.82$\pm$0.24  \\
    Fe5335 & 3.25$\pm$0.02 & 3.23$\pm$0.01 & 3.18$\pm$0.01 & 3.1$\pm$0.01 & 3.08$\pm$0.02 & 3.02$\pm$0.02 & 2.95$\pm$0.02 & 2.96$\pm$0.03 & 2.86$\pm$0.03 & 2.82$\pm$0.04 & 2.68$\pm$0.04  \\
    NaD & 4.58$\pm$0.01 & 4.53$\pm$0.01 & 4.34$\pm$0.01 & 4.07$\pm$0.01 & 3.87$\pm$0.02 & 3.75$\pm$0.02 & 3.58$\pm$0.02 & 3.49$\pm$0.02 & 3.37$\pm$0.03 & 3.18$\pm$0.03 & 3.03$\pm$0.04  \\
  \hline
  $10^{11.0 - 11.3}\;M_{\odot}$ & 0.0-0.1 & 0.1-0.2 & 0.2-0.3 & 0.3-0.4 & 0.4-0.5 & 0.5-0.6 & 0.6-0.7 & 0.7-0.8 & 0.8-1.0 & 1.1-1.2 & 1.2-1.5 \\
  \hline
    CN1 & 0.12$\pm$0.0 & 0.12$\pm$0.0 & 0.11$\pm$0.0 & 0.1$\pm$0.0 & 0.09$\pm$0.0 & 0.08$\pm$0.0 & 0.08$\pm$0.0 & 0.08$\pm$0.0 & 0.07$\pm$0.0 & 0.06$\pm$0.0 & 0.06$\pm$0.0  \\
    Ca4227 & 1.53$\pm$0.02 & 1.57$\pm$0.01 & 1.51$\pm$0.01 & 1.54$\pm$0.01 & 1.48$\pm$0.02 & 1.49$\pm$0.03 & 1.47$\pm$0.03 & 1.48$\pm$0.04 & 1.39$\pm$0.03 & 1.32$\pm$0.03 & 1.27$\pm$0.06  \\
    Fe4531 & 3.73$\pm$0.03 & 3.71$\pm$0.02 & 3.71$\pm$0.02 & 3.67$\pm$0.02 & 3.63$\pm$0.03 & 3.59$\pm$0.04 & 3.58$\pm$0.05 & 3.52$\pm$0.06 & 3.5$\pm$0.05 & 3.5$\pm$0.05 & 3.46$\pm$0.08  \\
    C24668 & 7.09$\pm$0.04 & 7.02$\pm$0.02 & 6.79$\pm$0.03 & 6.72$\pm$0.03 & 6.4$\pm$0.04 & 6.07$\pm$0.06 & 5.86$\pm$0.06 & 5.75$\pm$0.08 & 5.71$\pm$0.07 & 5.53$\pm$0.09 & 5.42$\pm$0.12  \\
    Hb & 1.67$\pm$0.02 & 1.7$\pm$0.01 & 1.72$\pm$0.01 & 1.78$\pm$0.02 & 1.82$\pm$0.02 & 1.79$\pm$0.02 & 1.74$\pm$0.03 & 1.75$\pm$0.03 & 1.75$\pm$0.03 & 1.85$\pm$0.04 & 1.77$\pm$0.05  \\
    Mgb & 4.67$\pm$0.02 & 4.61$\pm$0.02 & 4.58$\pm$0.01 & 4.54$\pm$0.02 & 4.42$\pm$0.02 & 4.38$\pm$0.03 & 4.31$\pm$0.04 & 4.17$\pm$0.05 & 4.08$\pm$0.04 & 3.89$\pm$0.05 & 3.68$\pm$0.06  \\
    Fe5270 & 3.15$\pm$0.02 & 3.16$\pm$0.01 & 3.15$\pm$0.02 & 3.1$\pm$0.02 & 3.05$\pm$0.03 & 3.0$\pm$0.03 & 3.0$\pm$0.04 & 3.01$\pm$0.05 & 2.94$\pm$0.04 & 2.76$\pm$0.05 & 2.68$\pm$0.06  \\
    Fe5335 & 3.08$\pm$0.07 & 3.07$\pm$0.04 & 3.09$\pm$0.04 & 3.07$\pm$0.07 & 3.04$\pm$0.08 & 2.99$\pm$0.11 & 2.96$\pm$0.11 & 2.86$\pm$0.14 & 2.79$\pm$0.14 & 2.7$\pm$0.16 & 2.53$\pm$0.22  \\
    NaD & 4.59$\pm$0.02 & 4.55$\pm$0.01 & 4.42$\pm$0.01 & 4.26$\pm$0.02 & 4.05$\pm$0.02 & 3.86$\pm$0.02 & 3.7$\pm$0.03 & 3.6$\pm$0.04 & 3.43$\pm$0.03 & 3.25$\pm$0.04 & 3.02$\pm$0.05  \\
  \hline
\end{longtable}

\newpage

\begin{longtable}{ccccccccccccc}
  \caption{Same as \autoref{tab:ind_measurements_ET} for LTGs.}
  \label{tab:ind_measurements_LT}
\endfirsthead
  \caption[]{(continued)}
  \endhead \\
 \hline
  $10^{8.6 - 9.2}\;M_{\odot}$ & 0.0-0.1 & 0.1-0.2 & 0.2-0.3 & 0.3-0.4 & 0.4-0.5 & 0.5-0.6 & 0.6-0.7 & 0.7-0.8 & 0.8-1.0 & 1.1-1.2 & 1.2-1.5 \\
  \hline
    CN1 & -0.09$\pm$0.01 & -0.08$\pm$0.0 & -0.09$\pm$0.0 & -0.09$\pm$0.0 & -0.09$\pm$0.0 & -0.09$\pm$0.0 & -0.09$\pm$0.0 & -0.09$\pm$0.0 & -0.09$\pm$0.0 & -0.09$\pm$0.0 & -0.09$\pm$0.0  \\
    Ca4227 & 0.48$\pm$0.08 & 0.48$\pm$0.06 & 0.47$\pm$0.05 & 0.52$\pm$0.04 & 0.5$\pm$0.04 & 0.49$\pm$0.05 & 0.49$\pm$0.05 & 0.46$\pm$0.07 & 0.47$\pm$0.05 & 0.47$\pm$0.07 & 0.48$\pm$0.09  \\
    Fe4531 & 2.0$\pm$0.2 & 2.09$\pm$0.12 & 1.97$\pm$0.09 & 1.79$\pm$0.09 & 1.87$\pm$0.09 & 1.74$\pm$0.09 & 1.85$\pm$0.1 & 1.83$\pm$0.11 & 1.74$\pm$0.1 & 1.75$\pm$0.12 & 1.7$\pm$0.14  \\
    C24668 & 1.68$\pm$0.31 & 1.53$\pm$0.2 & 1.32$\pm$0.15 & 1.36$\pm$0.13 & 1.3$\pm$0.14 & 1.43$\pm$0.13 & 1.41$\pm$0.14 & 1.41$\pm$0.18 & 1.55$\pm$0.15 & 1.25$\pm$0.21 & 1.23$\pm$0.25  \\
    Hb & 3.53$\pm$0.54 & 3.4$\pm$0.38 & 3.42$\pm$0.34 & 3.41$\pm$0.31 & 3.53$\pm$0.3 & 3.46$\pm$0.31 & 3.51$\pm$0.3 & 3.5$\pm$0.4 & 3.55$\pm$0.36 & 3.51$\pm$0.47 & 3.52$\pm$0.58  \\
    Mgb & 1.55$\pm$0.1 & 1.5$\pm$0.06 & 1.5$\pm$0.06 & 1.51$\pm$0.05 & 1.54$\pm$0.05 & 1.48$\pm$0.05 & 1.49$\pm$0.05 & 1.47$\pm$0.07 & 1.45$\pm$0.06 & 1.48$\pm$0.08 & 1.53$\pm$0.09  \\
    Fe5270 & 1.64$\pm$0.1 & 1.66$\pm$0.07 & 1.65$\pm$0.06 & 1.57$\pm$0.05 & 1.5$\pm$0.04 & 1.49$\pm$0.06 & 1.45$\pm$0.06 & 1.43$\pm$0.06 & 1.4$\pm$0.06 & 1.29$\pm$0.09 & 1.31$\pm$0.11  \\
    Fe5335 & 1.53$\pm$0.13 & 1.47$\pm$0.07 & 1.42$\pm$0.06 & 1.44$\pm$0.05 & 1.39$\pm$0.06 & 1.38$\pm$0.06 & 1.36$\pm$0.06 & 1.25$\pm$0.07 & 1.19$\pm$0.06 & 1.25$\pm$0.09 & 1.23$\pm$0.11  \\
    NaD & 1.06$\pm$0.13 & 1.01$\pm$0.08 & 1.02$\pm$0.07 & 1.05$\pm$0.06 & 0.98$\pm$0.06 & 0.94$\pm$0.06 & 0.96$\pm$0.07 & 0.97$\pm$0.08 & 0.94$\pm$0.07 & 0.91$\pm$0.1 & 0.92$\pm$0.13  \\
  \hline
  $10^{9.2 - 9.6}\;M_{\odot}$ & 0.0-0.1 & 0.1-0.2 & 0.2-0.3 & 0.3-0.4 & 0.4-0.5 & 0.5-0.6 & 0.6-0.7 & 0.7-0.8 & 0.8-1.0 & 1.1-1.2 & 1.2-1.5 \\
  \hline
    CN1 & -0.07$\pm$0.0 & -0.06$\pm$0.0 & -0.07$\pm$0.0 & -0.07$\pm$0.0 & -0.07$\pm$0.0 & -0.08$\pm$0.0 & -0.08$\pm$0.0 & -0.08$\pm$0.0 & -0.09$\pm$0.0 & -0.09$\pm$0.0 & -0.08$\pm$0.0  \\
    Ca4227 & 0.59$\pm$0.05 & 0.6$\pm$0.03 & 0.61$\pm$0.02 & 0.62$\pm$0.03 & 0.56$\pm$0.03 & 0.56$\pm$0.03 & 0.49$\pm$0.04 & 0.47$\pm$0.03 & 0.42$\pm$0.03 & 0.34$\pm$0.04 & 0.44$\pm$0.05  \\
    Fe4531 & 2.21$\pm$0.09 & 2.38$\pm$0.05 & 2.25$\pm$0.05 & 2.17$\pm$0.05 & 2.21$\pm$0.05 & 2.23$\pm$0.05 & 2.14$\pm$0.05 & 2.05$\pm$0.06 & 1.94$\pm$0.05 & 2.0$\pm$0.06 & 2.02$\pm$0.08  \\
    C24668 & 2.32$\pm$0.12 & 2.12$\pm$0.07 & 2.03$\pm$0.06 & 1.96$\pm$0.06 & 1.81$\pm$0.07 & 1.85$\pm$0.07 & 1.76$\pm$0.07 & 1.66$\pm$0.09 & 1.47$\pm$0.07 & 1.54$\pm$0.08 & 1.27$\pm$0.09  \\
    Hb & 3.08$\pm$0.05 & 3.06$\pm$0.03 & 3.07$\pm$0.02 & 3.13$\pm$0.02 & 3.18$\pm$0.02 & 3.28$\pm$0.02 & 3.26$\pm$0.03 & 3.37$\pm$0.03 & 3.38$\pm$0.03 & 3.39$\pm$0.03 & 3.36$\pm$0.04  \\
    Mgb & 1.91$\pm$0.04 & 1.94$\pm$0.03 & 1.89$\pm$0.03 & 1.8$\pm$0.03 & 1.66$\pm$0.02 & 1.69$\pm$0.03 & 1.66$\pm$0.03 & 1.63$\pm$0.03 & 1.62$\pm$0.03 & 1.58$\pm$0.04 & 1.69$\pm$0.05  \\
    Fe5270 & 1.97$\pm$0.05 & 2.03$\pm$0.03 & 2.02$\pm$0.03 & 1.96$\pm$0.03 & 1.88$\pm$0.03 & 1.72$\pm$0.03 & 1.69$\pm$0.04 & 1.66$\pm$0.04 & 1.7$\pm$0.03 & 1.61$\pm$0.04 & 1.64$\pm$0.05  \\
    Fe5335 & 1.84$\pm$0.07 & 1.82$\pm$0.04 & 1.83$\pm$0.04 & 1.76$\pm$0.04 & 1.71$\pm$0.04 & 1.68$\pm$0.05 & 1.65$\pm$0.05 & 1.65$\pm$0.06 & 1.5$\pm$0.04 & 1.43$\pm$0.06 & 1.38$\pm$0.07  \\
    NaD & 1.44$\pm$0.06 & 1.38$\pm$0.04 & 1.32$\pm$0.03 & 1.24$\pm$0.05 & 1.23$\pm$0.04 & 1.19$\pm$0.05 & 1.19$\pm$0.04 & 1.15$\pm$0.05 & 1.07$\pm$0.04 & 1.09$\pm$0.05 & 1.03$\pm$0.06  \\
  \hline
  $10^{9.6 - 9.9}\;M_{\odot}$ & 0.0-0.1 & 0.1-0.2 & 0.2-0.3 & 0.3-0.4 & 0.4-0.5 & 0.5-0.6 & 0.6-0.7 & 0.7-0.8 & 0.8-1.0 & 1.1-1.2 & 1.2-1.5 \\
  \hline
    CN1 & -0.05$\pm$0.0 & -0.05$\pm$0.0 & -0.05$\pm$0.0 & -0.06$\pm$0.0 & -0.07$\pm$0.0 & -0.08$\pm$0.0 & -0.08$\pm$0.0 & -0.08$\pm$0.0 & -0.08$\pm$0.0 & -0.08$\pm$0.0 & -0.08$\pm$0.0  \\
    Ca4227 & 0.77$\pm$0.08 & 0.8$\pm$0.05 & 0.79$\pm$0.04 & 0.84$\pm$0.04 & 0.79$\pm$0.04 & 0.77$\pm$0.04 & 0.75$\pm$0.05 & 0.72$\pm$0.05 & 0.63$\pm$0.04 & 0.61$\pm$0.05 & 0.52$\pm$0.07  \\
    Fe4531 & 2.76$\pm$0.11 & 2.68$\pm$0.07 & 2.54$\pm$0.05 & 2.56$\pm$0.07 & 2.45$\pm$0.07 & 2.4$\pm$0.06 & 2.45$\pm$0.07 & 2.36$\pm$0.08 & 2.32$\pm$0.06 & 2.13$\pm$0.08 & 2.23$\pm$0.1  \\
    C24668 & 3.1$\pm$0.16 & 2.92$\pm$0.09 & 2.77$\pm$0.09 & 2.85$\pm$0.09 & 2.56$\pm$0.1 & 2.5$\pm$0.1 & 2.43$\pm$0.1 & 2.21$\pm$0.12 & 2.05$\pm$0.1 & 2.07$\pm$0.11 & 2.35$\pm$0.15  \\
    Hb & 2.83$\pm$0.18 & 2.88$\pm$0.12 & 2.9$\pm$0.1 & 2.91$\pm$0.12 & 2.98$\pm$0.12 & 3.09$\pm$0.11 & 3.2$\pm$0.12 & 3.29$\pm$0.13 & 3.43$\pm$0.11 & 3.47$\pm$0.13 & 3.19$\pm$0.16  \\
    Mgb & 2.22$\pm$0.07 & 2.27$\pm$0.05 & 2.26$\pm$0.05 & 2.24$\pm$0.05 & 2.16$\pm$0.05 & 2.02$\pm$0.05 & 2.02$\pm$0.05 & 2.03$\pm$0.06 & 1.98$\pm$0.05 & 2.02$\pm$0.05 & 1.89$\pm$0.08  \\
    Fe5270 & 2.31$\pm$0.06 & 2.36$\pm$0.04 & 2.31$\pm$0.03 & 2.28$\pm$0.04 & 2.29$\pm$0.04 & 2.24$\pm$0.04 & 2.19$\pm$0.04 & 2.19$\pm$0.05 & 2.12$\pm$0.04 & 2.12$\pm$0.05 & 1.83$\pm$0.06  \\
    Fe5335 & 2.19$\pm$0.06 & 2.17$\pm$0.04 & 2.17$\pm$0.04 & 2.17$\pm$0.04 & 2.14$\pm$0.04 & 2.05$\pm$0.04 & 1.98$\pm$0.04 & 1.95$\pm$0.04 & 1.82$\pm$0.04 & 1.74$\pm$0.06 & 1.58$\pm$0.07  \\
    NaD & 1.77$\pm$0.05 & 1.76$\pm$0.04 & 1.69$\pm$0.03 & 1.57$\pm$0.04 & 1.55$\pm$0.04 & 1.52$\pm$0.04 & 1.44$\pm$0.04 & 1.4$\pm$0.04 & 1.28$\pm$0.04 & 1.29$\pm$0.05 & 1.24$\pm$0.06  \\
  \hline
    $10^{9.9 - 10.1}\;M_{\odot}$ & 0.0-0.1 & 0.1-0.2 & 0.2-0.3 & 0.3-0.4 & 0.4-0.5 & 0.5-0.6 & 0.6-0.7 & 0.7-0.8 & 0.8-1.0 & 1.1-1.2 & 1.2-1.5 \\
  \hline
    CN1 & -0.03$\pm$0.0 & -0.03$\pm$0.0 & -0.03$\pm$0.0 & -0.04$\pm$0.0 & -0.04$\pm$0.0 & -0.05$\pm$0.0 & -0.06$\pm$0.0 & -0.07$\pm$0.0 & -0.07$\pm$0.0 & -0.07$\pm$0.0 & -0.06$\pm$0.0  \\
    Ca4227 & 0.98$\pm$0.02 & 1.0$\pm$0.02 & 0.98$\pm$0.01 & 0.9$\pm$0.01 & 0.92$\pm$0.02 & 0.8$\pm$0.02 & 0.79$\pm$0.02 & 0.74$\pm$0.02 & 0.73$\pm$0.02 & 0.66$\pm$0.02 & 0.75$\pm$0.02  \\
    Fe4531 & 2.8$\pm$0.04 & 2.83$\pm$0.02 & 2.81$\pm$0.02 & 2.71$\pm$0.02 & 2.61$\pm$0.03 & 2.63$\pm$0.03 & 2.46$\pm$0.03 & 2.38$\pm$0.04 & 2.45$\pm$0.03 & 2.36$\pm$0.04 & 2.15$\pm$0.04  \\
    C24668 & 3.81$\pm$0.06 & 3.76$\pm$0.04 & 3.61$\pm$0.04 & 3.6$\pm$0.03 & 3.21$\pm$0.04 & 3.08$\pm$0.04 & 2.87$\pm$0.05 & 2.63$\pm$0.05 & 2.54$\pm$0.03 & 2.59$\pm$0.05 & 2.32$\pm$0.07  \\
    Hb & 2.57$\pm$0.04 & 2.54$\pm$0.03 & 2.64$\pm$0.03 & 2.67$\pm$0.03 & 2.75$\pm$0.03 & 2.79$\pm$0.03 & 2.85$\pm$0.04 & 2.99$\pm$0.04 & 3.08$\pm$0.03 & 3.21$\pm$0.04 & 3.17$\pm$0.05  \\
    Mgb & 2.84$\pm$0.02 & 2.79$\pm$0.01 & 2.76$\pm$0.01 & 2.71$\pm$0.01 & 2.69$\pm$0.01 & 2.47$\pm$0.02 & 2.37$\pm$0.02 & 2.35$\pm$0.02 & 2.22$\pm$0.02 & 2.24$\pm$0.02 & 2.2$\pm$0.02  \\
    Fe5270 & 2.69$\pm$0.02 & 2.6$\pm$0.02 & 2.6$\pm$0.01 & 2.57$\pm$0.01 & 2.57$\pm$0.02 & 2.48$\pm$0.02 & 2.44$\pm$0.02 & 2.38$\pm$0.02 & 2.29$\pm$0.02 & 2.21$\pm$0.02 & 2.1$\pm$0.03  \\
    Fe5335 & 2.44$\pm$0.02 & 2.45$\pm$0.02 & 2.43$\pm$0.02 & 2.38$\pm$0.01 & 2.39$\pm$0.02 & 2.4$\pm$0.02 & 2.4$\pm$0.02 & 2.27$\pm$0.02 & 2.02$\pm$0.02 & 1.97$\pm$0.02 & 1.9$\pm$0.02  \\
    NaD & 2.5$\pm$0.02 & 2.41$\pm$0.01 & 2.31$\pm$0.01 & 2.18$\pm$0.01 & 2.08$\pm$0.02 & 2.01$\pm$0.02 & 1.9$\pm$0.02 & 1.84$\pm$0.02 & 1.76$\pm$0.02 & 1.66$\pm$0.02 & 1.54$\pm$0.02  \\
  \hline
  \newpage
  \hline
  $10^{10.1 - 10.3}\;M_{\odot}$ & 0.0-0.1 & 0.1-0.2 & 0.2-0.3 & 0.3-0.4 & 0.4-0.5 & 0.5-0.6 & 0.6-0.7 & 0.7-0.8 & 0.8-1.0 & 1.1-1.2 & 1.2-1.5 \\
  \hline
    CN1 & 0.01$\pm$0.0 & 0.01$\pm$0.0 & -0.0$\pm$0.0 & -0.01$\pm$0.0 & -0.02$\pm$0.0 & -0.03$\pm$0.0 & -0.04$\pm$0.0 & -0.04$\pm$0.0 & -0.05$\pm$0.0 & -0.06$\pm$0.0 & -0.07$\pm$0.0  \\
    Ca4227 & 1.23$\pm$0.01 & 1.26$\pm$0.01 & 1.2$\pm$0.01 & 1.1$\pm$0.01 & 1.04$\pm$0.01 & 0.94$\pm$0.01 & 0.91$\pm$0.01 & 0.88$\pm$0.02 & 0.82$\pm$0.01 & 0.84$\pm$0.01 & 0.77$\pm$0.02  \\
    Fe4531 & 3.15$\pm$0.02 & 3.15$\pm$0.01 & 3.12$\pm$0.01 & 3.08$\pm$0.02 & 2.93$\pm$0.02 & 2.88$\pm$0.02 & 2.86$\pm$0.03 & 2.67$\pm$0.03 & 2.64$\pm$0.02 & 2.57$\pm$0.03 & 2.41$\pm$0.03  \\
    C24668 & 4.92$\pm$0.03 & 5.04$\pm$0.02 & 4.81$\pm$0.02 & 4.63$\pm$0.02 & 4.29$\pm$0.03 & 3.89$\pm$0.03 & 3.87$\pm$0.02 & 3.61$\pm$0.03 & 3.16$\pm$0.03 & 3.11$\pm$0.04 & 2.8$\pm$0.04  \\
    Hb & 2.22$\pm$0.03 & 2.28$\pm$0.02 & 2.33$\pm$0.02 & 2.4$\pm$0.03 & 2.49$\pm$0.03 & 2.61$\pm$0.03 & 2.64$\pm$0.04 & 2.77$\pm$0.04 & 2.91$\pm$0.03 & 2.91$\pm$0.04 & 2.88$\pm$0.05  \\
    Mgb & 3.26$\pm$0.01 & 3.34$\pm$0.01 & 3.29$\pm$0.01 & 3.18$\pm$0.01 & 3.06$\pm$0.01 & 2.88$\pm$0.01 & 2.83$\pm$0.01 & 2.69$\pm$0.01 & 2.61$\pm$0.01 & 2.44$\pm$0.02 & 2.26$\pm$0.02  \\
    Fe5270 & 2.91$\pm$0.01 & 2.92$\pm$0.01 & 2.82$\pm$0.01 & 2.81$\pm$0.01 & 2.71$\pm$0.01 & 2.67$\pm$0.01 & 2.63$\pm$0.01 & 2.56$\pm$0.02 & 2.46$\pm$0.01 & 2.3$\pm$0.02 & 2.04$\pm$0.02  \\
    Fe5335 & 2.76$\pm$0.01 & 2.75$\pm$0.01 & 2.71$\pm$0.01 & 2.67$\pm$0.01 & 2.61$\pm$0.01 & 2.55$\pm$0.01 & 2.47$\pm$0.02 & 2.4$\pm$0.02 & 2.25$\pm$0.02 & 2.23$\pm$0.02 & 2.12$\pm$0.03  \\
    NaD & 2.91$\pm$0.01 & 2.74$\pm$0.01 & 2.56$\pm$0.01 & 2.44$\pm$0.01 & 2.33$\pm$0.01 & 2.25$\pm$0.01 & 2.21$\pm$0.01 & 2.08$\pm$0.02 & 1.93$\pm$0.01 & 1.82$\pm$0.02 & 1.69$\pm$0.02  \\
  \hline
  $10^{10.3 - 10.6}\;M_{\odot}$ & 0.0-0.1 & 0.1-0.2 & 0.2-0.3 & 0.3-0.4 & 0.4-0.5 & 0.5-0.6 & 0.6-0.7 & 0.7-0.8 & 0.8-1.0 & 1.1-1.2 & 1.2-1.5 \\
  \hline
    CN1 & 0.03$\pm$0.0 & 0.02$\pm$0.0 & 0.02$\pm$0.0 & -0.0$\pm$0.0 & -0.0$\pm$0.0 & -0.01$\pm$0.0 & -0.03$\pm$0.0 & -0.04$\pm$0.0 & -0.04$\pm$0.0 & -0.07$\pm$0.0 & -0.06$\pm$0.0  \\
    Ca4227 & 1.28$\pm$0.01 & 1.33$\pm$0.01 & 1.28$\pm$0.01 & 1.23$\pm$0.01 & 1.07$\pm$0.01 & 1.03$\pm$0.01 & 0.98$\pm$0.02 & 0.9$\pm$0.02 & 0.79$\pm$0.02 & 0.81$\pm$0.02 & 0.83$\pm$0.02  \\
    Fe4531 & 3.42$\pm$0.02 & 3.4$\pm$0.01 & 3.32$\pm$0.02 & 3.19$\pm$0.02 & 3.13$\pm$0.02 & 3.2$\pm$0.02 & 2.98$\pm$0.03 & 2.96$\pm$0.03 & 2.83$\pm$0.02 & 2.55$\pm$0.03 & 2.45$\pm$0.04  \\
    C24668 & 5.56$\pm$0.03 & 5.45$\pm$0.02 & 5.16$\pm$0.02 & 4.98$\pm$0.03 & 4.65$\pm$0.03 & 4.54$\pm$0.03 & 4.23$\pm$0.04 & 3.86$\pm$0.04 & 3.48$\pm$0.04 & 2.95$\pm$0.04 & 2.58$\pm$0.06  \\
    Hb & 2.1$\pm$0.02 & 2.1$\pm$0.01 & 2.18$\pm$0.02 & 2.34$\pm$0.02 & 2.46$\pm$0.02 & 2.53$\pm$0.03 & 2.62$\pm$0.04 & 2.61$\pm$0.03 & 2.82$\pm$0.03 & 2.98$\pm$0.04 & 2.98$\pm$0.05  \\
    Mgb & 3.66$\pm$0.01 & 3.64$\pm$0.01 & 3.52$\pm$0.01 & 3.38$\pm$0.01 & 3.25$\pm$0.01 & 3.12$\pm$0.02 & 2.95$\pm$0.02 & 2.81$\pm$0.02 & 2.67$\pm$0.02 & 2.46$\pm$0.02 & 2.33$\pm$0.03  \\
    Fe5270 & 2.95$\pm$0.01 & 2.9$\pm$0.01 & 2.9$\pm$0.01 & 2.87$\pm$0.01 & 2.87$\pm$0.01 & 2.82$\pm$0.01 & 2.7$\pm$0.01 & 2.63$\pm$0.02 & 2.51$\pm$0.02 & 2.25$\pm$0.02 & 2.07$\pm$0.03  \\
    Fe5335 & 2.89$\pm$0.01 & 2.83$\pm$0.01 & 2.72$\pm$0.01 & 2.7$\pm$0.01 & 2.62$\pm$0.01 & 2.58$\pm$0.02 & 2.52$\pm$0.02 & 2.48$\pm$0.02 & 2.38$\pm$0.02 & 2.32$\pm$0.02 & 2.24$\pm$0.03  \\
    NaD & 3.14$\pm$0.01 & 3.0$\pm$0.01 & 2.83$\pm$0.01 & 2.65$\pm$0.01 & 2.54$\pm$0.01 & 2.41$\pm$0.02 & 2.28$\pm$0.02 & 2.2$\pm$0.02 & 2.04$\pm$0.02 & 1.9$\pm$0.02 & 1.73$\pm$0.02  \\
   \hline
  $10^{10.6 - 11.2}\;M_{\odot}$ & 0.0-0.1 & 0.1-0.2 & 0.2-0.3 & 0.3-0.4 & 0.4-0.5 & 0.5-0.6 & 0.6-0.7 & 0.7-0.8 & 0.8-1.0 & 1.1-1.2 & 1.2-1.5 \\
  \hline
    CN1 & 0.07$\pm$0.0 & 0.06$\pm$0.0 & 0.05$\pm$0.0 & 0.03$\pm$0.0 & 0.02$\pm$0.0 & 0.01$\pm$0.0 & -0.01$\pm$0.0 & -0.02$\pm$0.0 & -0.03$\pm$0.0 & -0.04$\pm$0.0 & -0.05$\pm$0.0  \\
    Ca4227 & 1.4$\pm$0.01 & 1.42$\pm$0.01 & 1.39$\pm$0.01 & 1.32$\pm$0.01 & 1.24$\pm$0.02 & 1.16$\pm$0.02 & 1.09$\pm$0.02 & 0.97$\pm$0.02 & 1.0$\pm$0.02 & 0.99$\pm$0.02 & 0.8$\pm$0.03  \\
    Fe4531 & 3.49$\pm$0.03 & 3.45$\pm$0.02 & 3.41$\pm$0.02 & 3.43$\pm$0.03 & 3.3$\pm$0.03 & 3.23$\pm$0.03 & 3.17$\pm$0.04 & 2.99$\pm$0.04 & 2.76$\pm$0.03 & 2.69$\pm$0.04 & 2.52$\pm$0.05  \\
    C24668 & 6.32$\pm$0.03 & 6.21$\pm$0.02 & 5.87$\pm$0.03 & 5.61$\pm$0.03 & 5.4$\pm$0.04 & 5.15$\pm$0.05 & 4.78$\pm$0.05 & 4.54$\pm$0.05 & 4.24$\pm$0.05 & 3.82$\pm$0.06 & 3.06$\pm$0.06  \\
    Hb & 1.83$\pm$0.01 & 1.79$\pm$0.01 & 1.92$\pm$0.01 & 2.1$\pm$0.01 & 2.2$\pm$0.02 & 2.32$\pm$0.02 & 2.51$\pm$0.02 & 2.57$\pm$0.02 & 2.67$\pm$0.02 & 2.79$\pm$0.02 & 2.7$\pm$0.03  \\
    Mgb & 4.0$\pm$0.01 & 3.94$\pm$0.01 & 3.86$\pm$0.01 & 3.73$\pm$0.01 & 3.56$\pm$0.02 & 3.39$\pm$0.02 & 3.24$\pm$0.02 & 3.13$\pm$0.02 & 2.95$\pm$0.02 & 2.72$\pm$0.02 & 2.46$\pm$0.03  \\
    Fe5270 & 3.07$\pm$0.01 & 3.04$\pm$0.01 & 3.01$\pm$0.01 & 2.96$\pm$0.01 & 2.9$\pm$0.02 & 2.8$\pm$0.02 & 2.72$\pm$0.02 & 2.72$\pm$0.02 & 2.63$\pm$0.02 & 2.44$\pm$0.03 & 2.19$\pm$0.04  \\
    Fe5335 & 3.08$\pm$0.01 & 3.03$\pm$0.01 & 2.93$\pm$0.01 & 2.82$\pm$0.02 & 2.77$\pm$0.02 & 2.62$\pm$0.02 & 2.58$\pm$0.02 & 2.54$\pm$0.03 & 2.5$\pm$0.02 & 2.31$\pm$0.03 & 2.14$\pm$0.03  \\
    NaD & 3.65$\pm$0.01 & 3.39$\pm$0.01 & 3.25$\pm$0.01 & 3.07$\pm$0.01 & 2.9$\pm$0.02 & 2.78$\pm$0.02 & 2.66$\pm$0.03 & 2.51$\pm$0.03 & 2.36$\pm$0.02 & 2.1$\pm$0.03 & 1.87$\pm$0.03  \\
  \hline 
\end{longtable}

\newpage

\begin{longtable}{ccccccccccccc}
  \caption{Stellar population parameters and errors derived using the TMJ models and a combination of optical indices for ETGs.}
  \label{tab:parameters_ET}
\endfirsthead
    \caption[]{(continued)}
  \endhead \\
  \hline
    log Age (Gyr)			& 0.0-0.1 & 0.1-0.2 & 0.2-0.3 & 0.3-0.4 & 0.4-0.5 & 0.5-0.6 & 0.6-0.7 & 0.7-0.8 & 0.8-1.0 & 1.1-1.2 & 1.2-1.5 \\
    \hline
    $8.8 - 9.8$ &  0.75$\pm$0.03 &  0.73$\pm$0.02 &  0.73$\pm$0.03 &  0.71$\pm$0.01 &  0.71$\pm$0.01 &  0.71$\pm$0.02 &  0.75$\pm$0.02 &  0.71$\pm$0.02 &  0.68$\pm$0.03 &  0.66$\pm$0.02 &  0.73$\pm$0.03  \\
    $9.8 - 10.3$ &  0.85$\pm$0.01 &  0.86$\pm$0.01 &  0.76$\pm$0.01 &  0.88$\pm$0.01 &  0.83$\pm$0.01 &  0.73$\pm$0.02 &  0.69$\pm$0.02 &  0.71$\pm$0.03 &  0.74$\pm$0.01 &  0.76$\pm$0.01 &  0.78$\pm$0.02  \\
    $10.3 - 10.6$ &  0.99$\pm$0.01 &  1.01$\pm$0.0 &  0.93$\pm$0.0 &  0.96$\pm$0.0 &  0.88$\pm$0.01 &  0.85$\pm$0.0 &  0.8$\pm$0.0 &  0.87$\pm$0.01 &  0.83$\pm$0.01 &  0.84$\pm$0.02 &  0.9$\pm$0.01  \\
    $10.6 - 10.8$ &  1.14$\pm$0.01 &  1.01$\pm$0.0 &  1.03$\pm$0.0 &  1.0$\pm$0.0 &  0.97$\pm$0.0 &  1.0$\pm$0.01 &  0.91$\pm$0.01 &  0.92$\pm$0.01 &  0.9$\pm$0.01 &  0.9$\pm$0.02 &  1.01$\pm$0.01  \\
    $10.8 - 11.0$ &  1.03$\pm$0.01 &  1.07$\pm$0.01 &  1.04$\pm$0.01 &  1.03$\pm$0.01 &  1.01$\pm$0.01 &  0.95$\pm$0.01 &  0.99$\pm$0.02 &  0.94$\pm$0.02 &  1.04$\pm$0.02 &  0.91$\pm$0.04 &  1.14$\pm$0.03  \\
    $11.0 - 11.3$ &  1.12$\pm$0.02 &  1.1$\pm$0.01 &  1.09$\pm$0.01 &  1.05$\pm$0.01 &  1.03$\pm$0.02 &  1.08$\pm$0.01 &  1.08$\pm$0.0 &  1.07$\pm$0.03 &  1.1$\pm$0.02 &  1.05$\pm$0.03 &  1.1$\pm$0.0  \\
     \hline
    [Z/H] (solar scaled)	& 0.0-0.1 & 0.1-0.2 & 0.2-0.3 & 0.3-0.4 & 0.4-0.5 & 0.5-0.6 & 0.6-0.7 & 0.7-0.8 & 0.8-1.0 & 1.1-1.2 & 1.2-1.5 \\
     \hline
    $8.8 - 9.8$ &  -0.24$\pm$0.02 &  -0.22$\pm$0.01 &  -0.24$\pm$0.02 &  -0.24$\pm$0.01 &  -0.25$\pm$0.01 &  -0.24$\pm$0.01 &  -0.28$\pm$0.01 &  -0.29$\pm$0.01 &  -0.27$\pm$0.02 &  -0.27$\pm$0.01 &  -0.32$\pm$0.02  \\
    $9.8 - 10.3$ &  0.07$\pm$0.01 &  0.06$\pm$0.01 &  0.1$\pm$0.01 &  0.01$\pm$0.01 &  0.04$\pm$0.01 &  0.03$\pm$0.01 &  0.04$\pm$0.01 &  -0.01$\pm$0.01 &  -0.05$\pm$0.01 &  -0.1$\pm$0.01 &  -0.14$\pm$0.01  \\
    $10.3 - 10.6$ &  0.07$\pm$0.01 &  0.05$\pm$0.0 &  0.09$\pm$0.0 &  0.05$\pm$0.0 &  0.08$\pm$0.0 &  0.05$\pm$0.0 &  0.06$\pm$0.0 &  0.02$\pm$0.01 &  0.01$\pm$0.01 &  -0.01$\pm$0.01 &  -0.09$\pm$0.01  \\
    $10.6 - 10.8$ &  0.03$\pm$0.01 &  0.12$\pm$0.0 &  0.07$\pm$0.0 &  0.06$\pm$0.0 &  0.05$\pm$0.0 &  0.0$\pm$0.01 &  0.04$\pm$0.01 &  0.01$\pm$0.01 &  0.01$\pm$0.01 &  -0.03$\pm$0.01 &  -0.15$\pm$0.01  \\
    $10.8 - 11.0$ &  0.15$\pm$0.01 &  0.11$\pm$0.01 &  0.1$\pm$0.01 &  0.07$\pm$0.01 &  0.05$\pm$0.01 &  0.07$\pm$0.01 &  0.03$\pm$0.02 &  0.04$\pm$0.02 &  -0.06$\pm$0.01 &  -0.01$\pm$0.02 &  -0.21$\pm$0.02  \\
    $11.0 - 11.3$ &  0.02$\pm$0.02 &  0.03$\pm$0.01 &  0.02$\pm$0.01 &  0.05$\pm$0.01 &  0.03$\pm$0.02 &  -0.03$\pm$0.01 &  -0.05$\pm$0.02 &  -0.07$\pm$0.02 &  -0.12$\pm$0.02 &  -0.15$\pm$0.02 &  -0.23$\pm$0.04  \\
    \hline
    [C/Fe]        			 & 0.0-0.1 & 0.1-0.2 & 0.2-0.3 & 0.3-0.4 & 0.4-0.5 & 0.5-0.6 & 0.6-0.7 & 0.7-0.8 & 0.8-1.0 & 1.1-1.2 & 1.2-1.5 \\
    \hline
    $8.8 - 9.8$ &  0.17$\pm$0.03 &  0.21$\pm$0.02 &  0.22$\pm$0.01 &  0.18$\pm$0.02 &  0.22$\pm$0.02 &  0.16$\pm$0.01 &  0.21$\pm$0.01 &  0.21$\pm$0.02 &  0.16$\pm$0.01 &  0.19$\pm$0.01 &  0.3$\pm$0.02  \\
    $9.8 - 10.3$ &  0.24$\pm$0.01 &  0.25$\pm$0.0 &  0.2$\pm$0.0 &  0.24$\pm$0.0 &  0.23$\pm$0.0 &  0.21$\pm$0.0 &  0.2$\pm$0.01 &  0.25$\pm$0.01 &  0.26$\pm$0.01 &  0.29$\pm$0.01 &  0.31$\pm$0.01  \\
    $10.3 - 10.6$ &  0.28$\pm$0.0 &  0.27$\pm$0.0 &  0.26$\pm$0.0 &  0.22$\pm$0.0 &  0.22$\pm$0.0 &  0.26$\pm$0.02 &  0.2$\pm$0.03 &  0.25$\pm$0.01 &  0.26$\pm$0.0 &  0.24$\pm$0.01 &  0.26$\pm$0.01  \\
    $10.6 - 10.8$ &  0.33$\pm$0.0 &  0.3$\pm$0.0 &  0.31$\pm$0.0 &  0.29$\pm$0.0 &  0.28$\pm$0.0 &  0.27$\pm$0.01 &  0.25$\pm$0.01 &  0.25$\pm$0.01 &  0.25$\pm$0.01 &  0.23$\pm$0.01 &  0.28$\pm$0.01  \\
    $10.8 - 11.0$ &  0.31$\pm$0.01 &  0.3$\pm$0.0 &  0.3$\pm$0.0 &  0.3$\pm$0.01 &  0.3$\pm$0.01 &  0.29$\pm$0.01 &  0.31$\pm$0.01 &  0.28$\pm$0.01 &  0.3$\pm$0.01 &  0.27$\pm$0.02 &  0.36$\pm$0.02  \\
    $11.0 - 11.3$ &  0.42$\pm$0.02 &  0.4$\pm$0.01 &  0.38$\pm$0.01 &  0.36$\pm$0.01 &  0.33$\pm$0.01 &  0.36$\pm$0.02 &  0.32$\pm$0.08 &  0.32$\pm$0.02 &  0.36$\pm$0.02 &  0.42$\pm$0.03 &  0.47$\pm$0.14  \\
    \hline
    [N/Fe]         			& 0.0-0.1 & 0.1-0.2 & 0.2-0.3 & 0.3-0.4 & 0.4-0.5 & 0.5-0.6 & 0.6-0.7 & 0.7-0.8 & 0.8-1.0 & 1.1-1.2 & 1.2-1.5 \\
    \hline
    $8.8 - 9.8$ &  -0.15$\pm$0.03 &  -0.09$\pm$0.02 &  -0.15$\pm$0.01 &  -0.15$\pm$0.02 &  -0.11$\pm$0.02 &  -0.03$\pm$0.01 &  -0.1$\pm$0.01 &  -0.1$\pm$0.01 &  -0.13$\pm$0.01 &  -0.09$\pm$0.01 &  0.02$\pm$0.02  \\
    $9.8 - 10.3$ &  0.11$\pm$0.01 &  0.14$\pm$0.0 &  0.15$\pm$0.0 &  0.09$\pm$0.0 &  0.09$\pm$0.01 &  0.04$\pm$0.01 &  0.09$\pm$0.01 &  0.09$\pm$0.01 &  0.08$\pm$0.01 &  0.11$\pm$0.01 &  0.06$\pm$0.01  \\
    $10.3 - 10.6$ &  0.15$\pm$0.0 &  0.08$\pm$0.0 &  0.01$\pm$0.0 &  0.08$\pm$0.0 &  0.04$\pm$0.0 &  0.11$\pm$0.02 &  0.12$\pm$0.03 &  0.13$\pm$0.01 &  0.1$\pm$0.0 &  0.17$\pm$0.01 &  0.15$\pm$0.01  \\
    $10.6 - 10.8$ &  0.03$\pm$0.0 &  0.09$\pm$0.0 &  0.06$\pm$0.0 &  0.06$\pm$0.0 &  0.07$\pm$0.0 &  -0.04$\pm$0.0 &  0.05$\pm$0.01 &  0.09$\pm$0.01 &  0.08$\pm$0.01 &  0.06$\pm$0.01 &  -0.01$\pm$0.01  \\
    $10.8 - 11.0$ &  0.21$\pm$0.01 &  0.2$\pm$0.0 &  0.19$\pm$0.0 &  0.17$\pm$0.01 &  0.19$\pm$0.01 &  0.19$\pm$0.01 &  0.16$\pm$0.01 &  0.13$\pm$0.01 &  0.11$\pm$0.01 &  0.23$\pm$0.02 &  0.17$\pm$0.02  \\
    $11.0 - 11.3$ &  0.3$\pm$0.02 &  0.3$\pm$0.01 &  0.28$\pm$0.01 &  0.23$\pm$0.01 &  0.25$\pm$0.01 &  0.24$\pm$0.02 &  0.23$\pm$0.08 &  0.24$\pm$0.03 &  0.23$\pm$0.02 &  0.24$\pm$0.03 &  0.26$\pm$0.14  \\
    \hline
    [Na/Fe]				& 0.0-0.1 & 0.1-0.2 & 0.2-0.3 & 0.3-0.4 & 0.4-0.5 & 0.5-0.6 & 0.6-0.7 & 0.7-0.8 & 0.8-1.0 & 1.1-1.2 & 1.2-1.5 \\
    \hline
    $8.8 - 9.8$ &  -0.11$\pm$0.03 &  -0.04$\pm$0.02 &  -0.07$\pm$0.01 &  -0.15$\pm$0.02 &  -0.11$\pm$0.02 &  -0.21$\pm$0.01 &  0.15$\pm$0.02 &  0.13$\pm$0.02 &  0.11$\pm$0.02 &  0.12$\pm$0.02 &  0.24$\pm$0.02  \\
    $9.8 - 10.3$ &  0.27$\pm$0.01 &  0.27$\pm$0.0 &  0.19$\pm$0.0 &  0.22$\pm$0.0 &  0.18$\pm$0.01 &  0.15$\pm$0.0 &  0.16$\pm$0.01 &  0.15$\pm$0.01 &  0.14$\pm$0.01 &  0.24$\pm$0.01 &  0.09$\pm$0.01  \\
    $10.3 - 10.6$ &  0.4$\pm$0.0 &  0.35$\pm$0.0 &  0.31$\pm$0.0 &  0.27$\pm$0.0 &  0.25$\pm$0.0 &  0.27$\pm$0.02 &  0.22$\pm$0.03 &  0.23$\pm$0.01 &  0.21$\pm$0.0 &  0.18$\pm$0.01 &  0.17$\pm$0.01  \\
    $10.6 - 10.8$ &  0.46$\pm$0.0 &  0.45$\pm$0.0 &  0.43$\pm$0.0 &  0.37$\pm$0.0 &  0.37$\pm$0.0 &  0.3$\pm$0.01 &  0.28$\pm$0.01 &  0.27$\pm$0.01 &  0.28$\pm$0.01 &  0.23$\pm$0.01 &  0.24$\pm$0.01  \\
    $10.8 - 11.0$ &  0.45$\pm$0.01 &  0.43$\pm$0.0 &  0.41$\pm$0.0 &  0.37$\pm$0.01 &  0.35$\pm$0.01 &  0.34$\pm$0.01 &  0.32$\pm$0.01 &  0.28$\pm$0.01 &  0.25$\pm$0.01 &  0.28$\pm$0.02 &  0.26$\pm$0.02  \\
    $11.0 - 11.3$ &  0.59$\pm$0.02 &  0.57$\pm$0.01 &  0.54$\pm$0.01 &  0.49$\pm$0.01 &  0.45$\pm$0.01 &  0.46$\pm$0.02 &  0.39$\pm$0.08 &  0.38$\pm$0.03 &  0.33$\pm$0.02 &  0.4$\pm$0.03 &  0.35$\pm$0.14  \\
    \hline
    [Mg/Fe]				& 0.0-0.1 & 0.1-0.2 & 0.2-0.3 & 0.3-0.4 & 0.4-0.5 & 0.5-0.6 & 0.6-0.7 & 0.7-0.8 & 0.8-1.0 & 1.1-1.2 & 1.2-1.5 \\
    \hline
    $8.8 - 9.8$ &  0.15$\pm$0.03 &  0.21$\pm$0.02 &  0.2$\pm$0.01 &  0.15$\pm$0.02 &  0.19$\pm$0.02 &  0.19$\pm$0.01 &  0.2$\pm$0.01 &  0.2$\pm$0.01 &  0.18$\pm$0.01 &  0.21$\pm$0.01 &  0.32$\pm$0.02  \\
    $9.8 - 10.3$ &  0.26$\pm$0.01 &  0.27$\pm$0.0 &  0.24$\pm$0.0 &  0.25$\pm$0.0 &  0.25$\pm$0.0 &  0.23$\pm$0.0 &  0.26$\pm$0.01 &  0.28$\pm$0.01 &  0.28$\pm$0.01 &  0.32$\pm$0.01 &  0.34$\pm$0.01  \\
    $10.3 - 10.6$ &  0.26$\pm$0.0 &  0.23$\pm$0.0 &  0.24$\pm$0.0 &  0.24$\pm$0.0 &  0.26$\pm$0.0 &  0.28$\pm$0.02 &  0.26$\pm$0.03 &  0.29$\pm$0.0 &  0.29$\pm$0.0 &  0.3$\pm$0.0 &  0.29$\pm$0.01  \\
    $10.6 - 10.8$ &  0.27$\pm$0.0 &  0.29$\pm$0.0 &  0.3$\pm$0.0 &  0.29$\pm$0.0 &  0.3$\pm$0.0 &  0.26$\pm$0.0 &  0.28$\pm$0.01 &  0.29$\pm$0.01 &  0.31$\pm$0.01 &  0.29$\pm$0.01 &  0.29$\pm$0.01  \\
    $10.8 - 11.0$ &  0.28$\pm$0.01 &  0.26$\pm$0.0 &  0.27$\pm$0.0 &  0.27$\pm$0.0 &  0.28$\pm$0.01 &  0.29$\pm$0.01 &  0.31$\pm$0.01 &  0.29$\pm$0.01 &  0.28$\pm$0.01 &  0.31$\pm$0.01 &  0.32$\pm$0.02  \\
    $11.0 - 11.3$ &  0.34$\pm$0.02 &  0.32$\pm$0.01 &  0.32$\pm$0.01 &  0.32$\pm$0.01 &  0.31$\pm$0.01 &  0.35$\pm$0.02 &  0.31$\pm$0.08 &  0.3$\pm$0.02 &  0.32$\pm$0.02 &  0.37$\pm$0.02 &  0.37$\pm$0.14  \\
    \hline
    [Ca/Fe]				& 0.0-0.1 & 0.1-0.2 & 0.2-0.3 & 0.3-0.4 & 0.4-0.5 & 0.5-0.6 & 0.6-0.7 & 0.7-0.8 & 0.8-1.0 & 1.1-1.2 & 1.2-1.5 \\
    \hline
    $8.8 - 9.8$ &  0.1$\pm$0.03 &  0.11$\pm$0.02 &  0.09$\pm$0.01 &  0.1$\pm$0.02 &  0.13$\pm$0.02 &  0.02$\pm$0.01 &  0.12$\pm$0.01 &  0.13$\pm$0.02 &  0.07$\pm$0.01 &  0.05$\pm$0.02 &  0.14$\pm$0.02  \\
    $9.8 - 10.3$ &  0.08$\pm$0.01 &  0.11$\pm$0.0 &  0.04$\pm$0.0 &  0.12$\pm$0.0 &  0.11$\pm$0.01 &  0.08$\pm$0.0 &  0.07$\pm$0.01 &  0.14$\pm$0.01 &  0.15$\pm$0.01 &  0.14$\pm$0.01 &  0.12$\pm$0.01  \\
    $10.3 - 10.6$ &  0.09$\pm$0.0 &  0.06$\pm$0.0 &  0.04$\pm$0.0 &  0.04$\pm$0.0 &  0.01$\pm$0.0 &  0.1$\pm$0.02 &  0.04$\pm$0.03 &  0.09$\pm$0.01 &  0.09$\pm$0.0 &  0.08$\pm$0.01 &  0.11$\pm$0.01  \\
    $10.6 - 10.8$ &  0.05$\pm$0.0 &  0.04$\pm$0.0 &  0.08$\pm$0.0 &  0.07$\pm$0.0 &  0.09$\pm$0.0 &  0.09$\pm$0.01 &  0.08$\pm$0.01 &  0.05$\pm$0.01 &  0.05$\pm$0.01 &  0.04$\pm$0.01 &  0.07$\pm$0.01  \\
    $10.8 - 11.0$ &  0.07$\pm$0.01 &  0.09$\pm$0.0 &  0.07$\pm$0.0 &  0.11$\pm$0.01 &  0.13$\pm$0.01 &  0.15$\pm$0.01 &  0.14$\pm$0.01 &  0.12$\pm$0.01 &  0.11$\pm$0.01 &  0.16$\pm$0.02 &  0.17$\pm$0.02  \\
    $11.0 - 11.3$ &  0.18$\pm$0.02 &  0.19$\pm$0.01 &  0.15$\pm$0.01 &  0.15$\pm$0.01 &  0.14$\pm$0.01 &  0.16$\pm$0.02 &  0.12$\pm$0.08 &  0.15$\pm$0.03 &  0.12$\pm$0.02 &  0.17$\pm$0.03 &  0.2$\pm$0.14  \\
    \hline
    [Ti/Fe]				& 0.0-0.1 & 0.1-0.2 & 0.2-0.3 & 0.3-0.4 & 0.4-0.5 & 0.5-0.6 & 0.6-0.7 & 0.7-0.8 & 0.8-1.0 & 1.1-1.2 & 1.2-1.5 \\
    \hline
    $8.8 - 9.8$ &  0.01$\pm$0.05 &  0.09$\pm$0.03 &  0.12$\pm$0.02 &  -0.0$\pm$0.02 &  0.11$\pm$0.02 &  0.03$\pm$0.02 &  0.14$\pm$0.02 &  0.07$\pm$0.03 &  0.04$\pm$0.02 &  0.05$\pm$0.02 &  0.28$\pm$0.04  \\
    $9.8 - 10.3$ &  0.17$\pm$0.01 &  0.2$\pm$0.01 &  0.14$\pm$0.01 &  0.15$\pm$0.01 &  0.18$\pm$0.01 &  0.15$\pm$0.01 &  0.18$\pm$0.01 &  0.2$\pm$0.01 &  0.21$\pm$0.01 &  0.28$\pm$0.01 &  0.27$\pm$0.01  \\
    $10.3 - 10.6$ &  0.21$\pm$0.01 &  0.16$\pm$0.0 &  0.18$\pm$0.0 &  0.16$\pm$0.0 &  0.18$\pm$0.01 &  0.24$\pm$0.02 &  0.14$\pm$0.03 &  0.19$\pm$0.01 &  0.21$\pm$0.01 &  0.22$\pm$0.01 &  0.18$\pm$0.01  \\
    $10.6 - 10.8$ &  0.24$\pm$0.01 &  0.22$\pm$0.0 &  0.22$\pm$0.0 &  0.25$\pm$0.01 &  0.25$\pm$0.01 &  0.2$\pm$0.01 &  0.2$\pm$0.01 &  0.19$\pm$0.01 &  0.22$\pm$0.01 &  0.24$\pm$0.02 &  0.24$\pm$0.02  \\
    $10.8 - 11.0$ &  0.22$\pm$0.01 &  0.2$\pm$0.01 &  0.19$\pm$0.01 &  0.24$\pm$0.01 &  0.23$\pm$0.01 &  0.24$\pm$0.01 &  0.25$\pm$0.01 &  0.2$\pm$0.02 &  0.18$\pm$0.02 &  0.2$\pm$0.02 &  0.22$\pm$0.03  \\
    $11.0 - 11.3$ &  0.33$\pm$0.03 &  0.3$\pm$0.01 &  0.31$\pm$0.01 &  0.28$\pm$0.01 &  0.27$\pm$0.02 &  0.31$\pm$0.02 &  0.26$\pm$0.08 &  0.25$\pm$0.03 &  0.25$\pm$0.03 &  0.4$\pm$0.04 &  0.42$\pm$0.15  \\
    \hline
\end{longtable}

\newpage

\begin{longtable}{ccccccccccccc}
  \caption{Same as \autoref{tab:parameters_ET} for LTGs.}
  \label{tab:parameters_LT}
  \endfirsthead
  \caption[]{(continued)}
  \endhead \\
  \hline
    log Age$_{LW}$ (Gyr)			& 0.0-0.1 & 0.1-0.2 & 0.2-0.3 & 0.3-0.4 & 0.4-0.5 & 0.5-0.6 & 0.6-0.7 & 0.7-0.8 & 0.8-1.0 & 1.1-1.2 & 1.2-1.5 \\
    \hline
    $9.2 - 9.6$ &  0.5$\pm$0.07 &  0.76$\pm$0.02 &  0.72$\pm$0.02 &  0.71$\pm$0.03 &  0.69$\pm$0.03 &  0.66$\pm$0.01 &  0.68$\pm$0.02 &  0.56$\pm$0.04 &  0.61$\pm$0.03 &  0.64$\pm$0.05 &  0.67$\pm$0.06 \\ 
$9.6 - 9.9$ &  0.49$\pm$0.15 &  0.25$\pm$0.16 &  0.26$\pm$0.13 &  0.41$\pm$0.12 &  0.44$\pm$0.11 &  0.4$\pm$0.12 &  0.27$\pm$0.04 &  0.52$\pm$0.03 &  0.51$\pm$0.02 &  0.42$\pm$0.03 &  0.61$\pm$0.15 \\ 
$9.9 - 10.1$ &  0.54$\pm$0.06 &  0.62$\pm$0.03 &  0.45$\pm$0.03 &  0.45$\pm$0.02 &  0.24$\pm$0.05 &  0.34$\pm$0.04 &  0.23$\pm$0.06 &  0.13$\pm$0.02 &  0.3$\pm$0.01 &  0.22$\pm$0.01 &  0.65$\pm$0.01 \\ 
$10.1 - 10.3$ &  0.78$\pm$0.03 &  0.69$\pm$0.02 &  0.75$\pm$0.02 &  0.69$\pm$0.03 &  0.52$\pm$0.06 &  0.41$\pm$0.03 &  0.47$\pm$0.04 &  0.3$\pm$0.06 &  0.13$\pm$0.01 &  0.35$\pm$0.05 &  0.45$\pm$0.06 \\ 
$10.3 - 10.6$ &  0.97$\pm$0.02 &  0.95$\pm$0.02 &  0.8$\pm$0.02 &  0.72$\pm$0.02 &  0.63$\pm$0.02 &  0.57$\pm$0.04 &  0.46$\pm$0.02 &  0.46$\pm$0.03 &  0.23$\pm$0.04 &  0.16$\pm$0.06 &  0.5$\pm$0.04 \\ 
$10.6 - 11.2$ &  1.0$\pm$0.01 &  1.06$\pm$0.01 &  0.99$\pm$0.02 &  0.83$\pm$0.02 &  0.81$\pm$0.01 &  0.74$\pm$0.02 &  0.66$\pm$0.01 &  0.43$\pm$0.02 &  0.4$\pm$0.02 &  0.31$\pm$0.03 &  0.57$\pm$0.03 \\ 
     \hline
    [Z/H] (solar scaled)	& 0.0-0.1 & 0.1-0.2 & 0.2-0.3 & 0.3-0.4 & 0.4-0.5 & 0.5-0.6 & 0.6-0.7 & 0.7-0.8 & 0.8-1.0 & 1.1-1.2 & 1.2-1.5 \\
     \hline
    $9.2 - 9.6$ &  -0.58$\pm$0.05 &  -0.73$\pm$0.03 &  -0.73$\pm$0.03 &  -0.77$\pm$0.03 &  -0.82$\pm$0.03 &  -0.83$\pm$0.02 &  -0.86$\pm$0.02 &  -0.82$\pm$0.04 &  -0.92$\pm$0.03 &  -0.96$\pm$0.04 &  -0.98$\pm$0.05   \\
    $9.6 - 9.9$ &  -0.37$\pm$0.05 &  -0.23$\pm$0.03 &  -0.24$\pm$0.03 &  -0.33$\pm$0.03 &  -0.39$\pm$0.03 &  -0.46$\pm$0.04 &  -0.43$\pm$0.05 &  -0.51$\pm$0.05 &  -0.57$\pm$0.05 &  -0.55$\pm$0.05 &  -0.76$\pm$0.11   \\
    $9.9 - 10.1$ &  -0.19$\pm$0.02 &  -0.22$\pm$0.01 &  -0.14$\pm$0.01 &  -0.21$\pm$0.01 &  -0.09$\pm$0.02 &  -0.24$\pm$0.02 &  -0.23$\pm$0.03 &  -0.23$\pm$0.02 &  -0.41$\pm$0.01 &  -0.39$\pm$0.02 &  -0.52$\pm$0.01   \\
    $10.1 - 10.3$ &  -0.05$\pm$0.02 &  -0.02$\pm$0.01 &  -0.07$\pm$0.01 &  -0.16$\pm$0.01 &  -0.12$\pm$0.02 &  -0.12$\pm$0.02 &  -0.21$\pm$0.03 &  -0.17$\pm$0.03 &  -0.13$\pm$0.01 &  -0.3$\pm$0.03 &  -0.42$\pm$0.03   \\
    $10.3 - 10.6$ &  -0.08$\pm$0.01 &  -0.1$\pm$0.01 &  -0.04$\pm$0.01 &  -0.09$\pm$0.01 &  -0.06$\pm$0.01 &  -0.12$\pm$0.01 &  -0.17$\pm$0.02 &  -0.2$\pm$0.01 &  -0.14$\pm$0.02 &  -0.21$\pm$0.03 &  -0.37$\pm$0.02   \\
    $10.6 - 11.2$ &  0.04$\pm$0.01 &  -0.07$\pm$0.01 &  -0.02$\pm$0.01 &  0.02$\pm$0.01 &  -0.01$\pm$0.01 &  -0.1$\pm$0.01 &  -0.12$\pm$0.01 &  -0.07$\pm$0.01 &  -0.16$\pm$0.01 &  -0.18$\pm$0.02 &  -0.36$\pm$0.01  \\
    \hline
    [C/Fe]        			 & 0.0-0.1 & 0.1-0.2 & 0.2-0.3 & 0.3-0.4 & 0.4-0.5 & 0.5-0.6 & 0.6-0.7 & 0.7-0.8 & 0.8-1.0 & 1.1-1.2 & 1.2-1.5 \\
    \hline
    $9.2 - 9.6$ &  0.26$\pm$0.03 &  0.21$\pm$0.02 &  0.18$\pm$0.02 &  0.21$\pm$0.02 &  0.15$\pm$0.02 &  0.23$\pm$0.05 &  0.21$\pm$0.06 &  0.16$\pm$0.02 &  0.24$\pm$0.02 &  0.32$\pm$0.03 &  0.33$\pm$0.03   \\
    $9.6 - 9.9$ &  0.17$\pm$0.04 &  0.12$\pm$0.02 &  0.21$\pm$0.02 &  0.22$\pm$0.02 &  0.16$\pm$0.02 &  0.13$\pm$0.03 &  0.13$\pm$0.02 &  0.14$\pm$0.02 &  0.2$\pm$0.02 &  0.27$\pm$0.03 &  0.33$\pm$0.05   \\
    $9.9 - 10.1$ &  0.19$\pm$0.02 &  0.16$\pm$0.02 &  0.19$\pm$0.01 &  0.19$\pm$0.01 &  0.11$\pm$0.01 &  0.09$\pm$0.01 &  0.08$\pm$0.01 &  0.05$\pm$0.01 &  0.23$\pm$0.01 &  0.28$\pm$0.01 &  0.24$\pm$0.01   \\
    $10.1 - 10.3$ &  0.18$\pm$0.01 &  0.2$\pm$0.01 &  0.08$\pm$0.01 &  0.18$\pm$0.01 &  0.17$\pm$0.01 &  0.14$\pm$0.01 &  0.09$\pm$0.01 &  0.17$\pm$0.01 &  0.12$\pm$0.01 &  0.2$\pm$0.01 &  0.16$\pm$0.01   \\
    $10.3 - 10.6$ &  0.23$\pm$0.01 &  0.22$\pm$0.0 &  0.24$\pm$0.0 &  0.22$\pm$0.0 &  0.21$\pm$0.01 &  0.21$\pm$0.01 &  0.16$\pm$0.01 &  0.16$\pm$0.01 &  0.12$\pm$0.01 &  0.13$\pm$0.01 &  0.17$\pm$0.01   \\
    $10.6 - 11.2$ &  0.25$\pm$0.01 &  0.3$\pm$0.0 &  0.27$\pm$0.0 &  0.23$\pm$0.01 &  0.24$\pm$0.01 &  0.25$\pm$0.01 &  0.22$\pm$0.01 &  0.16$\pm$0.01 &  0.19$\pm$0.01 &  0.27$\pm$0.01 &  0.24$\pm$0.02   \\
    \hline
    [N/Fe]         			& 0.0-0.1 & 0.1-0.2 & 0.2-0.3 & 0.3-0.4 & 0.4-0.5 & 0.5-0.6 & 0.6-0.7 & 0.7-0.8 & 0.8-1.0 & 1.1-1.2 & 1.2-1.5 \\
    \hline
    $9.2 - 9.6$ &  0.0$\pm$0.03 &  -0.19$\pm$0.02 &  -0.22$\pm$0.01 &  -0.19$\pm$0.02 &  -0.03$\pm$0.02 &  -0.14$\pm$0.05 &  -0.16$\pm$0.05 &  -0.11$\pm$0.02 &  -0.1$\pm$0.02 &  0.01$\pm$0.02 &  0.14$\pm$0.03   \\
    $9.6 - 9.9$ &  -0.23$\pm$0.04 &  -0.12$\pm$0.02 &  -0.04$\pm$0.02 &  -0.13$\pm$0.02 &  -0.17$\pm$0.02 &  -0.3$\pm$0.03 &  -0.3$\pm$0.02 &  -0.23$\pm$0.02 &  -0.12$\pm$0.02 &  -0.03$\pm$0.02 &  -0.15$\pm$0.05   \\
    $9.9 - 10.1$ &  -0.19$\pm$0.02 &  -0.23$\pm$0.02 &  -0.13$\pm$0.01 &  -0.29$\pm$0.01 &  -0.15$\pm$0.01 &  -0.26$\pm$0.01 &  -0.32$\pm$0.01 &  -0.31$\pm$0.01 &  -0.19$\pm$0.01 &  -0.18$\pm$0.01 &  -0.0$\pm$0.01   \\
    $10.1 - 10.3$ &  -0.17$\pm$0.0 &  -0.19$\pm$0.01 &  -0.3$\pm$0.01 &  -0.36$\pm$0.01 &  -0.33$\pm$0.0 &  -0.32$\pm$0.01 &  -0.41$\pm$0.01 &  -0.31$\pm$0.01 &  -0.26$\pm$0.01 &  -0.29$\pm$0.01 &  -0.32$\pm$0.01   \\
    $10.3 - 10.6$ &  -0.21$\pm$0.01 &  -0.21$\pm$0.0 &  -0.13$\pm$0.0 &  -0.28$\pm$0.0 &  -0.27$\pm$0.0 &  -0.3$\pm$0.01 &  -0.35$\pm$0.01 &  -0.34$\pm$0.01 &  -0.24$\pm$0.01 &  -0.29$\pm$0.01 &  -0.15$\pm$0.01   \\
    $10.6 - 11.2$ &  0.08$\pm$0.01 &  -0.03$\pm$0.0 &  -0.06$\pm$0.0 &  -0.04$\pm$0.01 &  -0.15$\pm$0.01 &  -0.18$\pm$0.01 &  -0.27$\pm$0.01 &  -0.33$\pm$0.01 &  -0.35$\pm$0.01 &  -0.22$\pm$0.01 &  -0.14$\pm$0.02   \\
    \hline
    [Na/Fe]				& 0.0-0.1 & 0.1-0.2 & 0.2-0.3 & 0.3-0.4 & 0.4-0.5 & 0.5-0.6 & 0.6-0.7 & 0.7-0.8 & 0.8-1.0 & 1.1-1.2 & 1.2-1.5 \\
    \hline
    $9.2 - 9.6$ &  -0.02$\pm$0.03 &  -0.15$\pm$0.02 &  -0.22$\pm$0.01 &  -0.28$\pm$0.02 &  -0.24$\pm$0.02 &  -0.19$\pm$0.05 &  -0.21$\pm$0.05 &  -0.28$\pm$0.02 &  -0.2$\pm$0.02 &  -0.12$\pm$0.02 &  -0.07$\pm$0.03   \\
    $9.6 - 9.9$ &  0.01$\pm$0.04 &  -0.06$\pm$0.02 &  -0.04$\pm$0.02 &  -0.11$\pm$0.02 &  -0.12$\pm$0.02 &  -0.12$\pm$0.03 &  -0.2$\pm$0.02 &  -0.25$\pm$0.02 &  -0.17$\pm$0.02 &  -0.08$\pm$0.02 &  -0.13$\pm$0.05   \\
    $9.9 - 10.1$ &  0.24$\pm$0.02 &  0.16$\pm$0.02 &  0.17$\pm$0.01 &  0.13$\pm$0.01 &  0.03$\pm$0.01 &  0.03$\pm$0.01 &  -0.02$\pm$0.01 &  -0.05$\pm$0.01 &  0.16$\pm$0.01 &  0.12$\pm$0.01 &  -0.03$\pm$0.01   \\
    $10.1 - 10.3$ &  0.19$\pm$0.01 &  0.14$\pm$0.01 &  -0.01$\pm$0.01 &  0.07$\pm$0.01 &  0.08$\pm$0.01 &  0.07$\pm$0.01 &  0.01$\pm$0.01 &  0.07$\pm$0.01 &  -0.03$\pm$0.01 &  0.03$\pm$0.01 &  -0.01$\pm$0.02   \\
    $10.3 - 10.6$ &  0.22$\pm$0.01 &  0.19$\pm$0.0 &  0.22$\pm$0.0 &  0.16$\pm$0.0 &  0.14$\pm$0.01 &  0.13$\pm$0.01 &  0.08$\pm$0.01 &  0.09$\pm$0.01 &  -0.0$\pm$0.01 &  0.03$\pm$0.01 &  0.01$\pm$0.01   \\
    $10.6 - 11.2$ &  0.29$\pm$0.01 &  0.22$\pm$0.0 &  0.24$\pm$0.0 &  0.23$\pm$0.01 &  0.21$\pm$0.01 &  0.23$\pm$0.01 &  0.21$\pm$0.01 &  0.11$\pm$0.01 &  0.12$\pm$0.01 &  0.14$\pm$0.01 &  0.11$\pm$0.02   \\
    \hline
    [Mg/Fe]				& 0.0-0.1 & 0.1-0.2 & 0.2-0.3 & 0.3-0.4 & 0.4-0.5 & 0.5-0.6 & 0.6-0.7 & 0.7-0.8 & 0.8-1.0 & 1.1-1.2 & 1.2-1.5 \\
    \hline
    $9.2 - 9.6$ &  0.3$\pm$0.03 &  0.26$\pm$0.02 &  0.23$\pm$0.01 &  0.22$\pm$0.02 &  0.19$\pm$0.02 &  0.26$\pm$0.05 &  0.24$\pm$0.05 &  0.22$\pm$0.02 &  0.3$\pm$0.02 &  0.33$\pm$0.02 &  0.43$\pm$0.03   \\
    $9.6 - 9.9$ &  0.17$\pm$0.04 &  0.21$\pm$0.02 &  0.31$\pm$0.02 &  0.27$\pm$0.02 &  0.23$\pm$0.02 &  0.18$\pm$0.03 &  0.18$\pm$0.02 &  0.25$\pm$0.02 &  0.33$\pm$0.02 &  0.42$\pm$0.02 &  0.3$\pm$0.05   \\
    $9.9 - 10.1$ &  0.22$\pm$0.02 &  0.18$\pm$0.02 &  0.28$\pm$0.01 &  0.21$\pm$0.01 &  0.25$\pm$0.01 &  0.15$\pm$0.01 &  0.16$\pm$0.01 &  0.17$\pm$0.01 &  0.29$\pm$0.01 &  0.32$\pm$0.01 &  0.37$\pm$0.01   \\
    $10.1 - 10.3$ &  0.2$\pm$0.0 &  0.21$\pm$0.01 &  0.1$\pm$0.01 &  0.14$\pm$0.01 &  0.17$\pm$0.0 &  0.18$\pm$0.01 &  0.09$\pm$0.01 &  0.2$\pm$0.01 &  0.21$\pm$0.01 &  0.21$\pm$0.01 &  0.18$\pm$0.01   \\
    $10.3 - 10.6$ &  0.19$\pm$0.01 &  0.19$\pm$0.0 &  0.27$\pm$0.0 &  0.2$\pm$0.0 &  0.23$\pm$0.0 &  0.2$\pm$0.01 &  0.16$\pm$0.01 &  0.17$\pm$0.01 &  0.19$\pm$0.01 &  0.21$\pm$0.01 &  0.26$\pm$0.01   \\
    $10.6 - 11.2$ &  0.22$\pm$0.0 &  0.23$\pm$0.0 &  0.24$\pm$0.0 &  0.26$\pm$0.01 &  0.25$\pm$0.01 &  0.22$\pm$0.01 &  0.2$\pm$0.01 &  0.17$\pm$0.01 &  0.15$\pm$0.01 &  0.25$\pm$0.01 &  0.26$\pm$0.02   \\
    \hline
    [Ca/Fe]				& 0.0-0.1 & 0.1-0.2 & 0.2-0.3 & 0.3-0.4 & 0.4-0.5 & 0.5-0.6 & 0.6-0.7 & 0.7-0.8 & 0.8-1.0 & 1.1-1.2 & 1.2-1.5 \\
    \hline
    $9.2 - 9.6$ &  -0.2$\pm$0.03 &  -0.25$\pm$0.02 &  -0.27$\pm$0.01 &  -0.18$\pm$0.02 &  -0.24$\pm$0.02 &  -0.22$\pm$0.05 &  -0.26$\pm$0.05 &  -0.28$\pm$0.02 &  -0.2$\pm$0.02 &  -0.17$\pm$0.02 &  -0.07$\pm$0.03   \\
    $9.6 - 9.9$ &  -0.15$\pm$0.04 &  -0.29$\pm$0.02 &  -0.19$\pm$0.02 &  -0.23$\pm$0.02 &  -0.27$\pm$0.02 &  -0.3$\pm$0.03 &  -0.32$\pm$0.02 &  -0.25$\pm$0.02 &  -0.17$\pm$0.02 &  -0.08$\pm$0.02 &  -0.13$\pm$0.05   \\
    $9.9 - 10.1$ &  -0.06$\pm$0.02 &  -0.08$\pm$0.02 &  -0.15$\pm$0.01 &  -0.09$\pm$0.01 &  -0.25$\pm$0.01 &  -0.36$\pm$0.01 &  -0.35$\pm$0.01 &  -0.33$\pm$0.01 &  -0.21$\pm$0.01 &  -0.18$\pm$0.01 &  -0.13$\pm$0.01   \\
    $10.1 - 10.3$ &  -0.0$\pm$0.01 &  0.07$\pm$0.01 &  -0.1$\pm$0.01 &  -0.01$\pm$0.01 &  -0.0$\pm$0.0 &  -0.15$\pm$0.01 &  -0.21$\pm$0.01 &  -0.13$\pm$0.01 &  -0.29$\pm$0.01 &  -0.14$\pm$0.01 &  -0.22$\pm$0.01   \\
    $10.3 - 10.6$ &  0.02$\pm$0.01 &  0.04$\pm$0.0 &  0.07$\pm$0.0 &  0.1$\pm$0.0 &  -0.02$\pm$0.0 &  0.0$\pm$0.01 &  -0.07$\pm$0.01 &  -0.11$\pm$0.01 &  -0.31$\pm$0.01 &  -0.29$\pm$0.01 &  -0.25$\pm$0.01   \\
    $10.6 - 11.2$ &  0.05$\pm$0.01 &  0.05$\pm$0.0 &  0.08$\pm$0.0 &  0.07$\pm$0.01 &  0.06$\pm$0.01 &  0.07$\pm$0.01 &  0.03$\pm$0.01 &  -0.06$\pm$0.01 &  -0.05$\pm$0.01 &  0.1$\pm$0.01 &  -0.16$\pm$0.02   \\
    \hline
    [Ti/Fe]				& 0.0-0.1 & 0.1-0.2 & 0.2-0.3 & 0.3-0.4 & 0.4-0.5 & 0.5-0.6 & 0.6-0.7 & 0.7-0.8 & 0.8-1.0 & 1.1-1.2 & 1.2-1.5 \\
    \hline
    $9.2 - 9.6$ &  0.0$\pm$0.03 &  0.18$\pm$0.02 &  0.06$\pm$0.01 &  0.05$\pm$0.02 &  0.15$\pm$0.03 &  0.33$\pm$0.06 &  0.24$\pm$0.07 &  0.05$\pm$0.03 &  0.13$\pm$0.02 &  0.44$\pm$0.05 &  0.45$\pm$0.05   \\
    $9.6 - 9.9$ &  0.07$\pm$0.06 &  -0.07$\pm$0.02 &  -0.09$\pm$0.02 &  -0.06$\pm$0.02 &  -0.19$\pm$0.02 &  -0.2$\pm$0.03 &  -0.17$\pm$0.02 &  -0.13$\pm$0.02 &  -0.02$\pm$0.02 &  -0.08$\pm$0.02 &  0.29$\pm$0.07   \\
    $9.9 - 10.1$ &  -0.11$\pm$0.02 &  -0.13$\pm$0.02 &  -0.03$\pm$0.01 &  -0.17$\pm$0.01 &  -0.25$\pm$0.01 &  -0.31$\pm$0.01 &  -0.35$\pm$0.01 &  -0.33$\pm$0.01 &  -0.01$\pm$0.01 &  -0.03$\pm$0.01 &  -0.13$\pm$0.01   \\
    $10.1 - 10.3$ &  -0.03$\pm$0.01 &  -0.01$\pm$0.01 &  -0.18$\pm$0.01 &  -0.06$\pm$0.01 &  -0.13$\pm$0.0 &  -0.12$\pm$0.01 &  -0.21$\pm$0.01 &  -0.21$\pm$0.01 &  -0.29$\pm$0.01 &  -0.16$\pm$0.01 &  -0.27$\pm$0.01   \\
    $10.3 - 10.6$ &  0.11$\pm$0.01 &  0.11$\pm$0.01 &  0.18$\pm$0.01 &  0.04$\pm$0.01 &  0.09$\pm$0.01 &  0.09$\pm$0.01 &  -0.05$\pm$0.01 &  -0.05$\pm$0.01 &  -0.14$\pm$0.01 &  -0.29$\pm$0.01 &  -0.25$\pm$0.01   \\
    $10.6 - 11.2$ &  0.09$\pm$0.01 &  0.09$\pm$0.01 &  0.12$\pm$0.01 &  0.17$\pm$0.01 &  0.12$\pm$0.01 &  0.11$\pm$0.01 &  0.06$\pm$0.02 &  -0.11$\pm$0.01 &  -0.23$\pm$0.01 &  -0.1$\pm$0.01 &  -0.16$\pm$0.02   \\
    \hline
\end{longtable}

\bsp	
\label{lastpage}
\end{landscape}


\end{document}